\documentclass[useAMS,usenatbib,usegraphicx]{mn2e}

\usepackage{psfig}

\def\aj{AJ}%
\def\apj{ApJ}%
\def\apjl{ApJ}%
\def\apjs{ApJS}%
\def\aap{A\&A}%
\def\mnras{MNRAS}%
\def\na{New A}%
\title[Infrared properties of blazars]
{Infrared properties of blazars: putting the GASP-WEBT sources into context
\thanks{The data collected by the GASP-WEBT collaboration are stored in the GASP-WEBT archive; for questions regarding their availability, please contact the WEBT President Massimo Villata ({\tt villata@oato.inaf.it}).}
}
\author[C. M. Raiteri et al.]
{C.~M.~Raiteri,$^{ 1}$\thanks{E-mail:raiteri@oato.inaf.it}
M.~Villata,$^{ 1}$
M.~I.~Carnerero,$^{1,2,3}$
J.~A.~Acosta-Pulido,$^{2,3}$
\newauthor
V.~M.~Larionov,$^{4,5,6}$
F.~D'Ammando,$^{7,8}$
M.~J.~Ar\'evalo,$^{2,3}$
A.~A.~Arkharov,$^{5}$
\newauthor
A.~Bueno~Bueno,$^{2,3}$
A.~Di~Paola,$^{9}$
N.~V.~Efimova,$^{5}$
P.~A.~Gonz\'alez-Morales,$^{2,3}$
\newauthor
D.~L.~Gorshanov,$^{5}$
A.~B.~Grinon-Marin,$^{2,3}$
C.~L\'azaro,$^{2,3}$
A.~Manilla-Robles,$^{3}$
\newauthor
A.~Pastor Yabar,$^{2,3}$
I. Puerto Gim\'enez,$^{2,3}$ 
and S.~Velasco,$^{2,3}$
\\
$^{ 1}$INAF, Osservatorio Astrofisico di Torino, via Osservatorio 20, I-10025 Pino Torinese, Italy\\
$^{ 2}$Instituto de Astrofisica de Canarias (IAC), E-38200 La Laguna, Tenerife, Canary Islands, Spain\\
$^{ 3}$Departamento de Astrofisica, Universidad de La Laguna, E-38205 La Laguna, Tenerife, Canary Islands, Spain\\
$^{ 4}$Astron.\ Inst., St.-Petersburg State Univ., Universitetsky pr., 28, 198504, Petrodvorets, St.Petersburg, Russia\\
$^{ 5}$Pulkovo Observatory, Pulkovskoe sh., 65, 196140, St.-Petersburg, Russia \\
$^{ 6}$Isaac Newton Institute of Chile, St.-Petersburg Branch\\
$^{ 7}$INAF, Istituto di Radioastronomia, I-40129 Bologna, Italy\\
$^{ 8}$DIFA, Università di Bologna, Viale B.\ Pichat 6/2, I-40127 Bologna, Italy\\
$^{ 9}$INAF, Osservatorio Astronomico di Roma, Via di Frascati, 33, I-00040 Monte Porzio Catone, Italy\\
}
\begin{document}

\maketitle

\label{firstpage}

\begin{abstract}
The infrared properties of blazars can be studied from the statistical point of view with the help of sky surveys, like that provided by the Wide-field Infrared Survey Explorer ({\em WISE}) and the Two Micron All Sky Survey (2MASS). However, these sources are known for their strong and unpredictable variability, which can be monitored for a handful of objects only. 
In this paper we consider the 28 blazars (14 BL Lac objects and 14 flat-spectrum radio quasars, FSRQs) that are regularly monitored by the GLAST-AGILE Support Program (GASP) of the Whole Earth Blazar Telescope (WEBT) since 2007. 
They show a variety of infrared colours, redshifts, and infrared--optical spectral energy distributions (SEDs), and thus represent an interesting mini-sample of bright blazars that can be investigated in more detail.
We present near-IR light curves and colours obtained by the GASP from 2007 to 2013, and discuss the infrared--optical SEDs.
These are analysed with the aim of understanding the interplay among different emission components.
BL Lac SEDs are accounted for by synchrotron emission plus an important contribution from the host galaxy in the closest objects, and dust signatures in 3C 66A and Mkn 421. FSRQ SEDs require synchrotron emission with the addition of a quasar-like contribution, which includes radiation from a generally bright accretion disc ($\nu L_\nu$ up to $\sim 4 \times 10^{46} \rm \, erg \, s^{-1}$), broad line region, and a relatively weak dust torus.
\end{abstract}

%

\section{Introduction}

BL Lacertae objects (BL Lacs) and flat-spectrum radio quasars (FSRQs) make up the two classes in which the active galactic nuclei (AGNs) known as ``blazars" are divided.
They share extreme properties, as noticeable flux variability at all wavelengths, from the radio to the $\gamma$-ray energies, and on a variety of timescales, ranging from hours to years. Their radio-to-UV (in some cases also X-ray) emission is mainly non-thermal, polarised synchrotron radiation from a relativistic jet pointing at a small angle to the line of sight, which explains the superluminal motions inferred from the radio images and the extremely high brightness temperature implied by fast radio variability.

According to the classical definition by \citet{sti91} and \citet{sto91}, the distinguishing feature of BL Lacs is that their spectra show optical emission lines with a rest-frame equivalent width smaller than 5 \AA, if any. This classification is however unsatisfactory, as objects can change class depending on the level of non-thermal continuum from the jet. Even BL Lacertae itself was found not to behave as a BL Lac in faint states \citep{ver95,cor00,cap10}.
New criteria to separate BL Lacs from FSRQs have recently been proposed  by \citet{ghi11} and \citet{gio12a}, which are based on the luminosity of the broad-line region in Eddington units and the ionization state, respectively.

Differences between the two blazar classes were observed in the wavelength-dependent behaviour of the optical polarisation, when present \citep[see e.g.][]{smi96}. BL Lacs show increasing polarisation toward the blue, likely due to causes intrinsic to the jet emitting region. An opposite behaviour characterises FSRQs, because of the dilution effect toward the blue produced by the unpolarised thermal radiation from the accretion disc \citep{smi96,rai12}.
Moreover, Very Long Baseline Interferometry (VLBI) observations revealed different polarisation structures, with BL Lacs exhibiting polarisation position angles in the radio knots parallel, while FSRQs perpendicular, to the jet structural axis \citep[e.g.][]{gab92}.

From the point of view of parsec-scale jet morphology and kinematics, \citet{kar12} found that BL Lacs have in general wider and more bent jets than FSRQs, and that transverse motion of inner knots in BL Lacs is more pronounced than the radial one, which would suggest a helical jet structure.
Variations of the innermost jet position angle in time are larger in FSRQs than in BL Lacs \citep{lister_2013}.

Besides the classification into FSRQs and BL Lacs, the latter sources are further divided into low-energy peaked (LBL) and high-energy peaked (HBL) BL Lacs, depending on the frequency at which the synchrotron emission component peaks in the spectral energy distribution (SED). The relation between LBLs and HBLs is still a matter of debate \citep[e.g.][]{lau99}.

Observations by the Wide-field Infrared Survey Explorer ({\em WISE}) satellite \citep{wri10} led to the development of a new diagnostic tool for AGN. In particular, it was noticed that blazars occupy a well-defined region of the $\rm W1-W2$ versus $\rm W2-W3$  plane\footnote{The {\em WISE} filters W1, W2, W3, and W4 have isophotal wavelengths of about 3.4, 4.6, 12, and 22 $\mu$m, respectively. These $\lambda_{\rm iso}$, together with the magnitude zero-points, are calibrated with respect to Vega \citep{wri10}.}: the so-called {\em WISE} blazar strip \citep{masf11}, which includes a subregion delineating $\gamma$-ray emitting blazars \citep{masf12}. This was used by \citet{dab13} and \citet{masf13} to recognise blazar candidates among the unidentified $\gamma$-ray sources detected by the Large Area Telescope (LAT) onboard the {\em Fermi} satellite. 

Using {\em WISE} data, \citet{plo12} concluded that there is no evidence of a dust torus in BL Lacs, which implies structural differences between the two types of blazars, possibly driven by different accretion rate regimes.

Statistical analysis of the blazar properties is facilitated by the presence of an on-line catalogue of blazars, the Roma-BZCAT\footnote{BZCAT is available on line at the ASDC website {http://www.asdc.asi.it/bzcat}.} \citep{mas09}, as well as of many catalogues resulting from multifrequency surveys, from both ground and space, as the Sloan Digital Sky Survey\footnote{http://www.sdss.org/} (SDSS), the Two Micron All Sky Survey\footnote{http://www.ipac.caltech.edu/2mass/} (2MASS), or the {\em WISE} All-Sky Database\footnote{http://irsa.ipac.caltech.edu/}. 
However, generally surveys give information on a single state of each source, while blazars are rapidly and unpredictably variable objects at all frequencies.

   \begin{figure*}
   \centering
   \includegraphics[width=15cm]{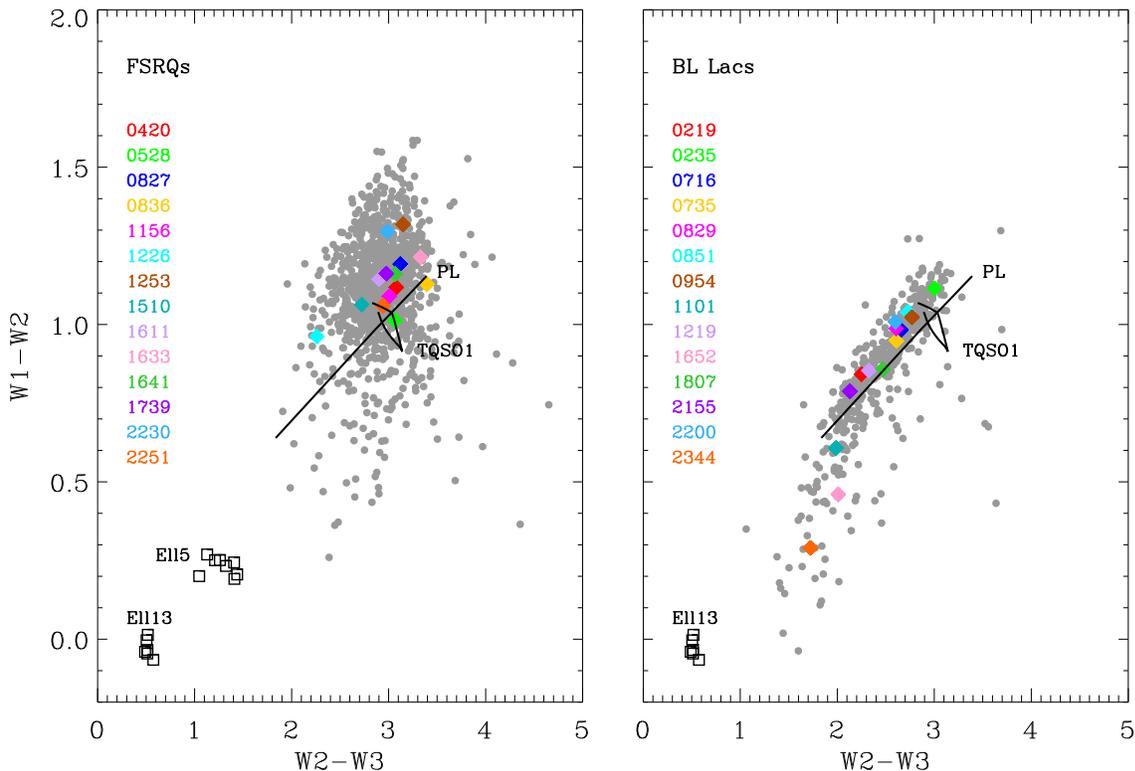}
   \caption{Colour-colour diagrams built with {\em WISE} data for FSRQs (left) and BL Lacs (right). Only objects with $S/N>3$ in all {\em WISE} bands were considered. The points corresponding to the 28 sources monitored by the GASP-WEBT are highlighted with coloured diamonds. The black squares in the bottom left represent the SWIRE templates of 5 and 13 Gyr old elliptical galaxies at different values of redshift.
The locations of Type 1 QSO spectra (TQSO1 SWIRE template) of various redshift, and power-law (PL) spectra with different spectral index are also plotted.}
    \label{wise_3}
    \end{figure*}

On the other hand, the long-term multifrequency behaviour of blazars can be studied in detail only for a handful of sources, and this is usually achieved by large collaborations that share the observing effort.
The Whole Earth Blazar Telescope\footnote{http://www.oato.inaf.it/blazars/webt/} (WEBT) was born in 1997 as an international collaboration of astronomers devoted to blazar studies.
The advent of the new-generation satellites for $\gamma$-ray astronomy, Astro-rivelatore Gamma ad Immagini Leggero (AGILE) and {\it Fermi} (formerly GLAST), led the WEBT to organise a long-term monitoring of 28 blazars, 14 BL Lacs and 14 FSRQs, which were selected among the $\gamma$-loud, optically brightest and potentially most interesting sources for coordinated low-energy and $\gamma$-ray studies. 
This long-term monitoring project was called the GLAST-AGILE Support Program (GASP) and started its activity in 2007 \citep[see e.g.][]{vil08}.

In this paper we first put into context the GASP sources by comparing their infrared properties with those of the other classified blazars.
Then we present the results of the near-IR monitoring of the GASP sources performed at the Campo Imperatore (Italy) and Teide (Canary Islands, Spain) observatories in 2008--2013.
We analyse their flux density variability, colours and SEDs with special attention to the differences between BL Lac objects and FSRQs.

\section{The GASP sources into context}
\label{compa}
Tables \ref{gasp_bllacs} and \ref{gasp_fsrqs} list the GASP sources, divided into the two blazars classes, according to the classical definition adopted in the BZCAT catalogue.
For each source we give in Column 1 the IAU name, in Column 2 the name with which the source is most often indicated in the literature, and in Column 3 the BZCAT identification. The source redshift is reported in Column 4. We adopted the redshift values of the BZCAT catalogue, but in the case of 0716+714, for which we set $z=0.31$ following \citet{nil08}, and of 0735+178, for which we set $z=0.45$ after \citet{nil12}. Finally, Column 5 gives the Galactic reddening obtained from the Galactic Dust Extinction Service at IRSA\footnote{http://irsa.ipac.caltech.edu/applications/DUST/}.
Reddening is particularly important ($>0.1$) for the BL Lacs 0954+658, 2200+420, and 2344+514, and for the FSRQs 0420$-$014, 0528+134, 1510$-$089, and 2251+158.

\begin{table*}
\begin{minipage}{130mm}
\caption{The BL Lac objects in the GASP target list.}
\label{gasp_bllacs}
\begin{tabular}{lllcc}
\hline
IAU Name & Other Name  & BZCAT Name    & $z$ & $E(B-V)$\\
\hline
0219+428 & 3C 66A      & BZBJ0222+4302 &  0.444 & 0.0847\\
0235+164 & AO 0235+16  & BZBJ0238+1636 &  0.940 & 0.0797\\
0716+714 & S5 0716+71  & BZBJ0721+7120 &  0.310 & 0.0312\\
0735+178 & PKS 0735+17 & BZBJ0738+1742 &  0.450 & 0.0339\\
0829+046 & OJ 49       & BZBJ0831+0429 &  0.174 & 0.0329\\
0851+202 & OJ 287      & BZBJ0854+2006 &  0.306 & 0.0283\\
0954+658 & S4 0954+65  & BZBJ0958+6533 &  0.367 & 0.1197\\
1101+384 & Mkn 421     & BZBJ1104+3812 &  0.030 & 0.0153\\
1219+285 & ON 231      & BZBJ1221+2813 &  0.102 & 0.0233\\
1652+398 & Mkn 501     & BZBJ1653+3945 &  0.033 & 0.0190\\
1807+698 & 3C 371      & BZBJ1806+6949 &  0.046 & 0.0340\\
2155$-$304&PKS 2155$-$304& BZBJ2158$-$3013 &  0.116 & 0.0219\\
2200+420 & BL Lacertae & BZBJ2202+4216 &  0.069 & 0.3280\\
2344+514 & 1ES 2344+514& BZBJ2347+5142 &  0.044 & 0.2097\\
\hline
\end{tabular}
\end{minipage}
\end{table*}

\begin{table*}
\begin{minipage}{130mm}
\caption{The FSRQs in the GASP target list.}
\label{gasp_fsrqs}
\begin{tabular}{lllcc}
\hline
IAU Name & Other Name     & BZCAT Name    & $z$ & $E(B-V)$ \\
\hline
0420$-$014 &PKS 0420$-$01& BZQJ0423$-$0120 &  0.916 & 0.1258\\
0528+134 & PKS 0528+134 & BZQJ0530+1331 &  2.070 & 0.8450\\
0827+243 & OJ 248       & BZQJ0830+2410 &  0.939 & 0.0333\\
0836+710 & 4C 71.07     & BZQJ0841+7053 &   2.218 & 0.0301\\
1156+295 & 4C 29.45     & BZQJ1159+2914 &  0.729 & 0.0199\\
1226+023 & 3C 273       & BZQJ1229+0203 &  0.158 & 0.0206\\
1253$-$055 & 3C 279     & BZQJ1256$-$0547 &  0.536 & 0.0286\\
1510$-$089 &PKS 1510$-$08& BZQJ1512$-$0905 &  0.360 & 0.1010\\
1611+343 & DA 406       & BZQJ1613+3412 &  1.397 & 0.0178\\
1633+382 & 4C 38.41     & BZQJ1635+3808 &  1.814 & 0.0122\\
1641+399 & 3C 345       & BZQJ1642+3948 &   0.593 & 0.0131\\
1739+522 & 4C 51.37     & BZQJ1740+5211 &  1.381 & 0.0355\\
2230+114 & CTA 102      & BZQJ2232+1143 &  1.037 & 0.0718\\
2251+158 & 3C 454.3     & BZQJ2253+1608 &   0.859 & 0.1078\\
\hline
\end{tabular}
\end{minipage}
\end{table*}

As mentioned in the Introduction, the IR survey undertaken by the {\em WISE} satellite provided a new tool to identify blazars.
We searched the {\em WISE} All-Sky Source Catalog for the BZCAT sources. 
A cone search radius of 3 arcsec was found to be the best choice between having reliable identifications and completeness.
Among the 1180 BL Lacs present in the BZCAT catalogue, we obtained 1122 {\em WISE} counterparts (95\%), while we found 1608 identifications for the 1676 BZCAT FSRQS\footnote{Actually, there were 1610 FSRQs {\em WISE} counterparts, but two sources, BZQJ0607-0834 and BZQJ0917-2131, had double identifications. For these two objects we kept only the closest {\em WISE} counterpart.} (96\%).
All of the GASP sources have a {\em WISE} counterpart.

Figure \ref{wise_3} shows the colour-colour {\em WISE} diagram for the BZCAT blazars, separating FSRQs from BL Lac objects.
We assume that Galactic extinction is negligible at {\em WISE} wavelengths. 
The points referring to the GASP sources are overplotted in colour.
The GASP FSRQs cover the region where most sources cluster and the GASP BL Lacs distribute along almost the whole BL Lac sequence extending toward the elliptical galaxies location. 
All the GASP sources were detected in $\gamma$ rays by {\em Fermi}-LAT \citep{nol12} 
and they all lie within the {\em WISE} Gamma-ray Strip \citep[][see also \citealt{dab12}]{masf12}, when considering the refined analysis by \citet{dab13} and \citet{masf13b}.
In the figure the position of elliptical galaxies is indicated, using the SWIRE template\footnote{http://www.iasf-milano.inaf.it/$\sim$polletta/templates/ swire\_templates.html} \citep{polletta2007} of a 13 Gyr old elliptical galaxy for redshifts between 0 and 0.25, and that of a 5 Gyr galaxy for redshifts from 0.3 to 1. We notice that for a given galaxy age, i.e.\ a given stellar population, redshift variations result in a small scatter of the points, while changing the model produces a larger shift in the diagram.
We also show the trace left by the SWIRE TQSO1 template, i.e.\ a type 1 QSO with broad emission lines in the optical spectrum and prominent torus, for redshift values in the range $z=0$--3.
Finally, the straight line is obtained from a power-law spectrum $F_\nu \propto \nu^{-\alpha}$ with spectral index $\alpha=0$--1.5, and represents the typical non-thermal, synchrotron emission from blazar jets. The range of considered $\alpha$ includes both flat ($\alpha < 1$) spectra, as most BL Lac objects show, and steep ($\alpha >1$) spectra, which characterize most FSRQs (see below). 
Deviation from the power-law line means that the source spectrum is not a pure synchrotron spectrum, but that either it is contaminated by other emission contributions (from host galaxy and QSO-like nucleus) or it presents a curvature in the {\em WISE} bands. This will be studied in depth in Sect.\ \ref{sed}.

   \begin{figure}
   \centering
   \resizebox{\hsize}{!}{\includegraphics{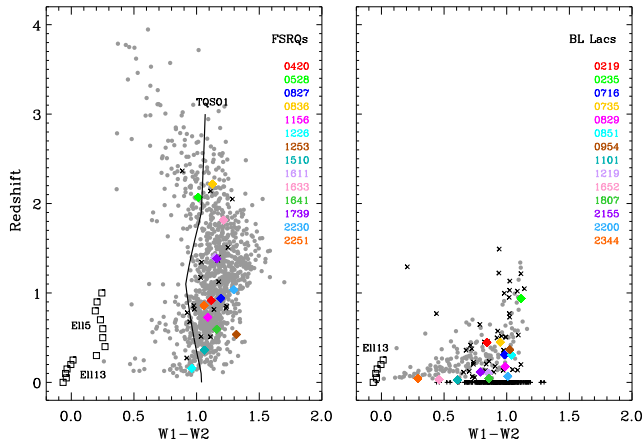}}
   \caption{The redshift versus {\em WISE} W1-W2 colour for FSRQs (left) and BL Lacs (right). Only objects with $S/N>3$ in all {\em WISE} bands were considered. Black crosses indicate sources with uncertain redshift; black plus signs those with undetermined $z$. The points corresponding to the 28 sources monitored by the GASP-WEBT are highlighted with coloured diamonds. The black squares in the bottom left represent the SWIRE templates of 5 and 13 Gyr old elliptical galaxies at different values of redshift.
The location of Type 1 QSO spectra (TQSO1 SWIRE template) of various redshift is also plotted.}
    \label{wise_z}
    \end{figure}

In Fig.\ \ref{wise_z} we plotted the redshift versus the {\em WISE} W1$-$W2 colour. We took into account new BL Lac redshift measurements by \citet{landoni2012}, \citet{landoni2013}, \citet{sandrinelli2013}, and \citet{pita2013}. The GASP FSRQs cover fairly well the strip formed by all the FSRQs, apart for the upper part, corresponding to the most distant objects. As for BL Lacs, the GASP source with higher redshift is 0235+164, but there are only few objects that are more distant, most of them with uncertain $z$.
Here again we plotted the position of elliptical galaxies and TQSO1 spectra of different redshifts.

   \begin{figure}
   \centering
   \resizebox{\hsize}{!}{\includegraphics{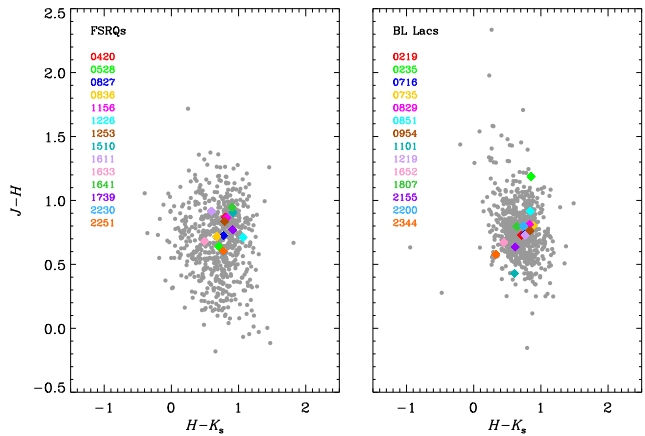}}
   \caption{Colour-colour diagram built with 2MASS data for FSRQs (left) and BL Lac objects (right). 
The 2MASS magnitudes have been corrected for Galactic reddening according to \citet{sch98}.
GASP sources are overplotted with coloured diamonds.}
    \label{2mass_ext}
    \end{figure}

   \begin{figure}
   \centering
   \resizebox{\hsize}{!}{\includegraphics{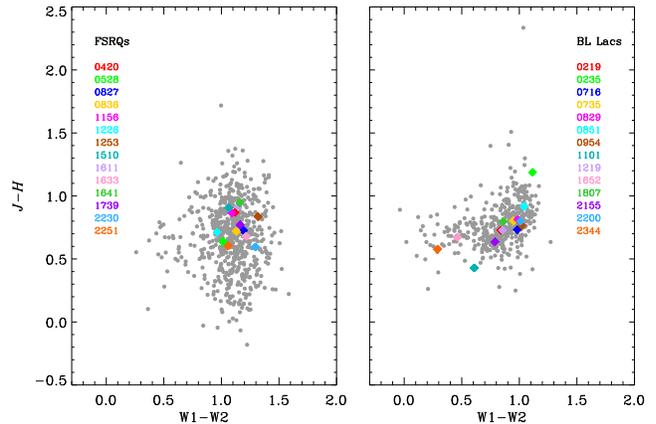}}
   \caption{Colour-colour diagram built with {\em WISE} and 2MASS data for FSRQs (left) and BL Lac objects (right).
The 2MASS magnitudes have been corrected for Galactic reddening according to \citet{sch98}.
GASP sources are overplotted with coloured diamonds.}
    \label{wise_2mass_ext}
    \end{figure}

The same search radius of 3 arcsec was adopted to query the 2MASS catalogue for the BZCAT objects. This is the maximum radial offset between the {\em WISE} sources and associated 2MASS sources in the {\em WISE} catalogue.
The 2MASS catalogue includes sources brighter than about 1 mJy in the $J$, $H$, and $K_{\rm s}$ bands, with signal-to-noise ratio ($S/N$) greater than 10.
We found 815 2MASS counterparts in the case of the BL Lac objects (69\%), and 686 for the FSRQs (41\%).
All the GASP sources have a 2MASS counterpart.
We corrected for the Galactic extinction according to \citet{sch98}, i.e.\ using the relative extinction values $A/E(B-V)=0.902$, 0.576, and 0.367 for the $J$, $H$, and $K_{\rm s}$ bands, respectively, with $E(B-V)$ downloaded from the Galactic Dust Extinction Service at IRSA mentioned above.

Figure  \ref{2mass_ext} shows the colour-colour diagram built with 2MASS data for both FSRQs and BL Lac objects. Combining {\em WISE} and 2MASS information, we obtained the diagram of Fig.\ \ref{wise_2mass_ext}.
In both plots the GASP FSRQs are concentrated in the central region of the diagram, while BL Lac objects are more distributed. 

It is interesting to compare our results on the 2MASS BL Lac colours with those by \citet{che05}. These authors started from the \citet{ver03} catalogue and found 511 BL Lacs with 2MASS counterparts, including uncertain sources. 
In the $J-H$ versus $H-K$ plot, the majority of both the \citet{che05} and our sources lie in the region where both the colour indices values are in the range 0.2--1.2. However, Fig.\ \ref{2mass_ext} shows much more objects with $J-H > 1.2$, and a lack of sources in the region with $H-K$ from $-0.2$ to 0.2 and $J-H$ in the range 0--0.5, where instead there are several objects in the \citet{che05} plot. The latter difference in the colour distribution likely reflects the presence of misclassified objects in the \citet{ver03} catalogue.

\begin{figure*}
   \centering
   \includegraphics[width=15cm]{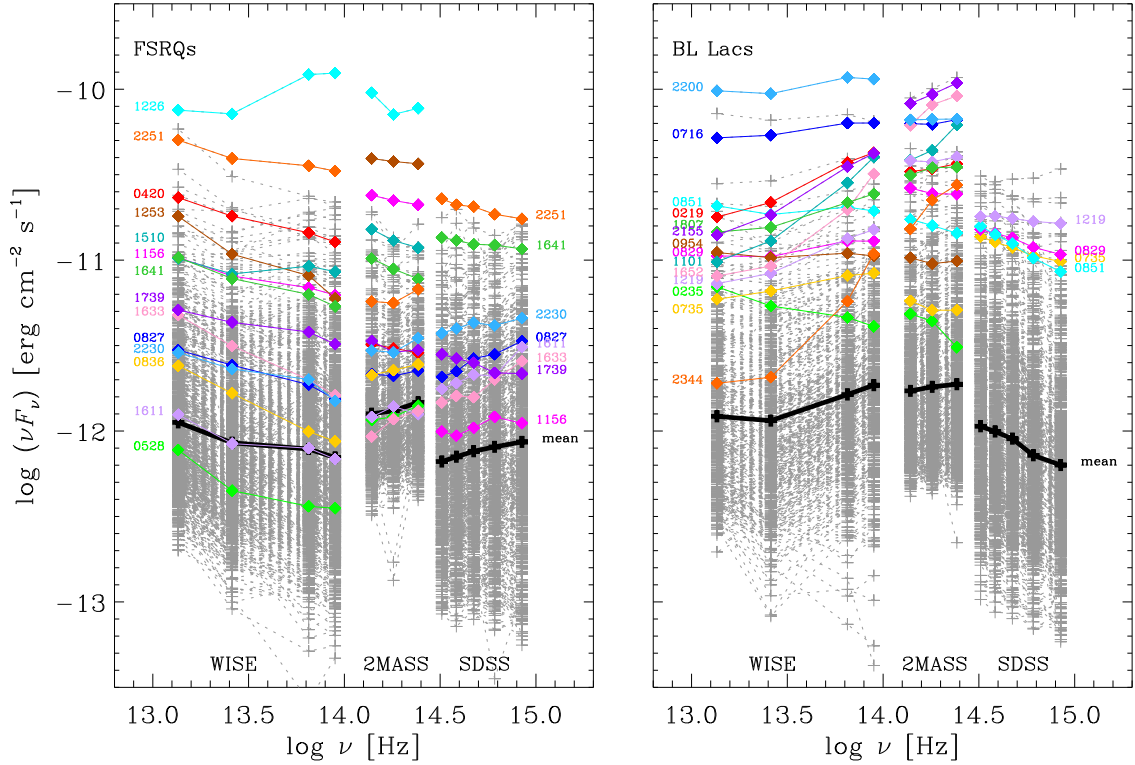}
   \caption{Spectral energy distributions of the BZCAT FSRQs (left) and BL Lac objects (right) built with infrared and optical data from {\em WISE}, 2MASS and SDSS. {\em WISE} and SDSS data are plotted only for objects with $S/N>3$ and magnitudes brighter than the survey limits, respectively, in all bands. The 2MASS and SDSS data are corrected for the Galactic extinction according to \citet{sch98}. Mean SEDs are highlighted with black plus signs; GASP sources are overplotted with coloured diamonds.}
    \label{sed_bzcat}
    \end{figure*}

Finally, we searched the BZCAT FSRQs and BL Lacs in the Sloan Digital Sky Survey (SDSS). To be sure that our objects photometry passed all tests
to be used for science, we set the {\tt CLEAN} flag to one in  the query form. We found 765 BL Lac objects (65\% of the sample) and 794 FSRQs (47\%) for which clean photometric data in the $u$, $g$, $r$, $i$, and $z$ bands are available.
Among them, only 4 BL Lacs and 8 FSRQs belong to the GASP source list.
In Fig.\ \ref{sed_bzcat} we plot the spectral energy distribution (SED) of the BZCAT FSRQs and BL Lacs built with {\em WISE}, 2MASS, and SDSS data. These latter have been corrected for the Galactic extinction in the same way as for the 2MASS data, i.e.\ adopting the $A/E(B-V)$ values given by \citet{sch98} (5.155, 3.793, 2.751, 2.086, and 1.479 from $u$ to $z$), and $E(B-V)$ as specified above. 
Transformation of de-reddened magnitudes into fluxes has been done by using the specific zero-mag flux densities.
To avoid very uncertain data, we plotted only objects for which {\em WISE} data have $S/N>3$ and SDSS magnitudes are less than the survey limits in all bands, which are 22.0, 22.2, 22.2, 21.3, and 20.5 mag in the $u$, $g$, $r$, $i$, and $z$ bands, respectively. 
The SEDs corresponding to the GASP sources are highlighted in colour. They show discontinuities between different datasets as a consequence of variability, since the observations by {\em WISE}, 2MASS, and SDSS were carried out at different epochs, when the sources were evidently in different brightness states.
We also display average SEDs. These are within the GASP SEDs or just below them for FSRQs, while they are quite below the SEDs of the GASP BL Lacs, meaning that the GASP choice of BL Lacs is more biased towards the brightest objects than in the case of FSRQs. 
The average FSRQs SED shows a steep spectrum in the {\em WISE} bands, while it turns into a flat spectrum in the 2MASS and SDSS bands. The upturn signs the transition from a dominant synchrotron emission contribution to the ``big blue bump", which is ascribed to thermal emission from the accretion disc.
In contrast, the average BL Lacs SED follows the synchrotron hump, with the peak between the 2MASS and SDSS bands.

In conclusion, the GASP sources show a variety of infrared colours, redshifts and infrared--optical SEDs, and thus form an interesting mini-sample of blazars that can be analysed in detail and that is likely representative of the bright objects.

\section{Near-IR flux variability}

Regular monitoring of the GASP sources in the near-IR bands is carried out at the Campo Imperatore (Italy) and Teide (Canary Islands, Spain) observatories.

The observations at Campo Imperatore were obtained by using the 1.1~m AZT-24 telescope and the SWIRCAM camera \citep{dal00}. SWIRCAM is based on a Rockwell PICNIC array having 256 $\times$ 256 pixels with a size of 40 $\mu m$, that corresponds to 1.04\arcsec\ on the sky. Every final image is composed from five dithered raw images after preliminary processing, that includes sky subtraction, flat-fielding and recentering\footnote{http://www.oa-roma.inaf.it/preprocess/preprocess.html}. Aperture photometry was made using properly calibrated field stars\footnote{http://www.astro.spbu.ru/staff/vlar/NIRlist.html}.

The observations at Teide Observatory (Canary Islands) were obtained with the 1.52~m Carlos Sanchez Telescope (TCS), using the near-infrared camera CAIN. This camera is equipped with a $256 \times 256$ pixels NICMOS-3 detector and it provides a scale of 1\arcsec/pixel with the wide field optics ($4.2\arcmin \times 4.2\arcmin$). Data were acquired in the three filters $J$, $H$ and $K_{\rm s}$.
Observations were performed using a 5-point dither pattern in order to facilitate a proper sky background subtraction. 
At each point, the exposure time was about 1~min, split in individual exposures of 10~s in the $J$ filter and 6~s in the $H$ and $K_{\rm s}$ filters to avoid saturation by sky brightness. The dither cycle was repeated twice, except for faint sources where the number of cycles was increased. 

Image reduction was performed with the {\tt caindr} package developed by Rafael Barrena-Delgado and Jose Acosta-Pulido under the IRAF environment\footnote{Image Reduction and Analysis Facility -- http://iraf.noao.edu/}. Data reduction includes flat-fielding, sky subtraction, and the shift and combination of all frames taken in the same dither cycle.  
Photometric calibration was performed using a IDL routine kindly provided by P. \'Abrah\'am (Obs. Konkoly - Hungary), which 
uses the 2MASS catalogue \citep{cut03}. The photometric zero point for each combined image was determined by averaging the 
offset between the instrumental and the 2MASS magnitudes of catalogue sources within the field of view. Whenever deviant targets appear they are excluded before computing the average.

\begin{table*}
\label{frac_bllacs}
\centering
\begin{minipage}{108mm}
\caption{Mean fractional variation of the near-IR flux densities, $f_i$, average de-reddened colour indices $<J-H>$ and $<H-K>$ with their standard deviations ($\sigma$) and their mean fractional variations $f_{J-H}$ and $f_{H-K}$ for the BL Lac objects in the GASP target list.}
\begin{tabular}{llllllll}
\hline
IAU Name  & $f_J$ & $f_H$ & $f_K$ & $<J-H>$ & $<H-K>$ & $f_{J-H}$ & $f_{H-K}$\\
\hline
0219+428  & 0.39  & 0.38  & 0.37  & 0.70 (0.04)  & 0.74 (0.04) & -    & -   \\ 
0235+164  & 1.50  & 1.44  & 1.39  & 1.07 (0.09)  & 0.98 (0.07) & 0.07 & 0.05\\ 
0716+714  & 0.49  & 0.35  & 0.33  & 0.83 (0.04)  & 0.89 (0.04) & -    & -   \\ 
0735+178  & 0.24  & 0.27  & 0.27  & 0.86 (0.05)  & 0.83 (0.05) & 0.04 & 0.05\\ 
0829+046  & 0.23  & 0.22  & 0.22  & 0.88 (0.03)  & 0.85 (0.05) & -    & 0.04\\ 
0851+202  & 0.39  & 0.41  & 0.45  & 0.89 (0.05)  & 0.91 (0.05) & 0.04 & 0.01\\ 
0954+658  & 0.45  & 0.41  & 0.40  & 0.90 (0.04)  & 0.93 (0.05) & -    & -   \\ 
1101+384  & 0.27  & 0.26  & 0.27  & 0.77 (0.02)  & 0.25 (0.03) & -    & -   \\ 
1219+285  & 0.32  & 0.33  & 0.31  & 0.72 (0.03)  & 0.75 (0.05) & -    & 0.05\\ 
1652+398  & 0.01  & 0.01  & 0.01  & 0.81 (0.01)  & 0.40 (0.01) & -    & -   \\ 
1807+698  & 0.08  & 0.08  & 0.07  & 0.83 (0.01)  & 0.66 (0.04) & -    & -   \\ 
2155$-$304& 0.20  & 0.20  & 0.21  & 0.61 (0.02)  & 0.80 (0.01) & -    & -   \\ 
2200+420  & 0.39  & 0.38  & 0.39  & 0.83 (0.04)  & 0.79 (0.04) & -    & 0.01\\ 
2344+514  & 0.09  & 0.09  & 0.10  & 0.75 (0.05)  & 0.33 (0.03) & 0.04 & -   \\ 
\hline
\end{tabular}
\end{minipage}
\end{table*}

\begin{table*}
\label{frac_fsrqs}
\centering
\begin{minipage}{108mm}
\caption{Mean fractional variation of the near-IR flux densities, $f_i$, average de-reddened colour indices $<J-H>$ and $<H-K>$ with their standard deviations ($\sigma$) and their mean fractional variations $f_{J-H}$ and $f_{H-K}$ for the FSRQs in the GASP target list.}
\begin{tabular}{llllllllll}
\hline
IAU Name  & $f_J$ & $f_H$ & $f_K$ & $<J-H>$ & $<H-K>$ & $f_{J-H}$ & $f_{H-K}$\\
\hline
0420$-$014& 1.00  & 1.01  & 0.89  & 0.88 (0.08)  & 0.97 (0.09) & 0.07 & 0.07\\ 
0528+134  & 0.55  & 0.65  & 0.64  & 0.63 (0.16)  & 0.90 (0.09) & 0.23 & 0.08\\ 
0827+243  & 0.85  & 0.87  & 0.93  & 0.58 (0.12)  & 0.89 (0.15) & 0.19 & 0.16\\ 
0836+710  & 0.47  & 0.60  & 0.65  & 0.63 (0.13)  & 0.83 (0.13) & 0.20 & 0.13\\ 
1156+295  & 0.73  & 0.68  & 0.69  & 0.88 (0.08)  & 0.90 (0.11) & 0.08 & 0.11\\ 
1226+023  & 0.04  & 0.06  & 0.05  & 0.81 (0.04)  & 1.15 (0.04) & 0.03 & 0.02\\ 
1253$-$055& 0.57  & 0.58  & 0.59  & 0.91 (0.09)  & 0.94 (0.07) & 0.09 & 0.05\\ 
1510$-$089& 0.33  & 0.33  & 0.31  & 0.90 (0.06)  & 1.02 (0.07) & 0.04 & 0.05\\ 
1611+343  & 0.24  & 0.23  & 0.32  & 0.85 (0.08)  & 0.64 (0.16) & 0.08 & 0.23\\ 
1633+382  & 0.64  & 0.63  & 0.58  & 0.90 (0.11)  & 0.94 (0.07) & 0.12 & 0.05\\ 
1641+399  & 0.38  & 0.40  & 0.40  & 0.98 (0.05)  & 0.94 (0.07) & 0.03 & 0.05\\ 
1739+522  & 0.21  & 0.24  & 0.26  & 0.87 (0.09)  & 0.72 (0.18) & 0.09 & 0.23\\ 
2230+114  & 1.12  & 1.27  & 1.25  & 0.59 (0.17)  & 0.86 (0.18) & 0.28 & 0.20\\ 
2251+158  & 1.10  & 1.31  & 1.32  & 0.60 (0.30)  & 0.87 (0.18) & 0.49 & 0.19\\ 
\hline
\end{tabular}
\end{minipage}
\end{table*}

\begin{figure*}
    \vspace{0.5cm}
    \centerline{
    \psfig{figure=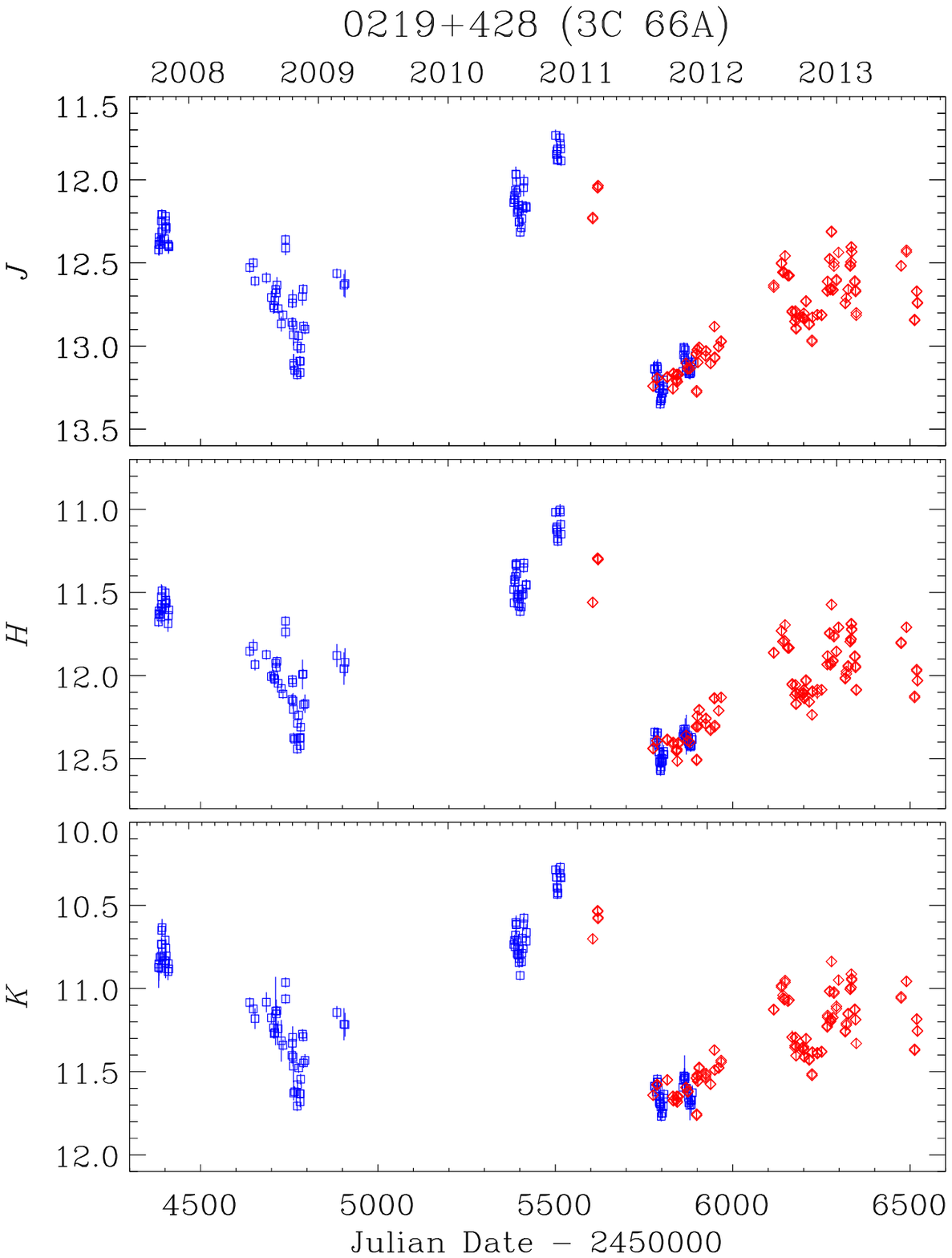,width=0.30\linewidth}
    \psfig{figure=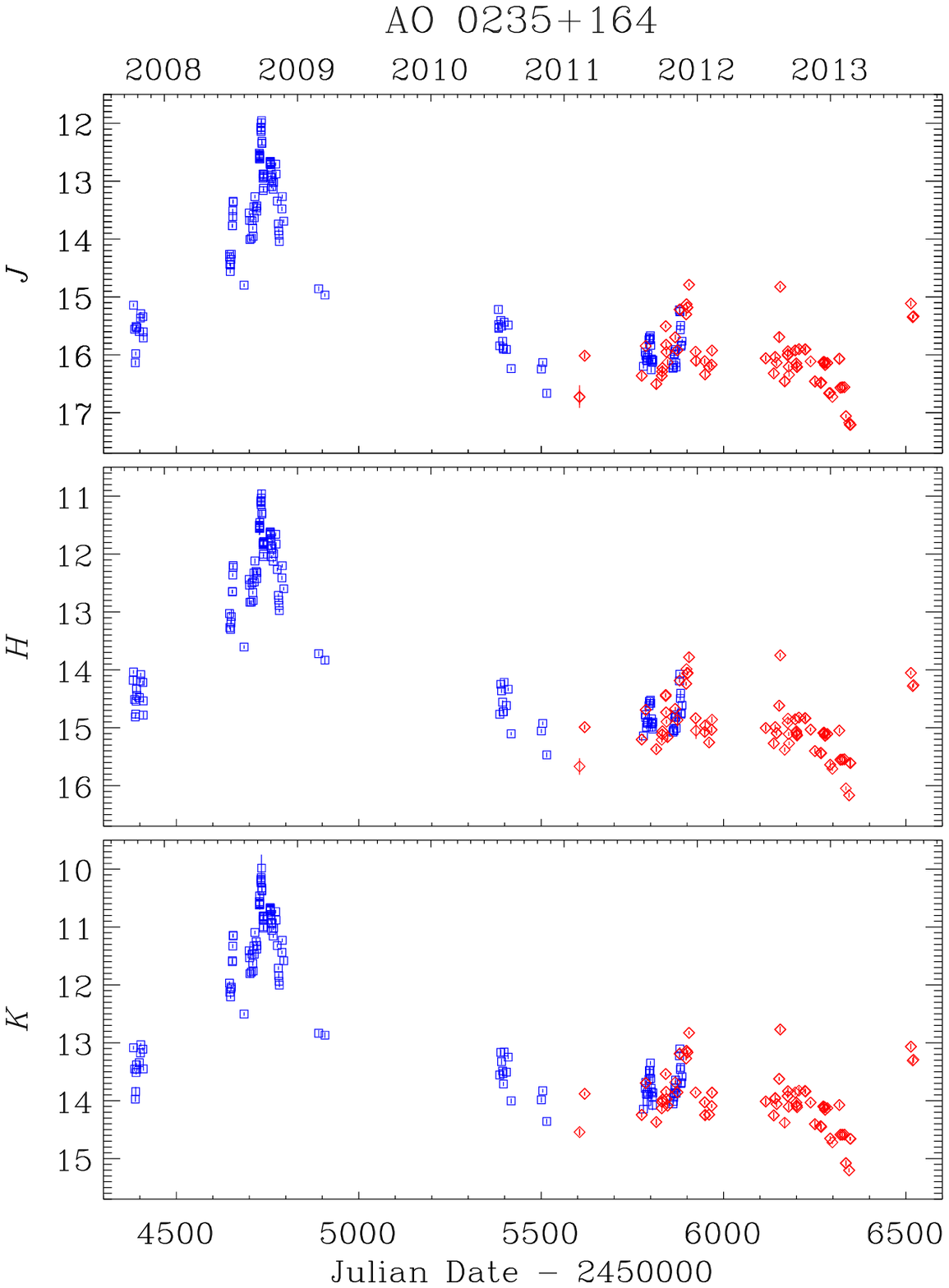,width=0.30\linewidth}
    \psfig{figure=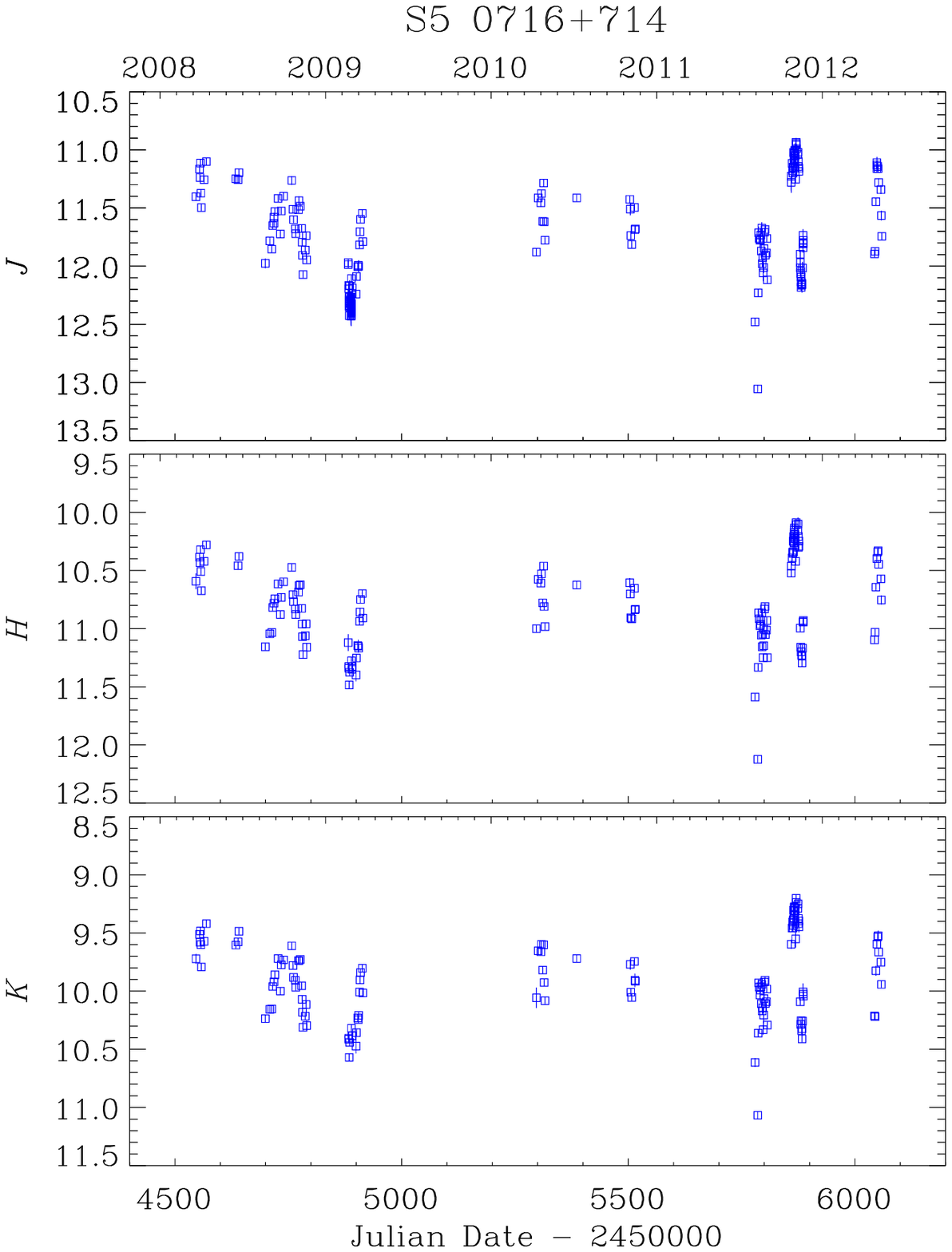,width=0.30\linewidth}}
    \vspace{0.5cm}
    \centerline{
    \psfig{figure=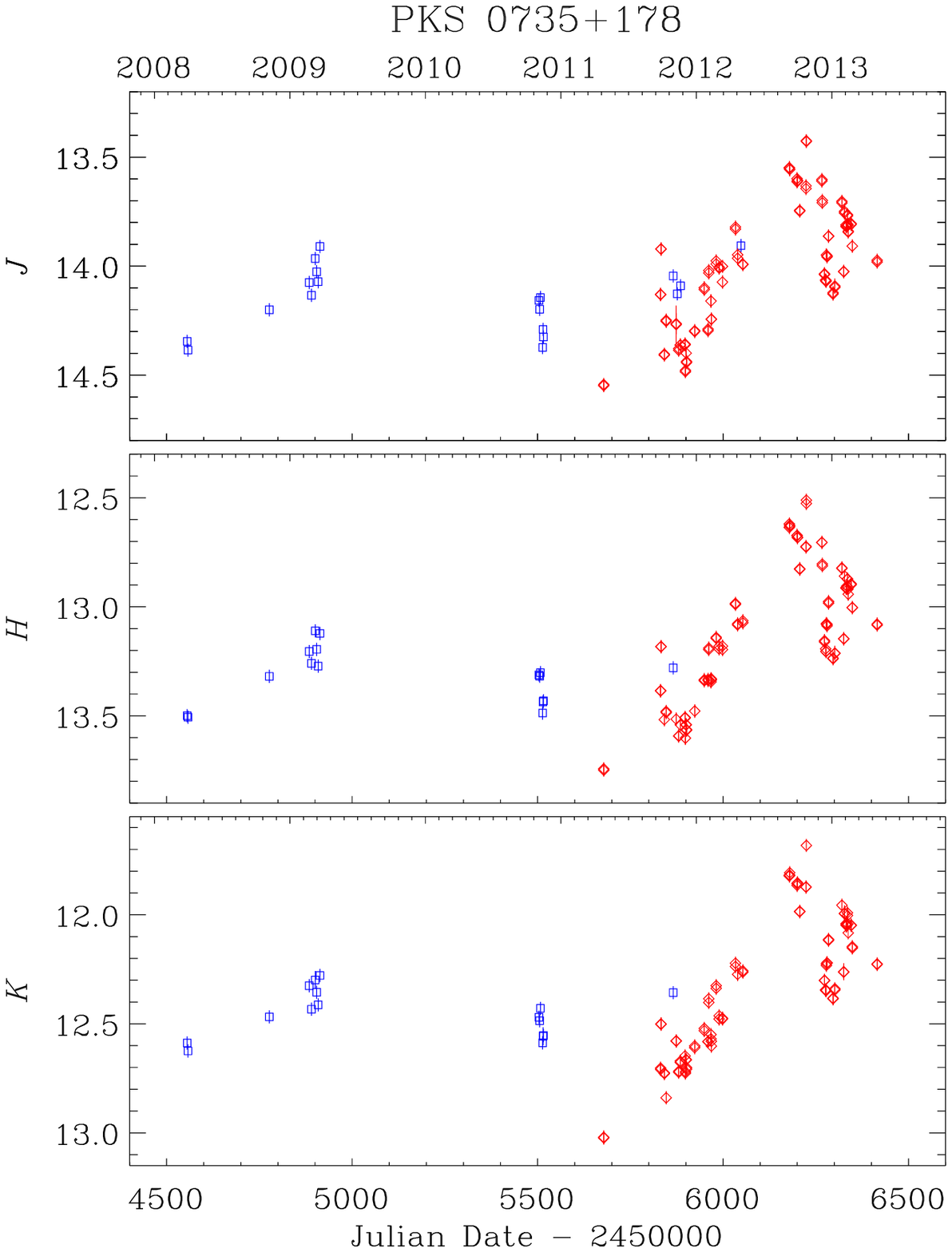,width=0.30\linewidth}
    \psfig{figure=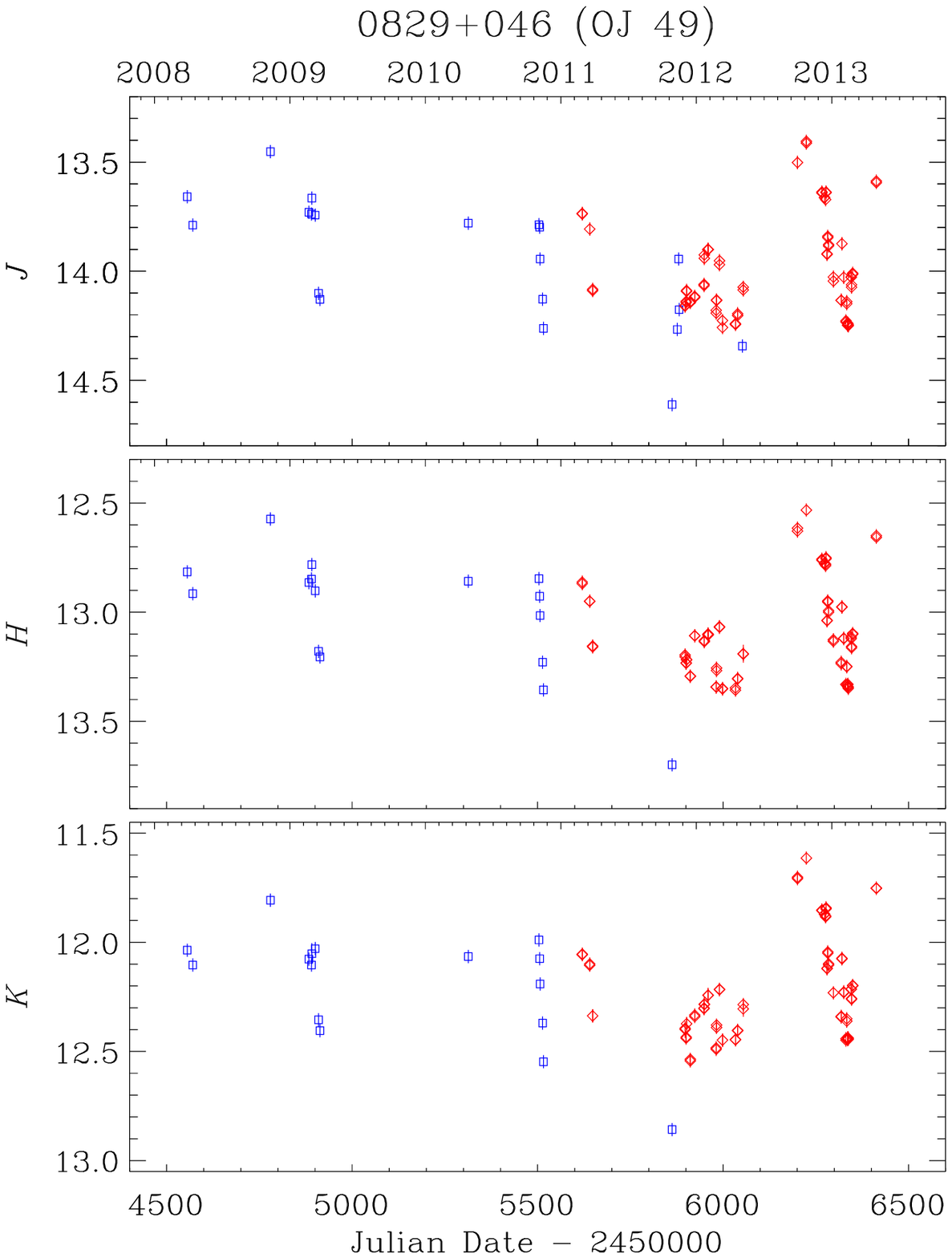,width=0.30\linewidth}
    \psfig{figure=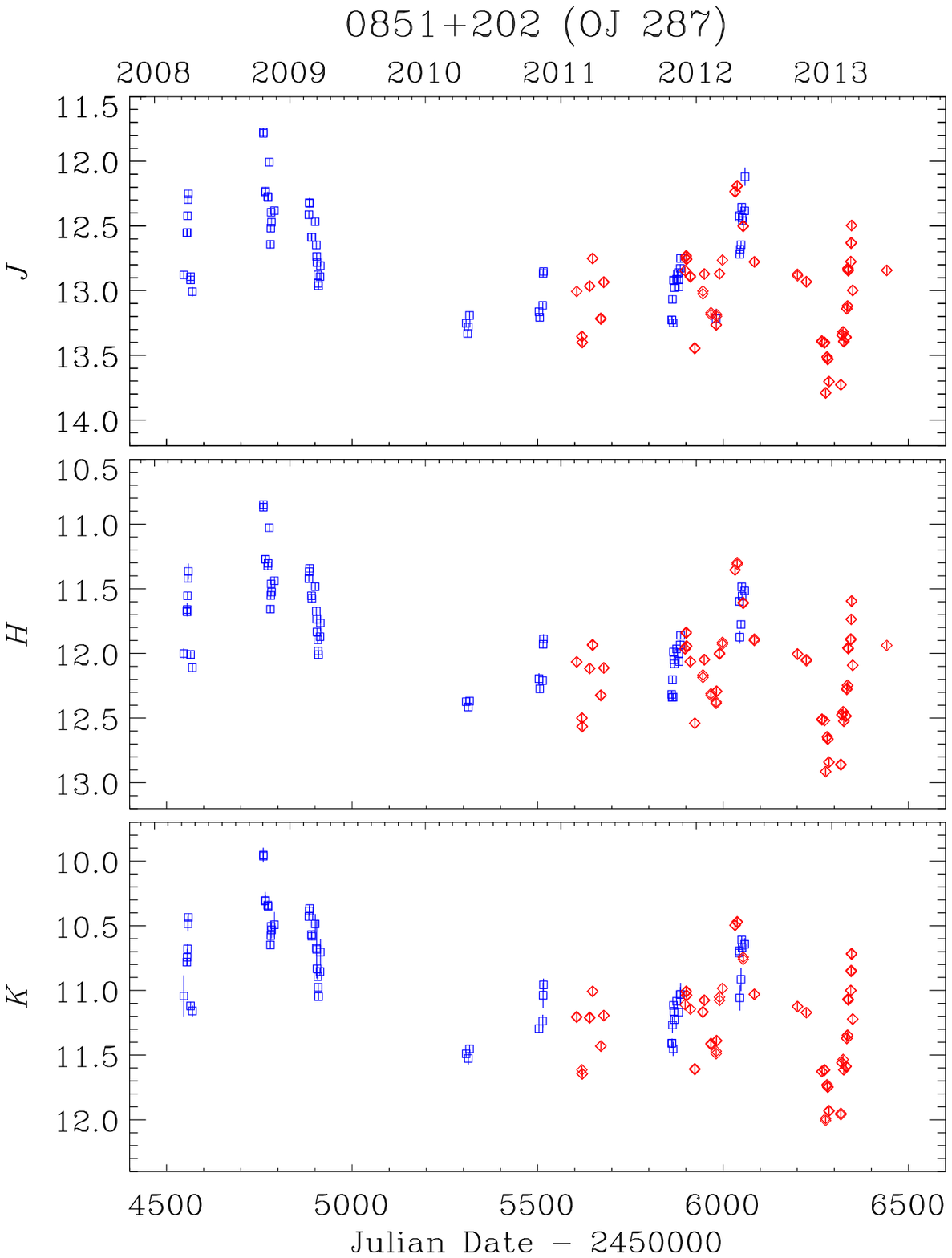,width=0.30\linewidth}}
    \vspace{0.5cm}
    \centerline{
    \psfig{figure=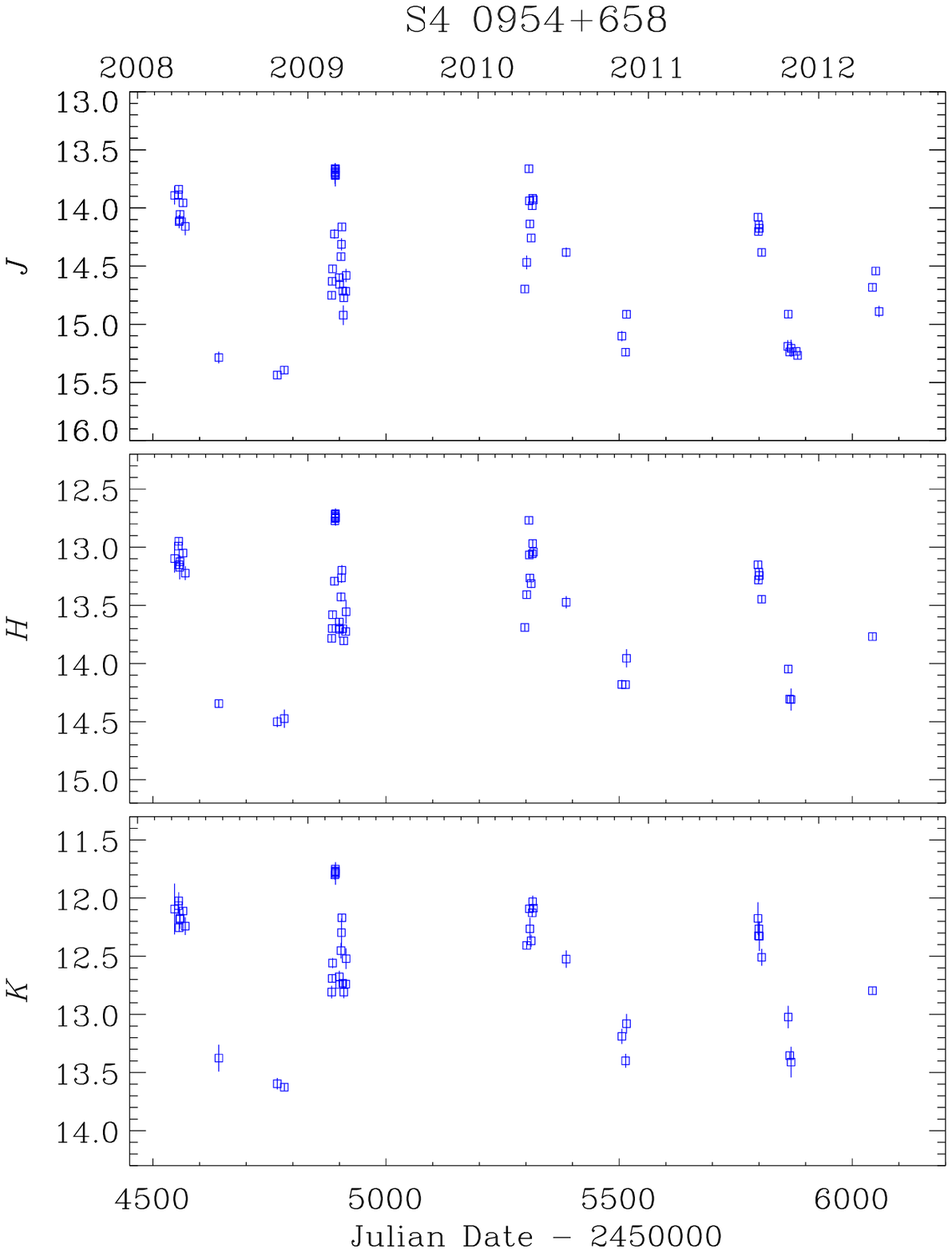,width=0.30\linewidth}
    \psfig{figure=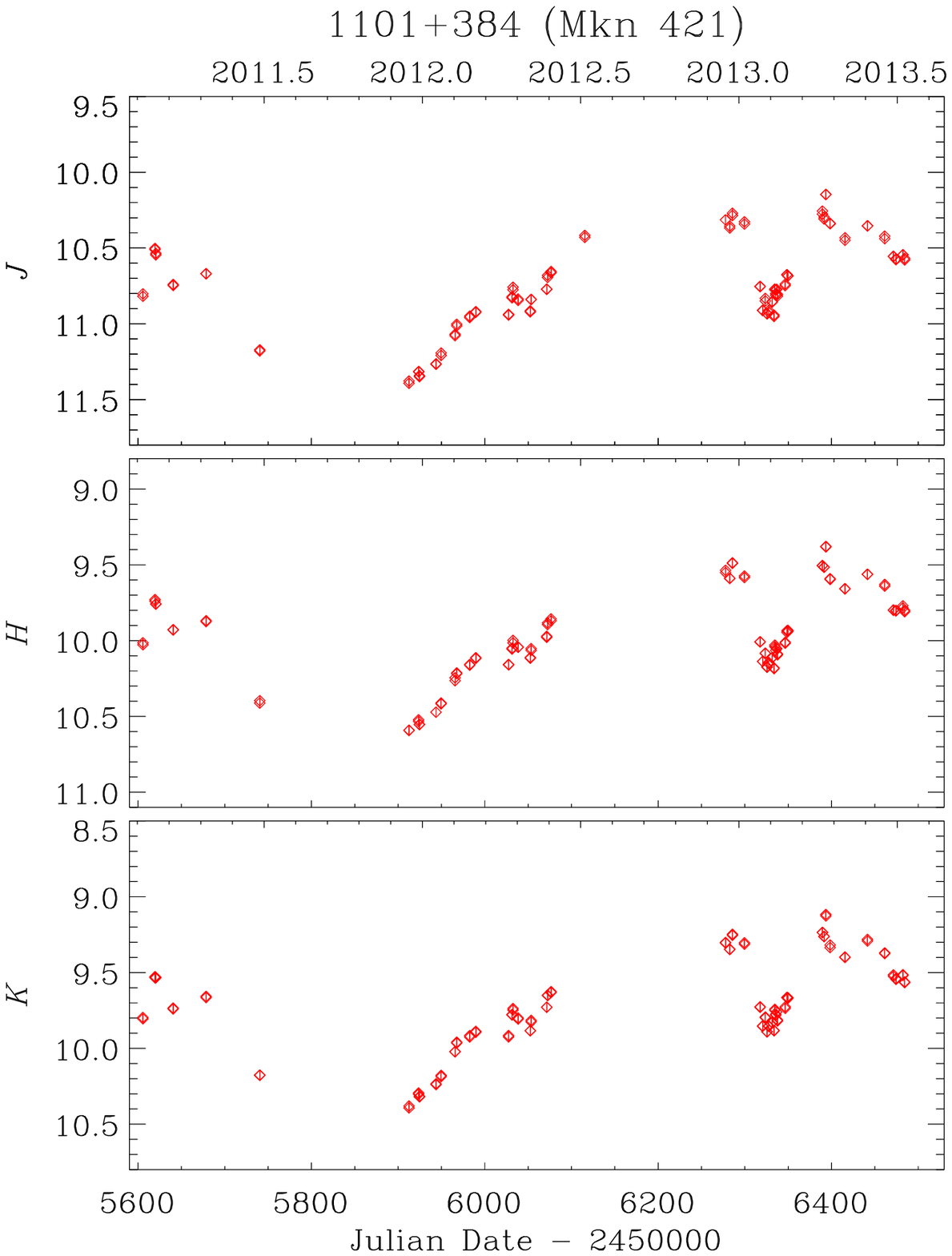,width=0.30\linewidth}
    \psfig{figure=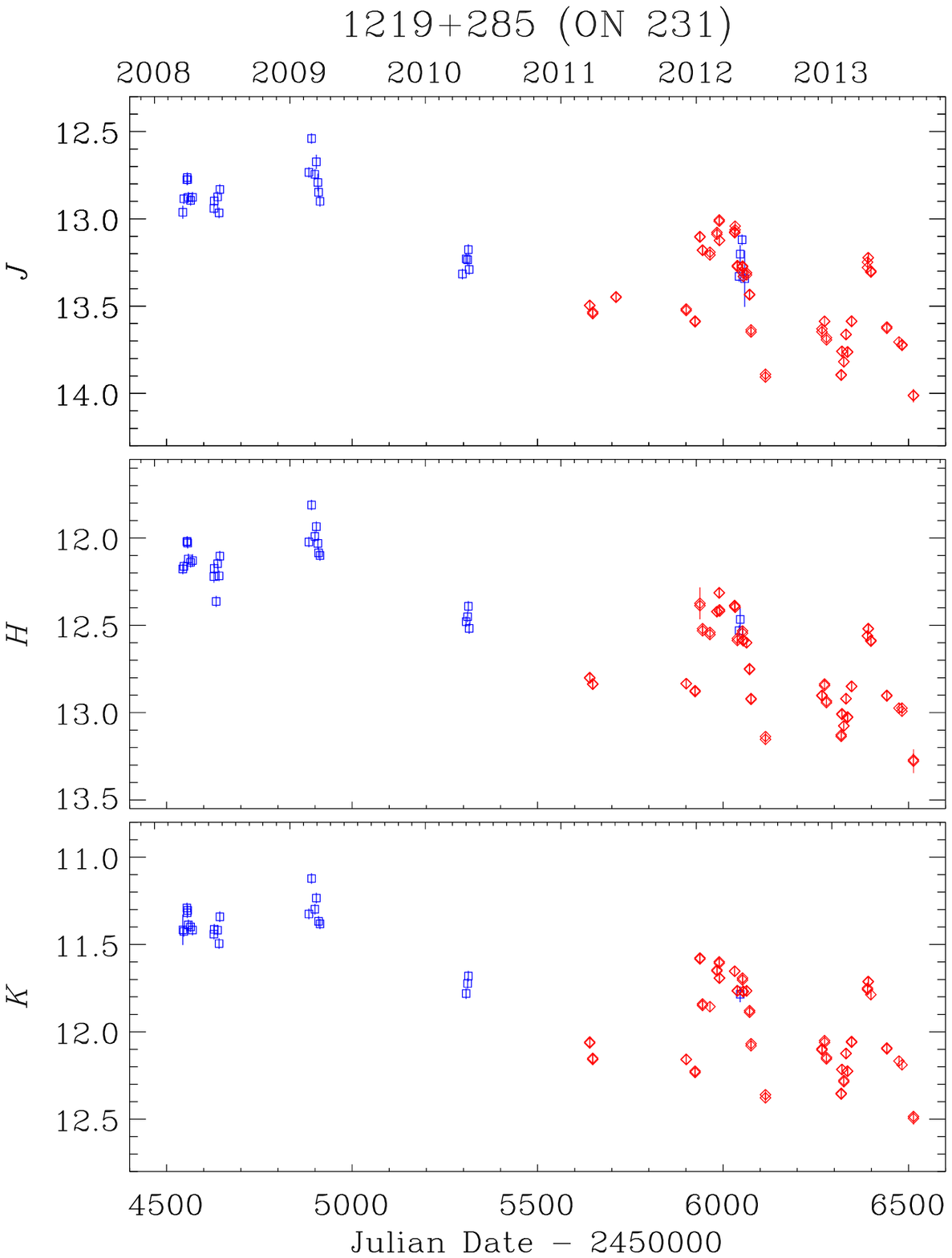,width=0.30\linewidth}}
    \caption{Near-IR light curves of the GASP-WEBT BL Lac objects. Blue squares indicate data from Campo Imperatore, red diamonds those from Teide.} 
    \label{teide_bllacs1}
   \end{figure*}

\begin{figure*}
    \vspace{0.5cm}
    \centerline{
    \psfig{figure=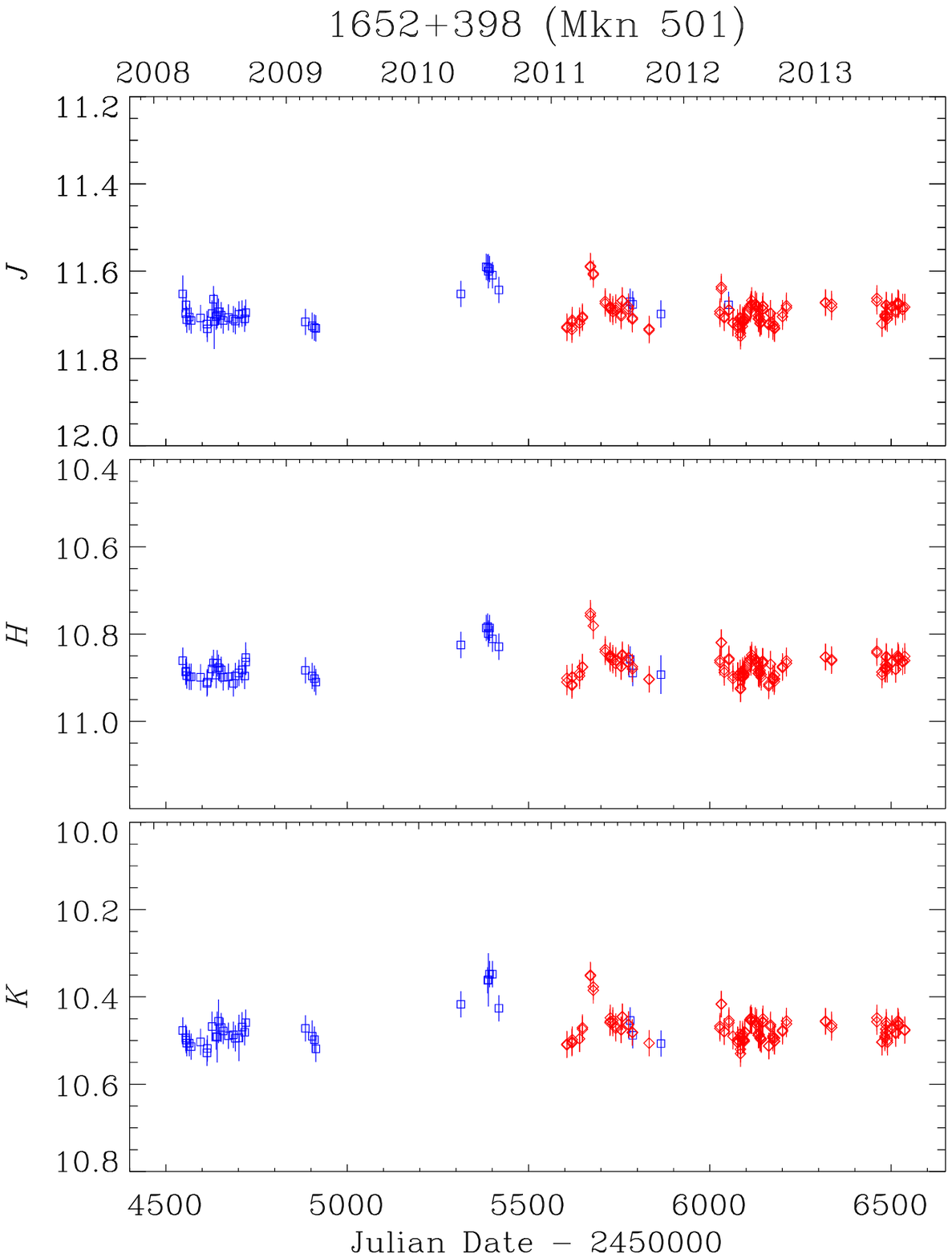,width=0.30\linewidth}
    \psfig{figure=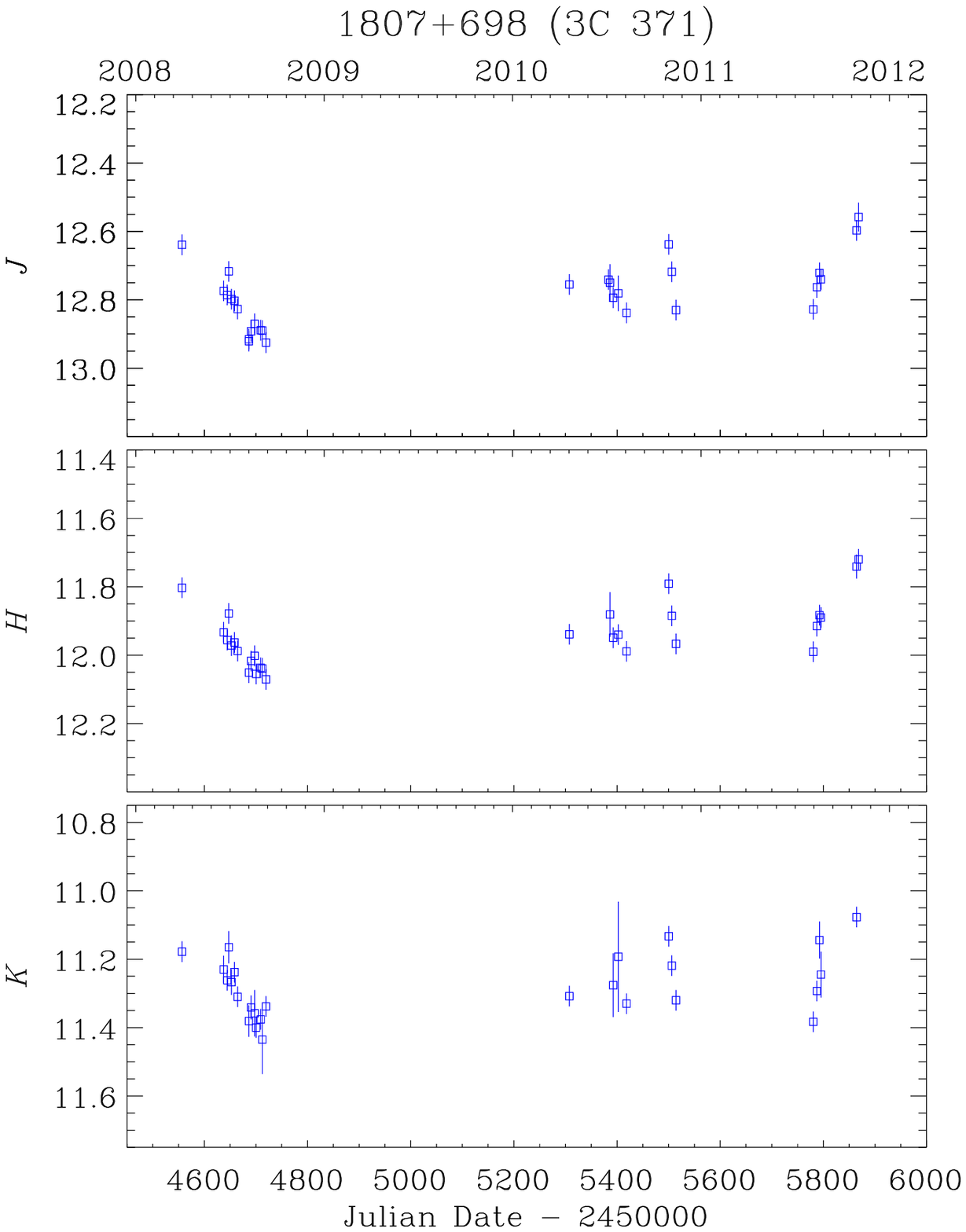,width=0.30\linewidth}
    \psfig{figure=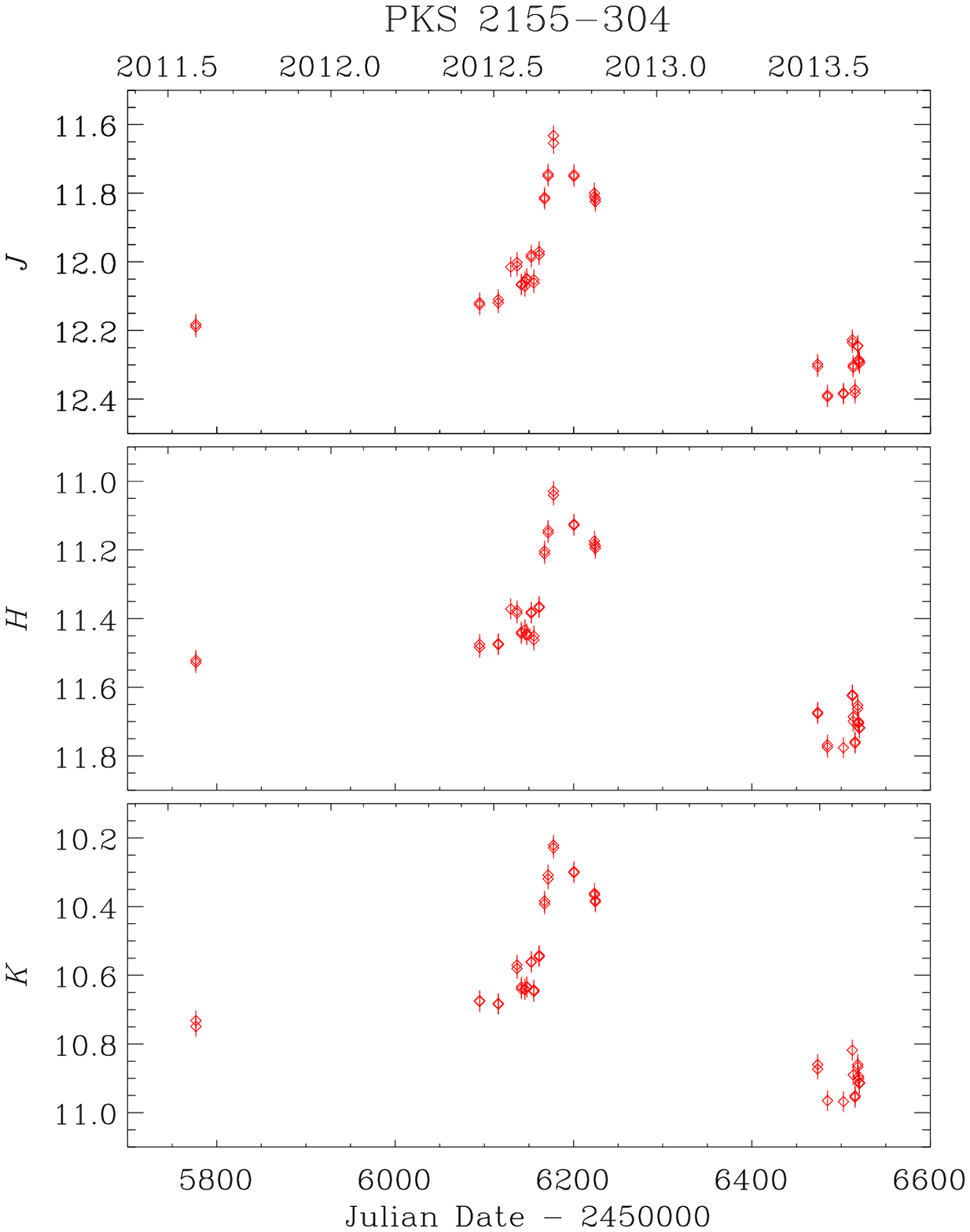,width=0.30\linewidth}}
    \vspace{0.5cm}
    \centerline{
    \psfig{figure=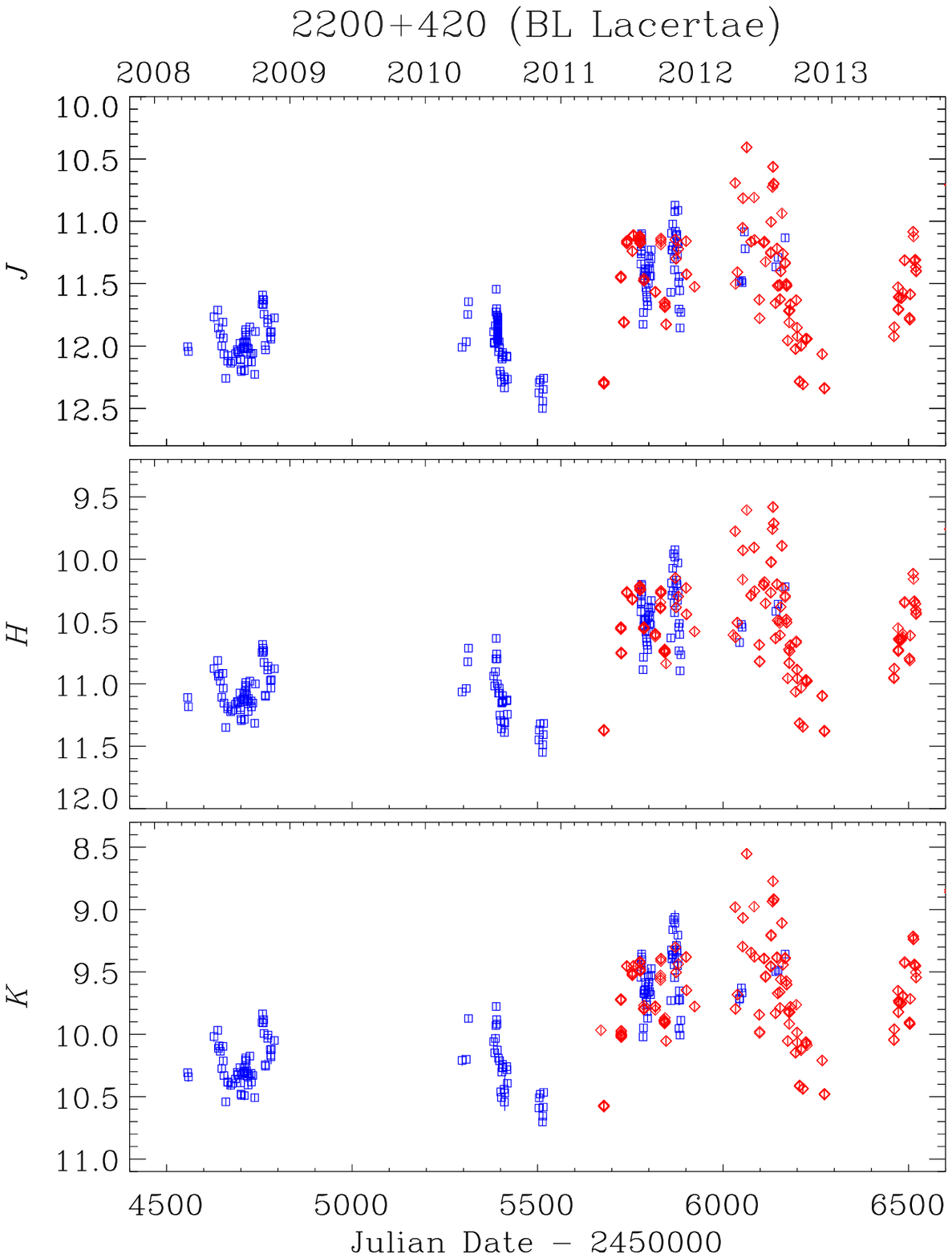,width=0.30\linewidth}
    \psfig{figure=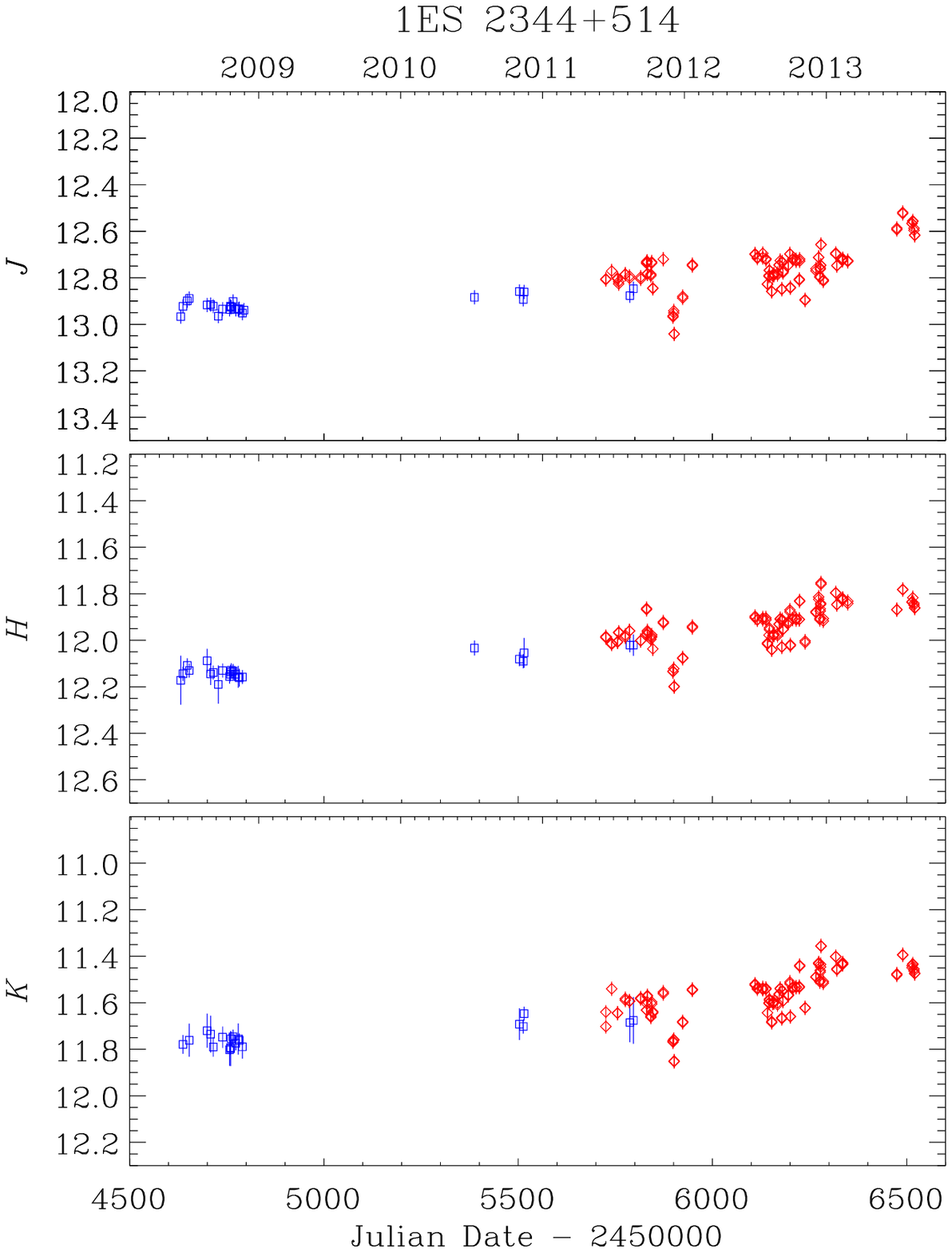,width=0.30\linewidth}}
    \caption{Near-IR light curves of the GASP-WEBT BL Lac objects. Blue squares indicate data from Campo Imperatore, red diamonds those from Teide.} 
    \label{teide_bllacs2}   
   \end{figure*}

\begin{figure*}
    \vspace{0.5cm}
    \centerline{
    \psfig{figure=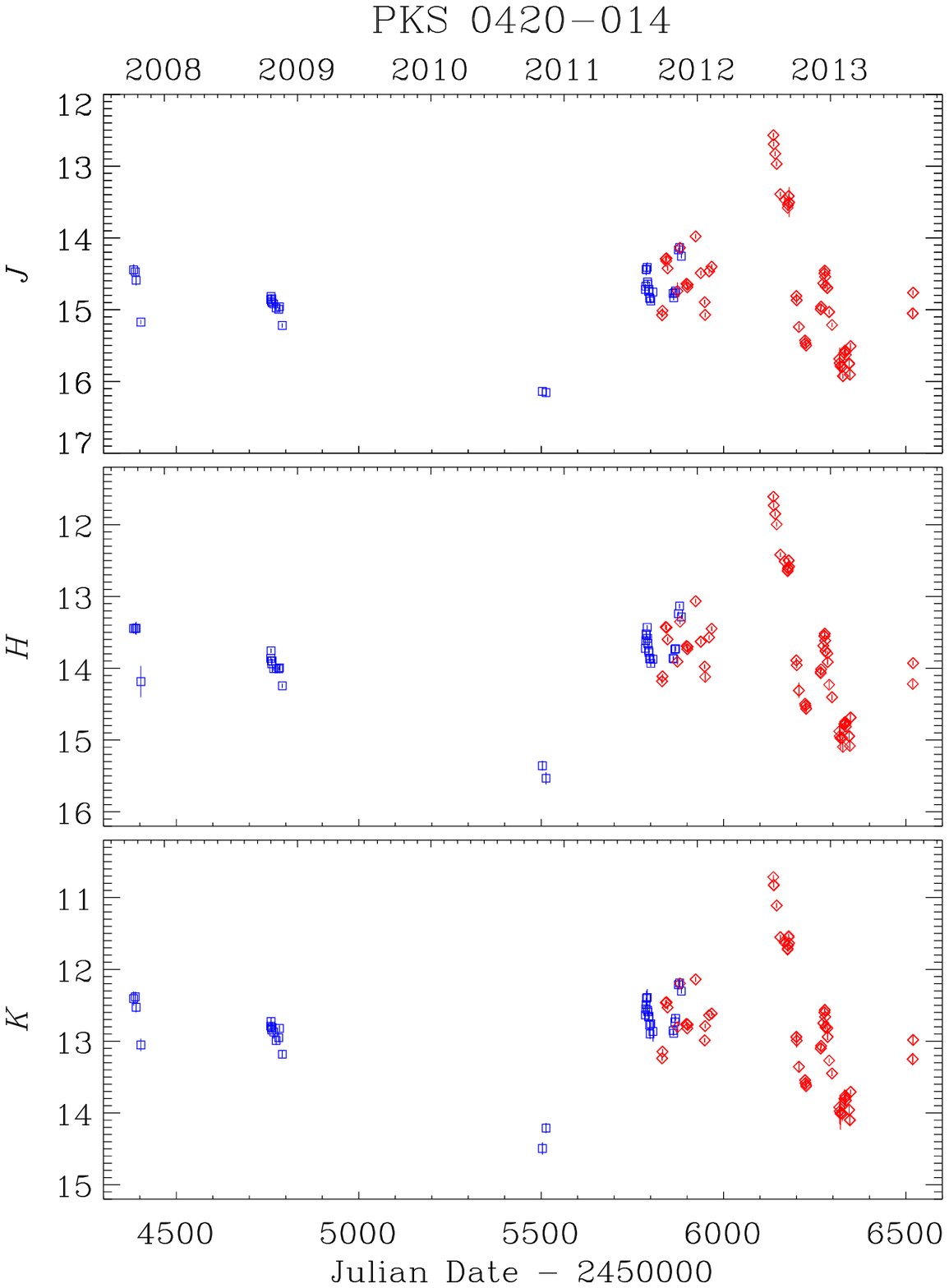,width=0.30\linewidth}
    \psfig{figure=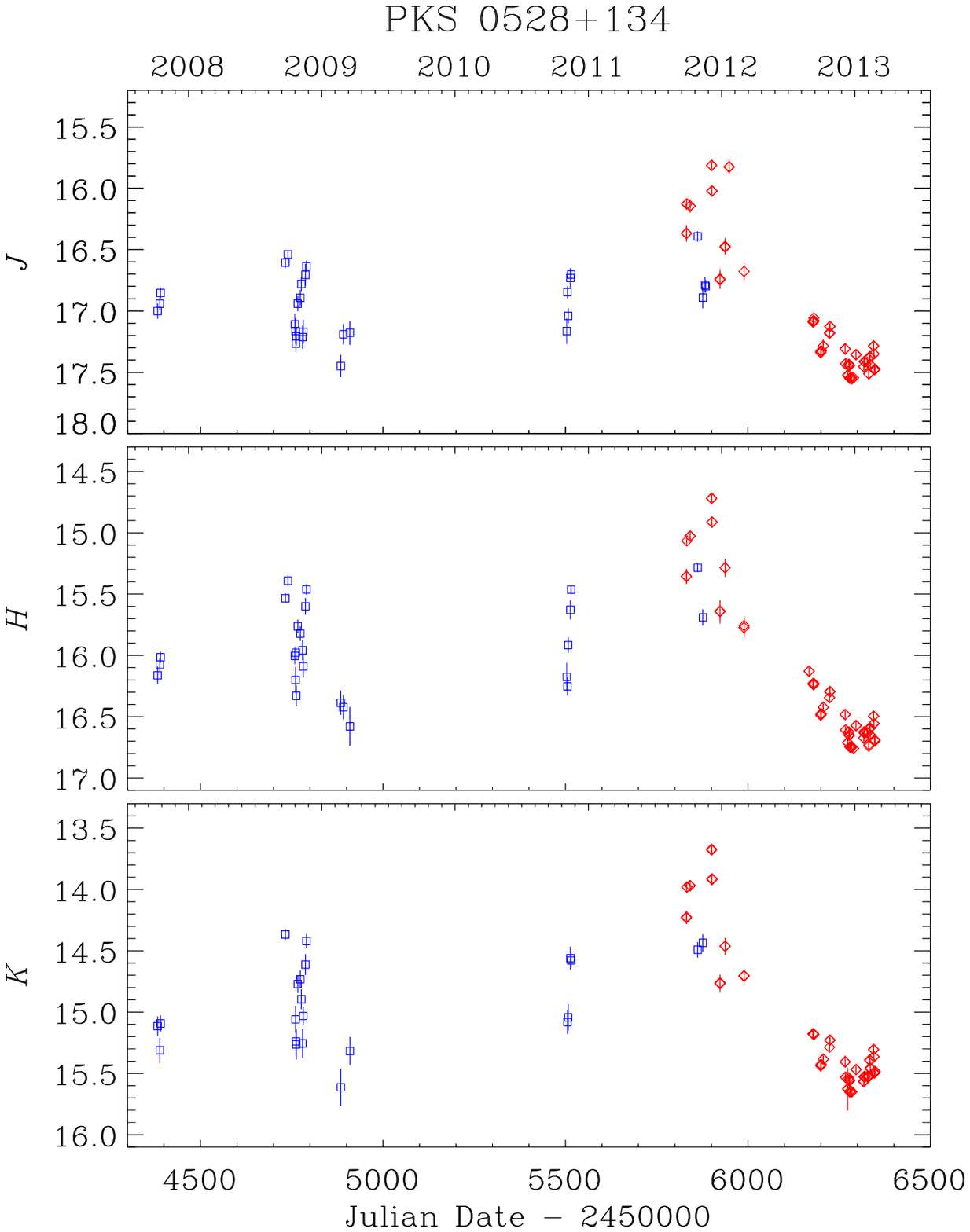,width=0.30\linewidth}
    \psfig{figure=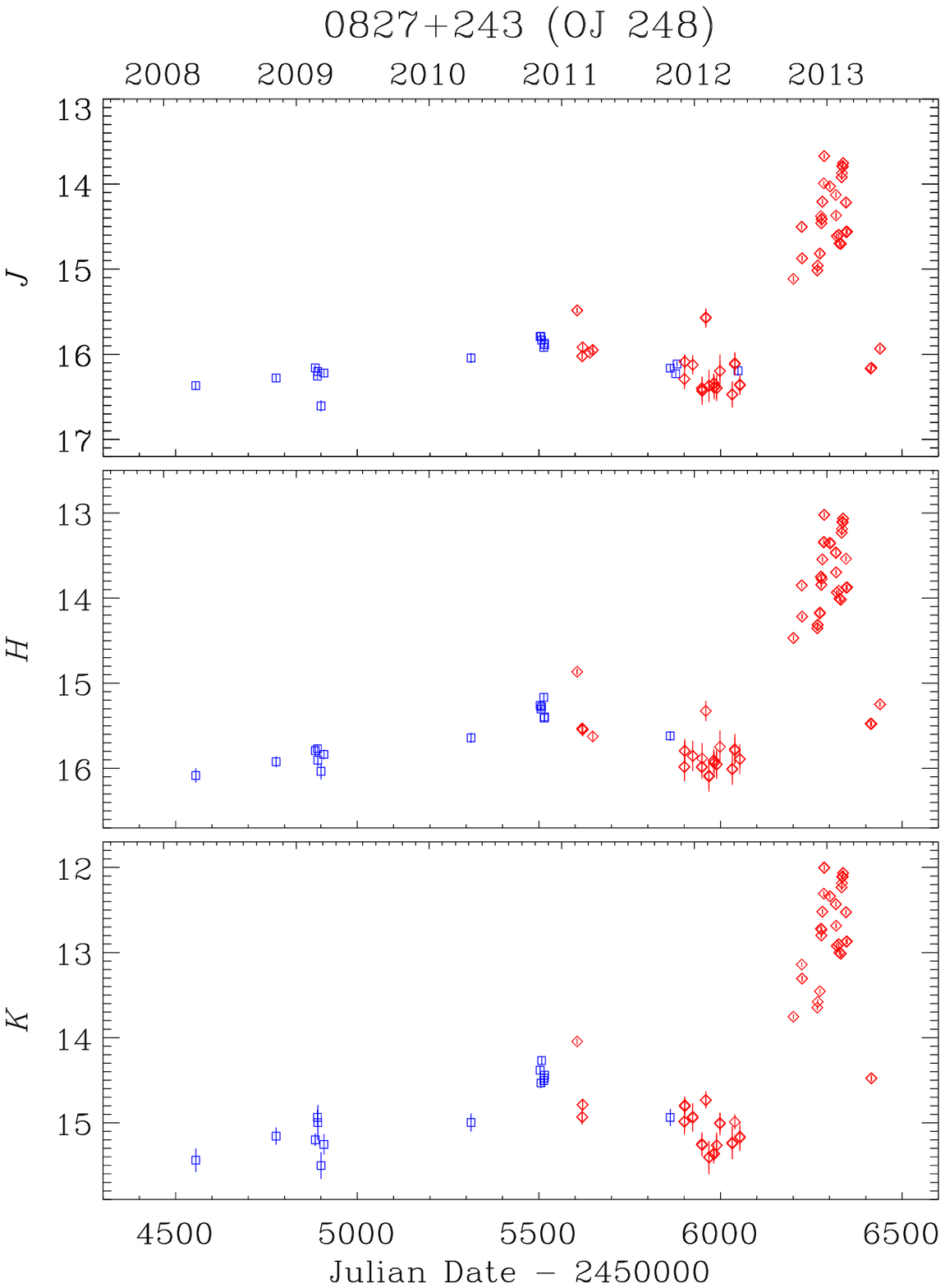,width=0.30\linewidth}}
    \vspace{0.5cm}
    \centerline{
    \psfig{figure=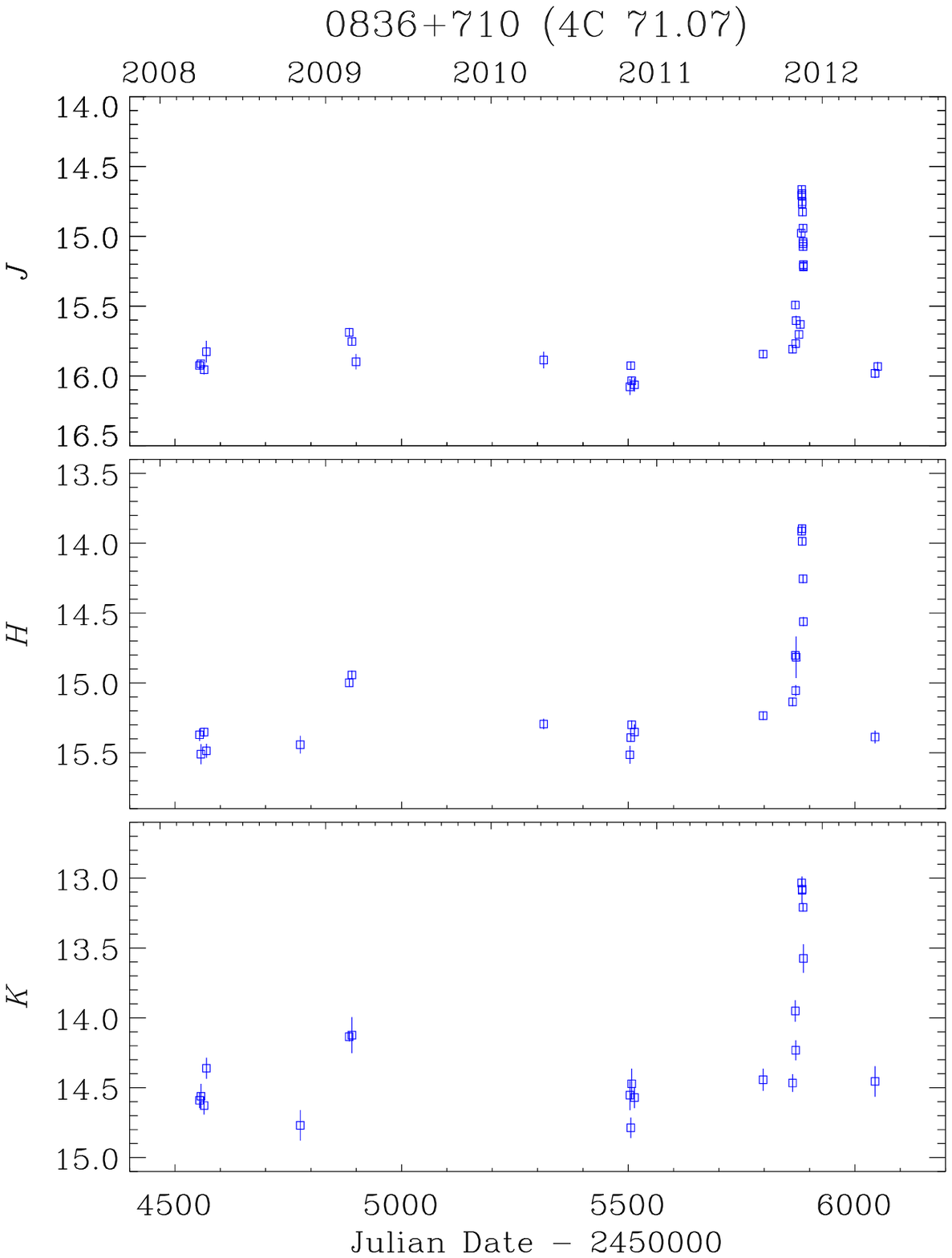,width=0.30\linewidth}
    \psfig{figure=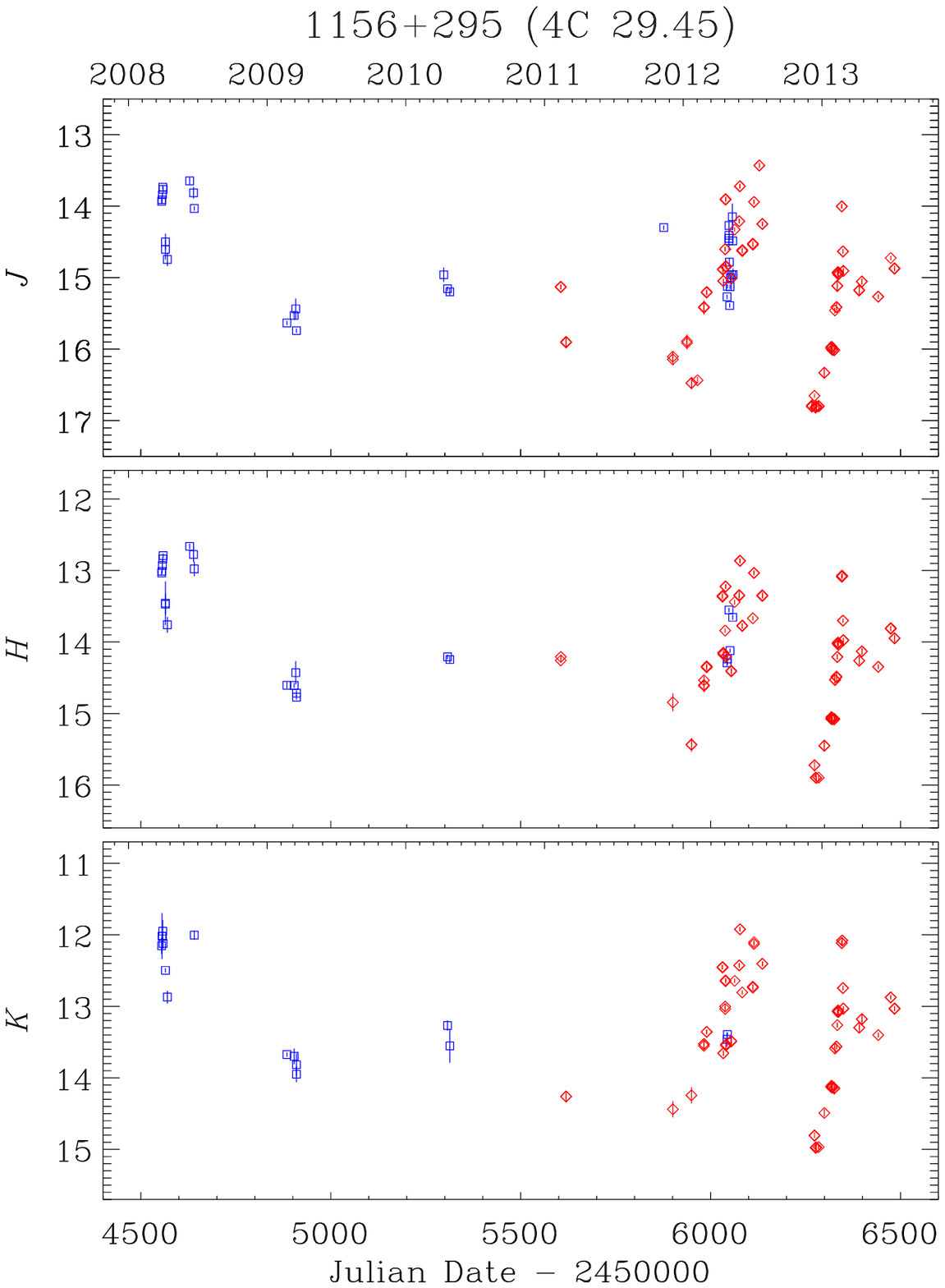,width=0.30\linewidth}
    \psfig{figure=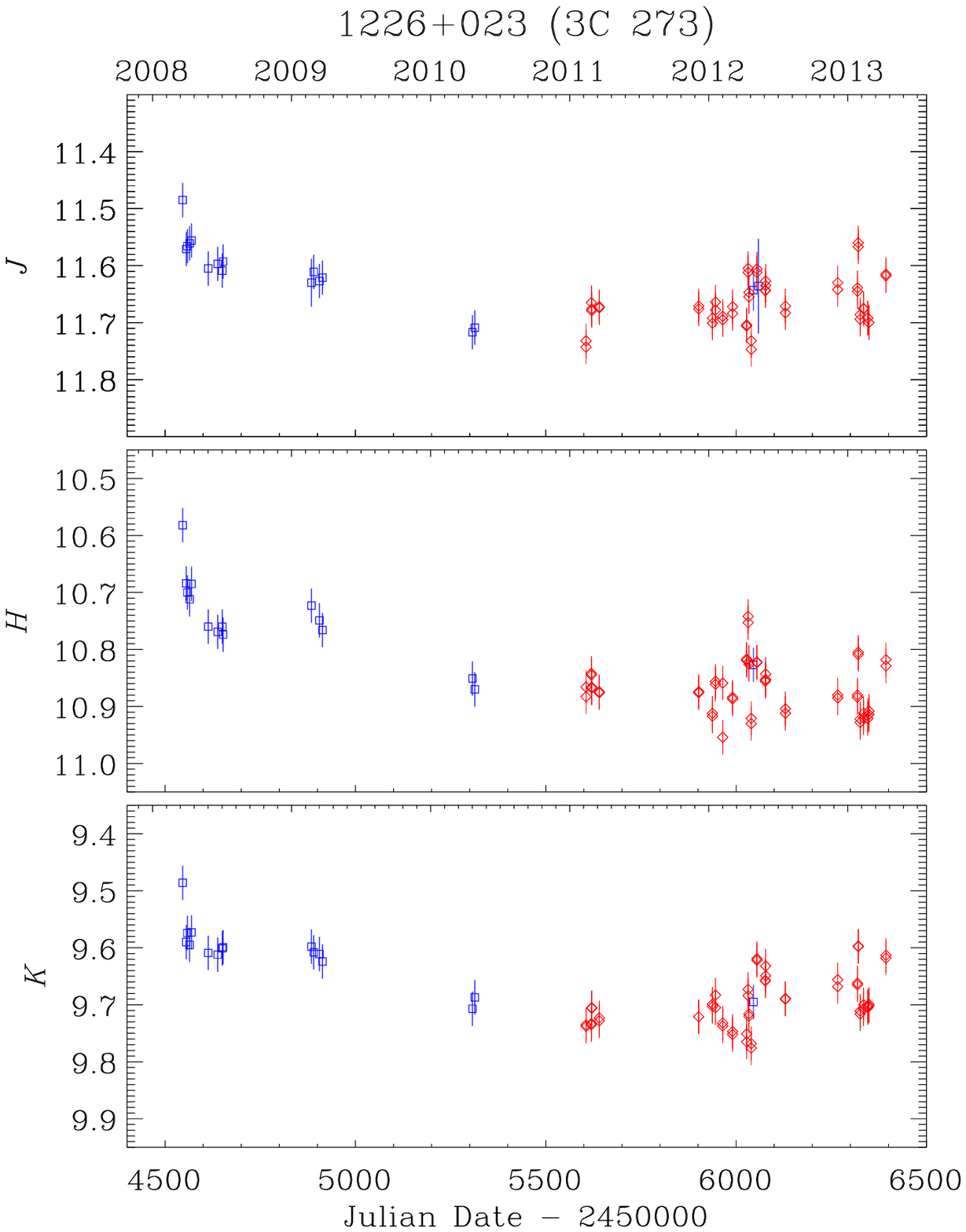,width=0.30\linewidth}}
    \vspace{0.5cm}
    \centerline{
    \psfig{figure=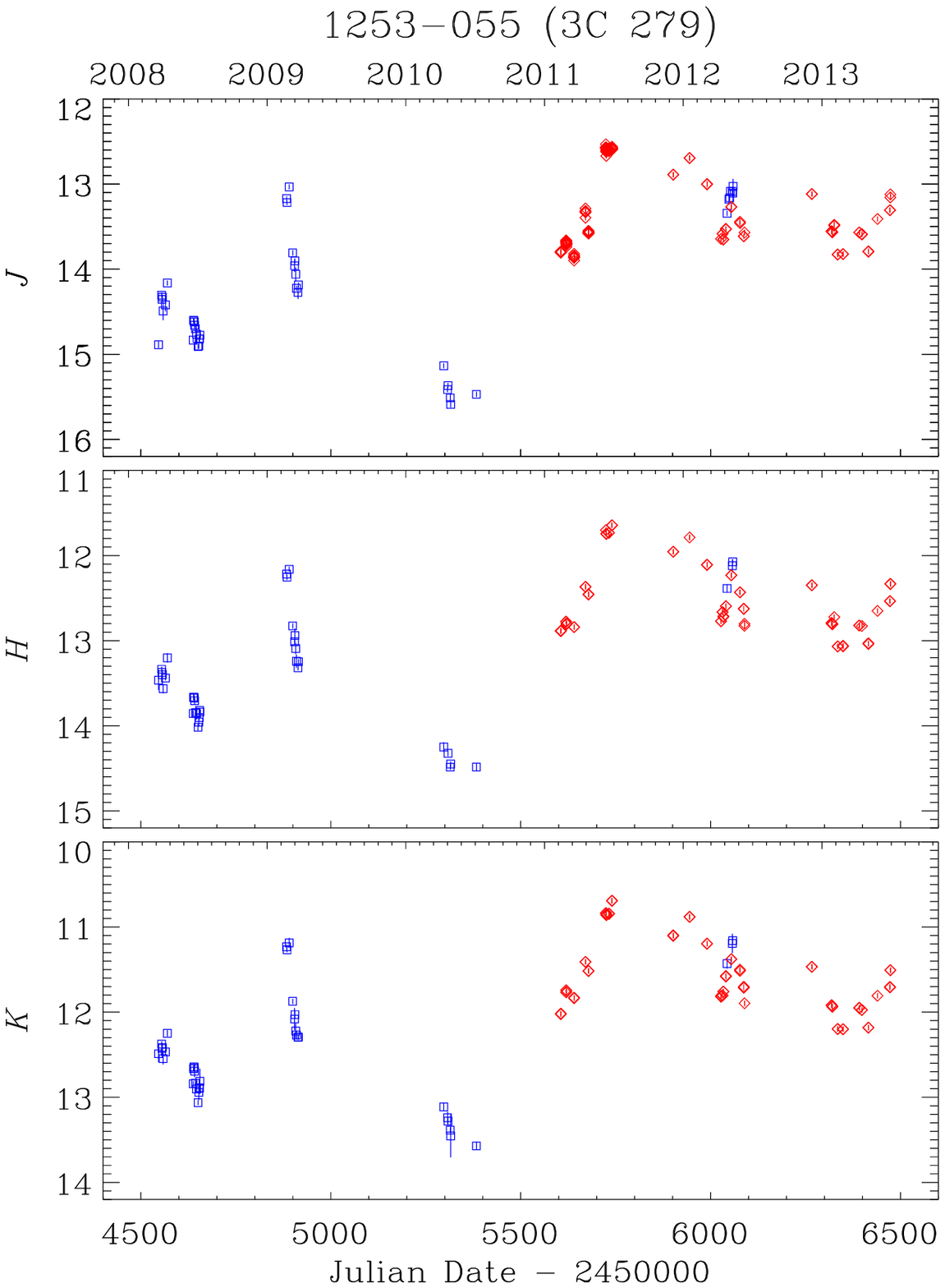,width=0.30\linewidth}
    \psfig{figure=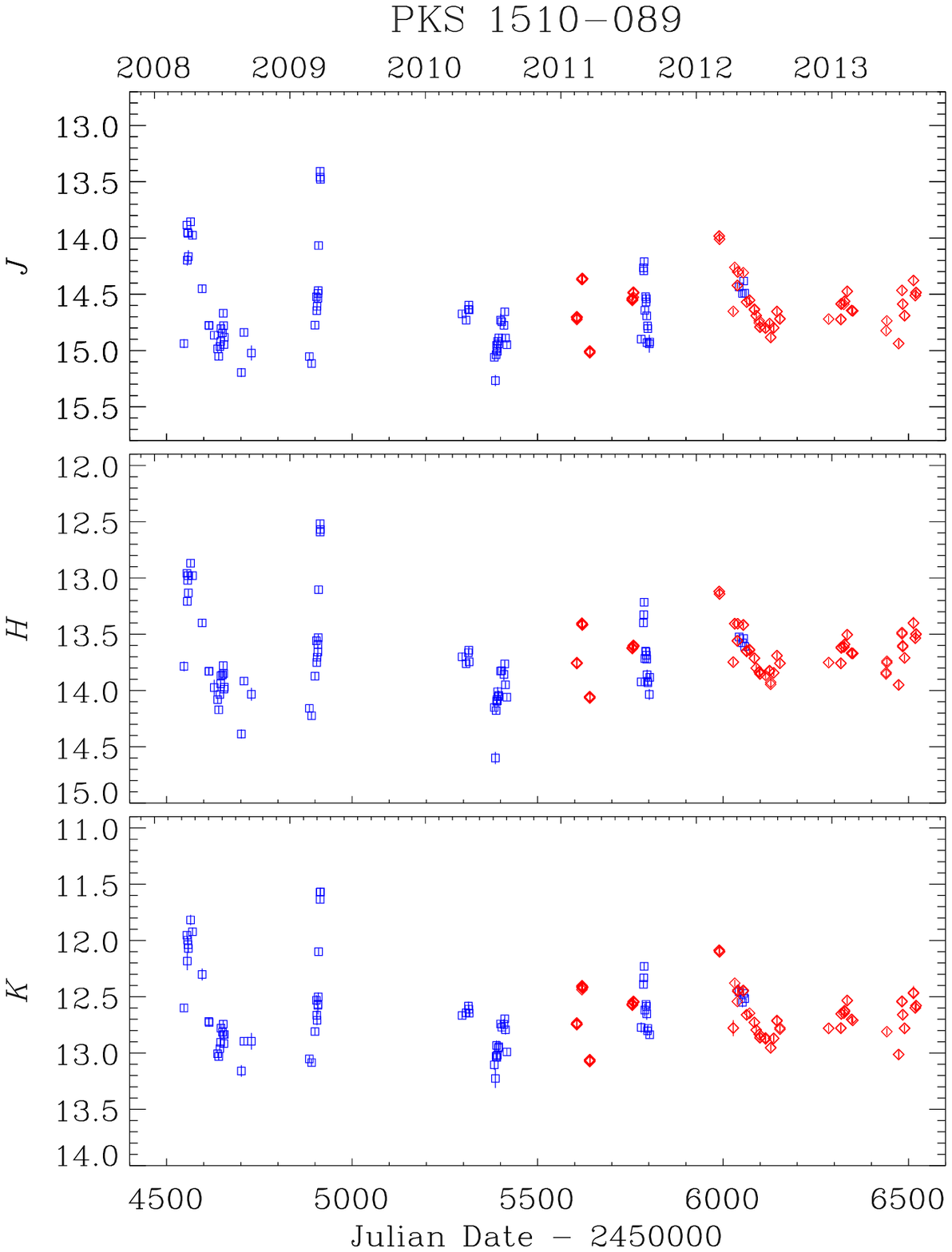,width=0.30\linewidth}
    \psfig{figure=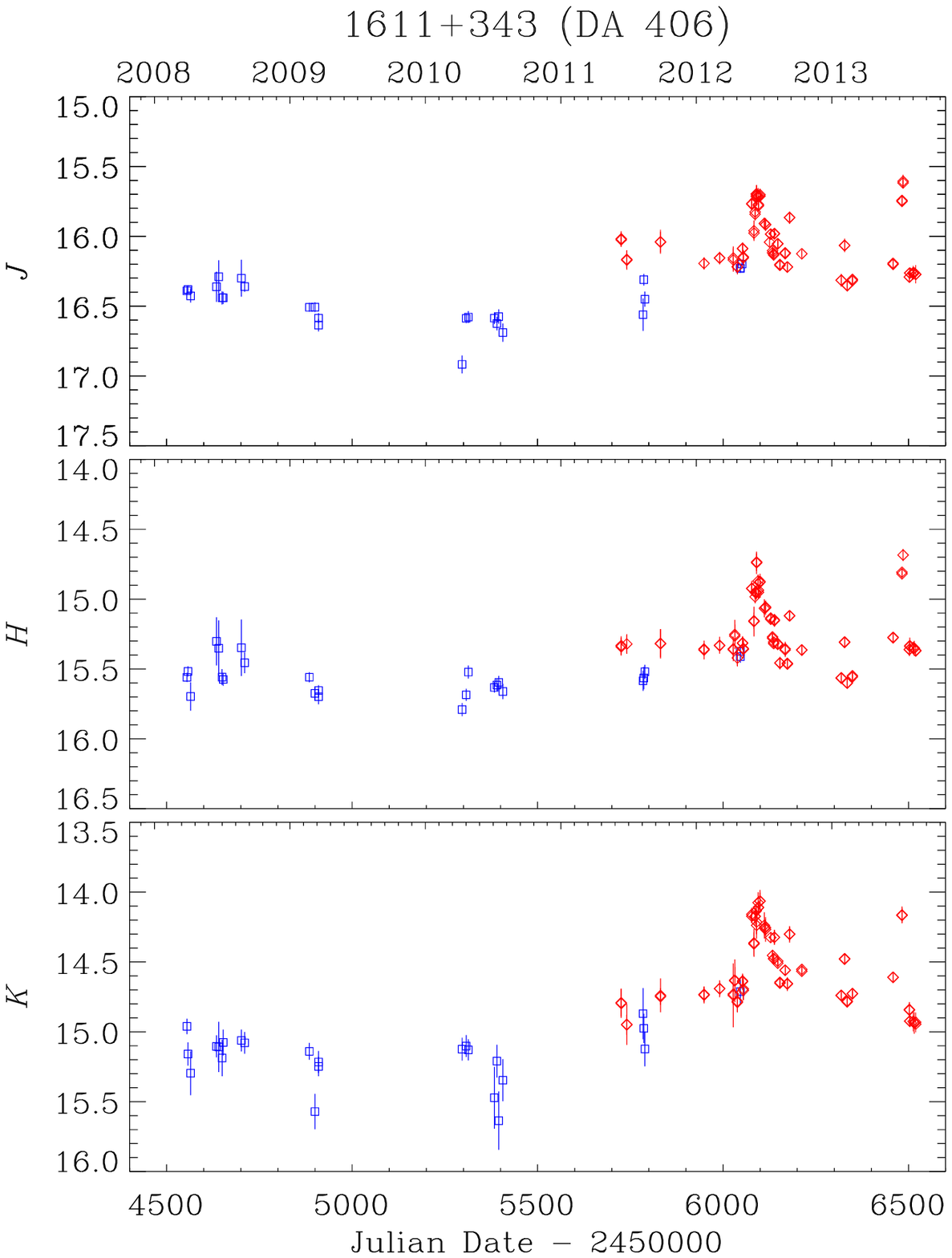,width=0.30\linewidth}}
    \caption{Near-IR light curves of the GASP-WEBT FSRQs. Blue squares indicate data from Campo Imperatore, red diamonds those from Teide.} 
    \label{teide_fsrqs1}
   \end{figure*}

\begin{figure*}
    \vspace{0.5cm}
    \centerline{
    \psfig{figure=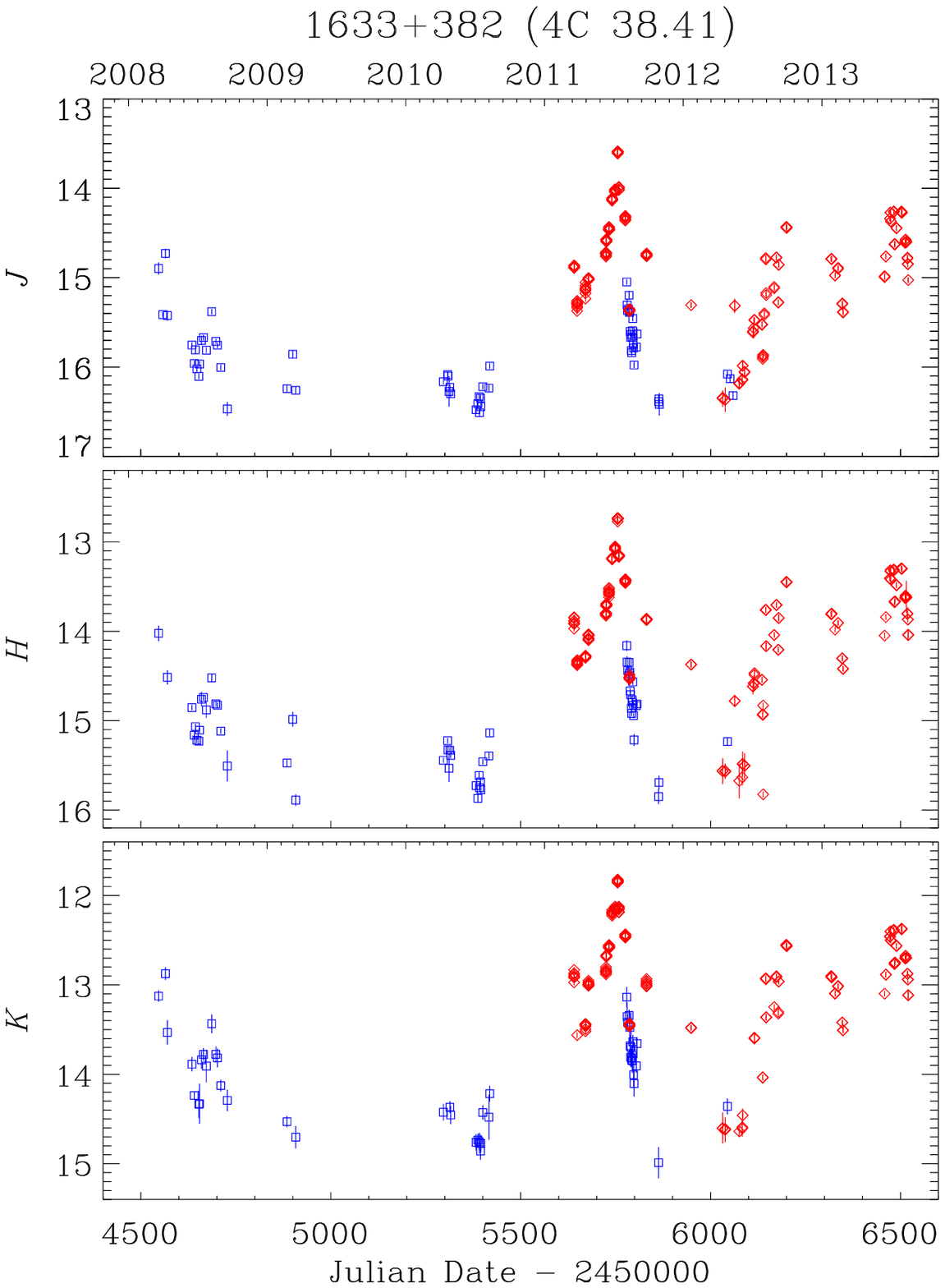,width=0.30\linewidth}
    \psfig{figure=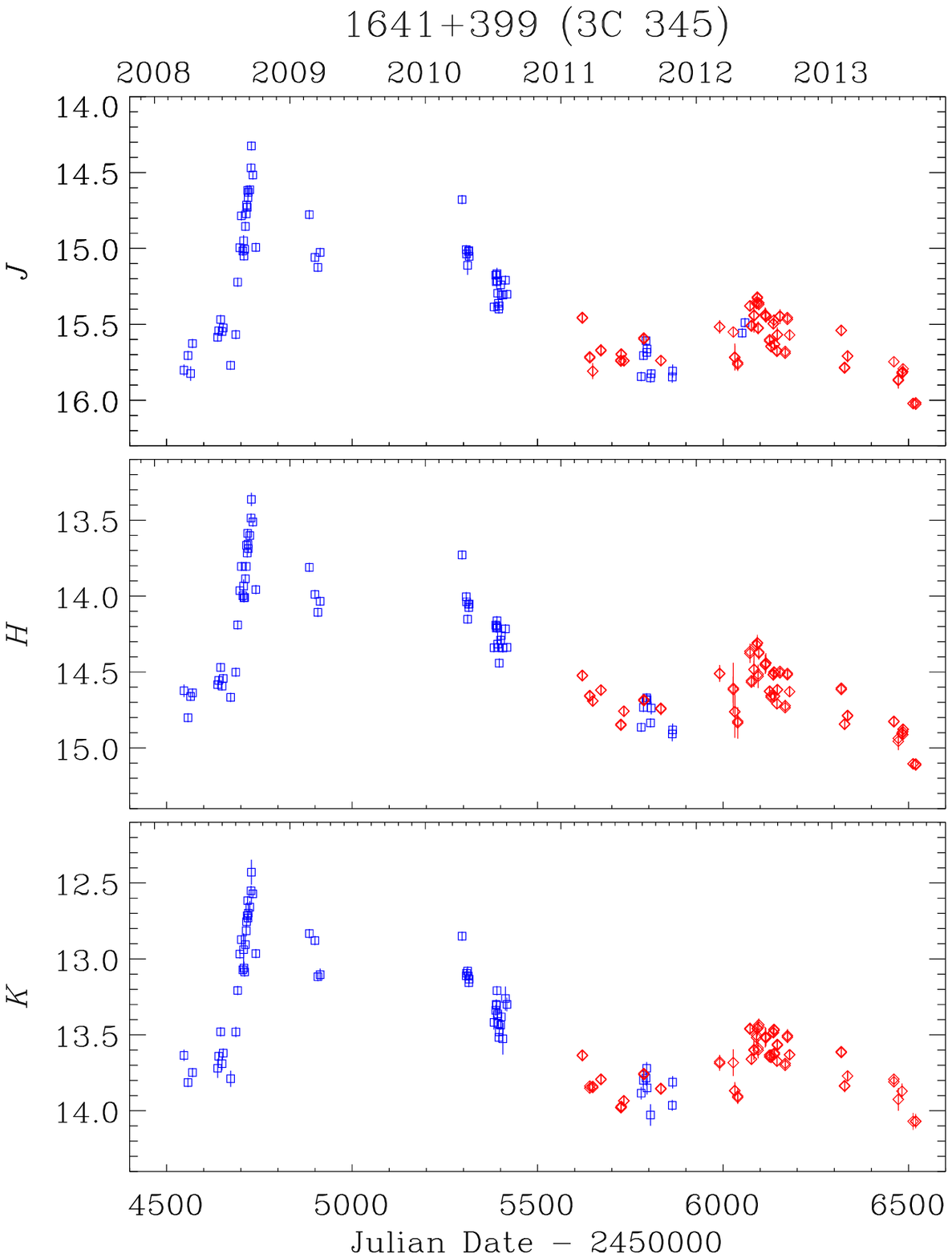,width=0.30\linewidth}
    \psfig{figure=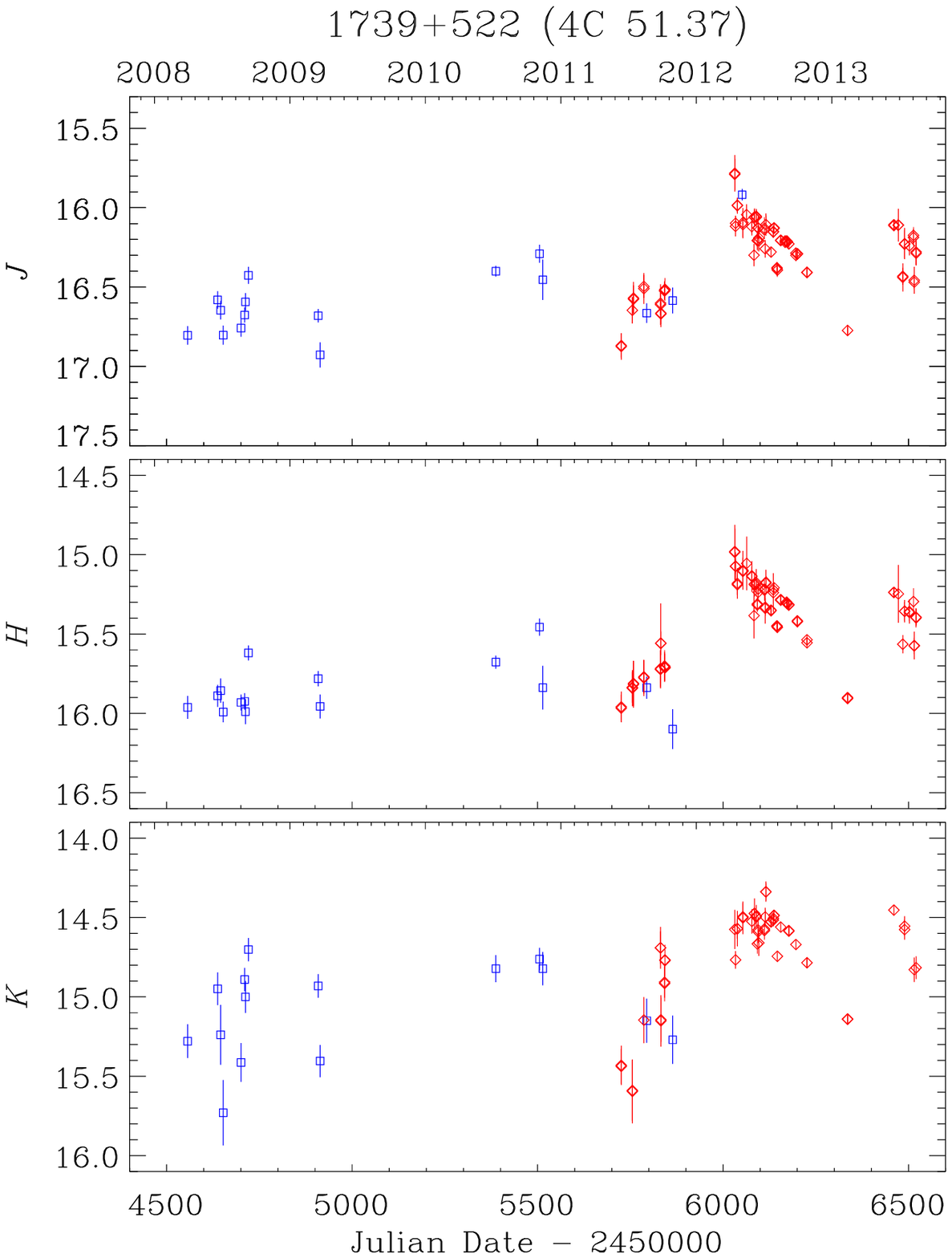,width=0.30\linewidth}}
    \vspace{0.5cm}
    \centerline{
    \psfig{figure=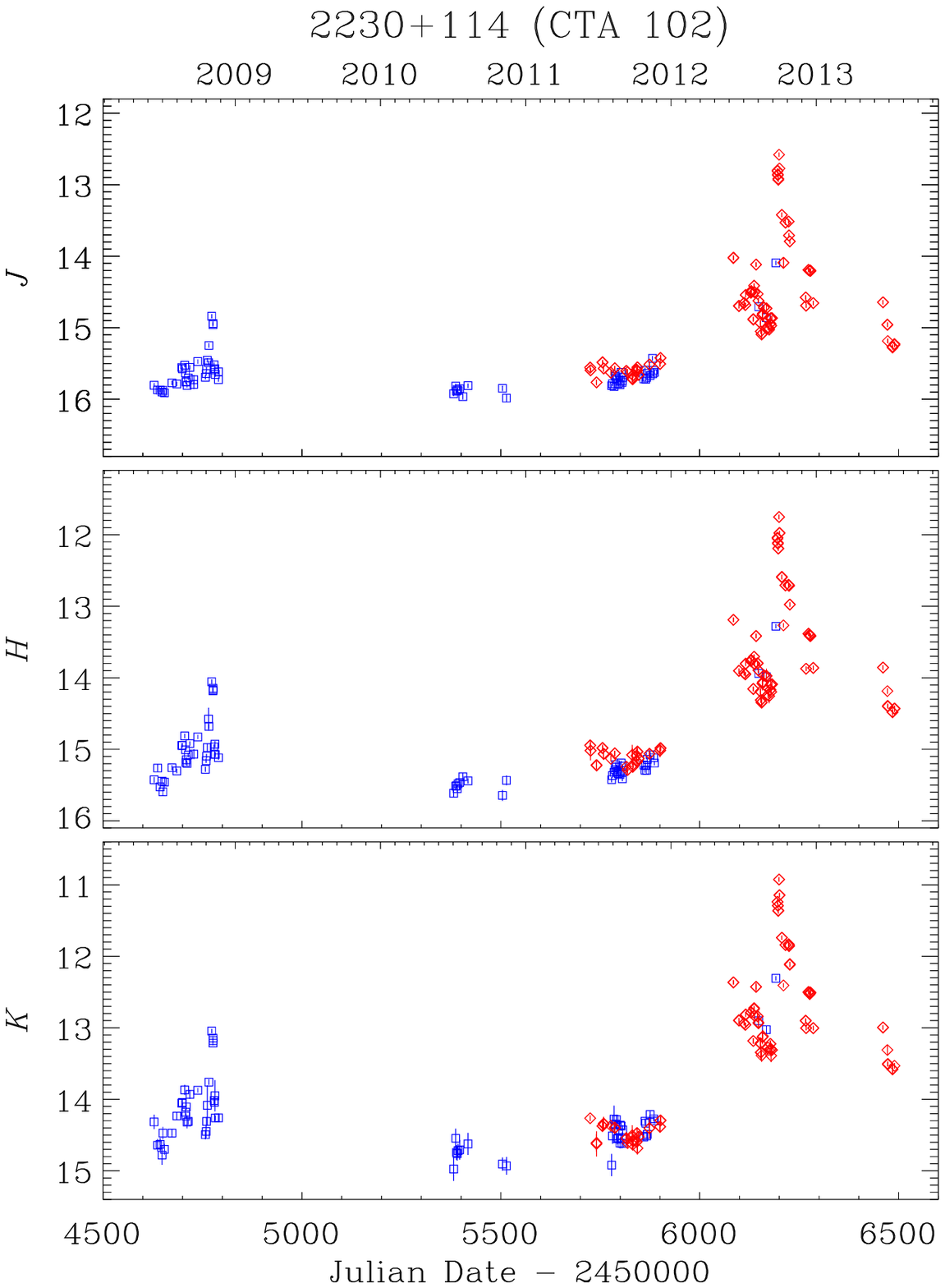,width=0.30\linewidth}
    \psfig{figure=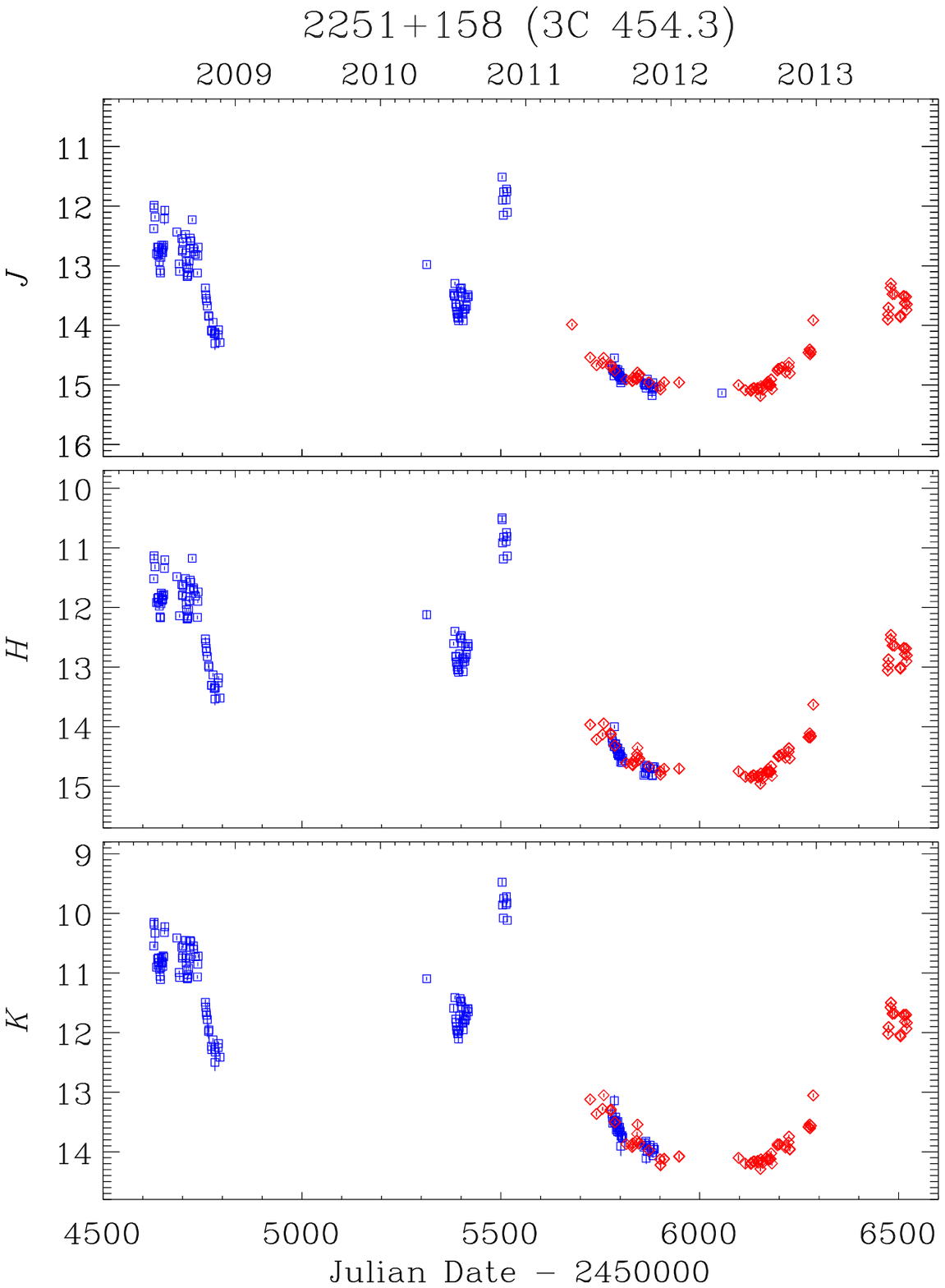,width=0.30\linewidth}}
    \caption{Near-IR light curves of the GASP-WEBT FSRQs. Blue squares indicate data from Campo Imperatore, red diamonds those from Teide.} 
    \label{teide_fsrqs2}   
   \end{figure*}

In Figs.\ \ref{teide_bllacs1} and \ref{teide_bllacs2} we show the near-IR light curves of the GASP BL Lacs, while in Figs.\ \ref{teide_fsrqs1} and \ref{teide_fsrqs2} those of the GASP FSRQs are displayed. In a few cases (0219+428, 0851+202, 1226+023, and 2344+514) we found offsets due to different source calibration between the Campo Imperatore and Teide data, and corrected for them. 

To quantify variability, magnitudes are converted into flux densities as explained in Sect.\ \ref{compa} and then the mean fractional variation is calculated. This is defined as: 
\begin{equation}
f={{\sqrt{\sigma^2-\delta^2}} \over {<F>}}, 
\end{equation}
where $<F>$ is the mean flux, $\sigma^2$ the variance of the flux, and $\delta^2$ the mean square uncertainty of the fluxes \citep[e.g.][]{pet01}.
The advantage of this quantity is that it takes into account the effect of errors, which produce apparent variability.
Tables 3 and 4 report the mean fractional variation for the GASP BL Lacs and FSRQs, respectively.
For a given object, this quantity obviously depends on the considered period and on the amount of data collected.
Only one BL Lac (0235+164) and 3 FSRQs (0420$-$014, 2230+114, and 2251+158) have $f \ge 1$, and 0235+164 is the most variable source. Notice that 0235+164 has sometimes been claimed to belong to the FSRQs \citep[e.g.][]{rai07a,ghi11}.
The values of the mean fractional variation indicate that FSRQs are typically more variable than BL Lacs.
Average $f$ values for the whole BL Lac sample in each of the near-IR bands are around 0.35\%, while they are $\sim 60$\% in the case of FSRQs. Moreover, in general BL Lacs are slightly more variable in the $J$ band and FSRQs in the $K$ band.
This recalls what happens in the optical domain, where BL Lacs are generally more variable in the blue than in the red, while for FSRQs the opposite is true. The smaller variability of FSRQs toward the blue is a clue in favour of an additional emission component, i.e.\ thermal emission from the disc (showing as a ``big blue bump" in the SED) and broad line region (BLR, producing a ``little blue bump" in the SED). Indeed, the contribution from this ``blue" component is expected to extend into the near-IR band, behaving as a base-level under which the flux cannot fall, and affecting the $J$ flux density more than the $K$ one.

\section{Near-IR colour variability}
\label{colour}

\begin{figure*}
    \vspace{0.5cm}
    \centerline{
    \psfig{figure=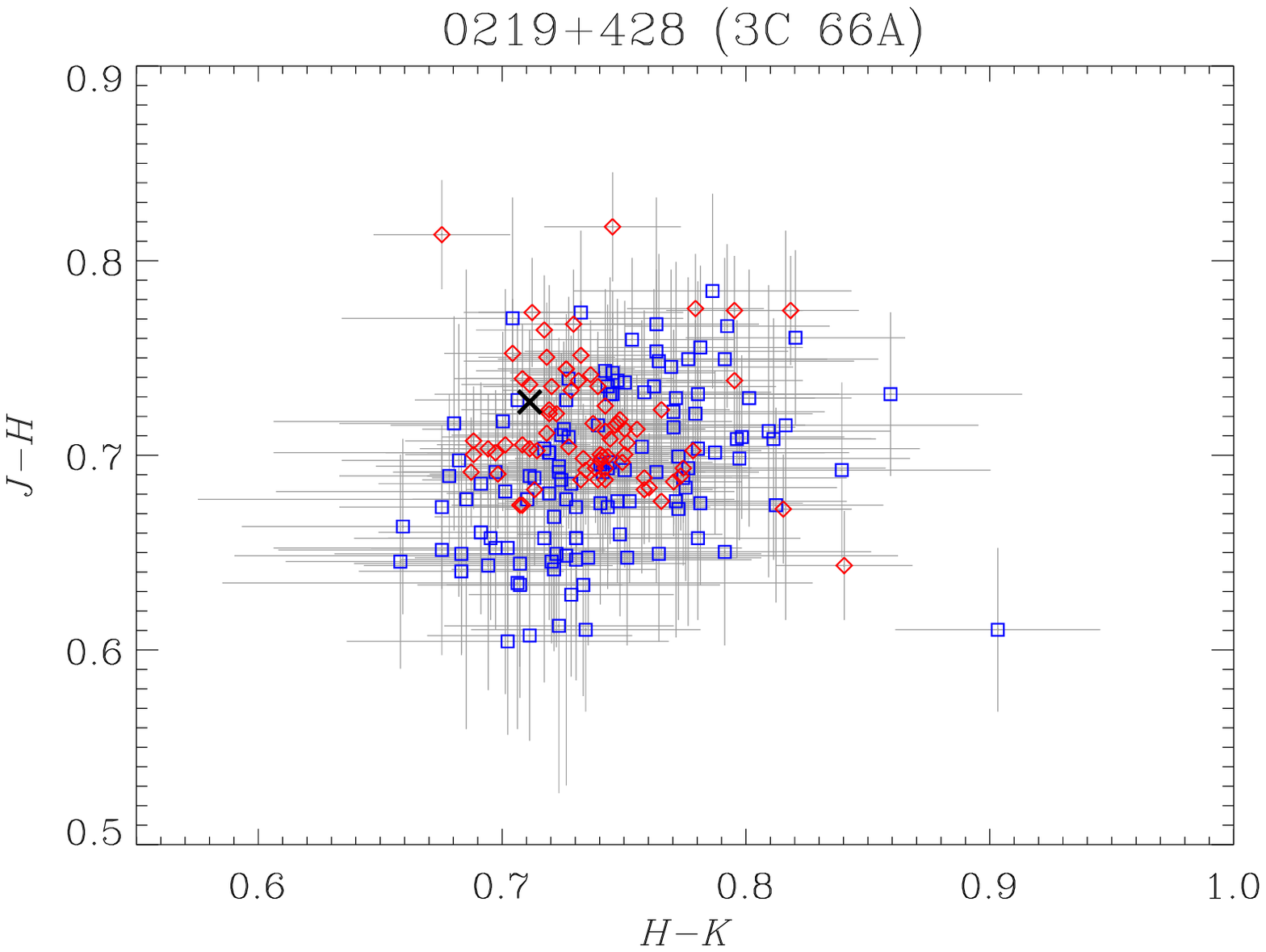,width=0.30\linewidth}
    \psfig{figure=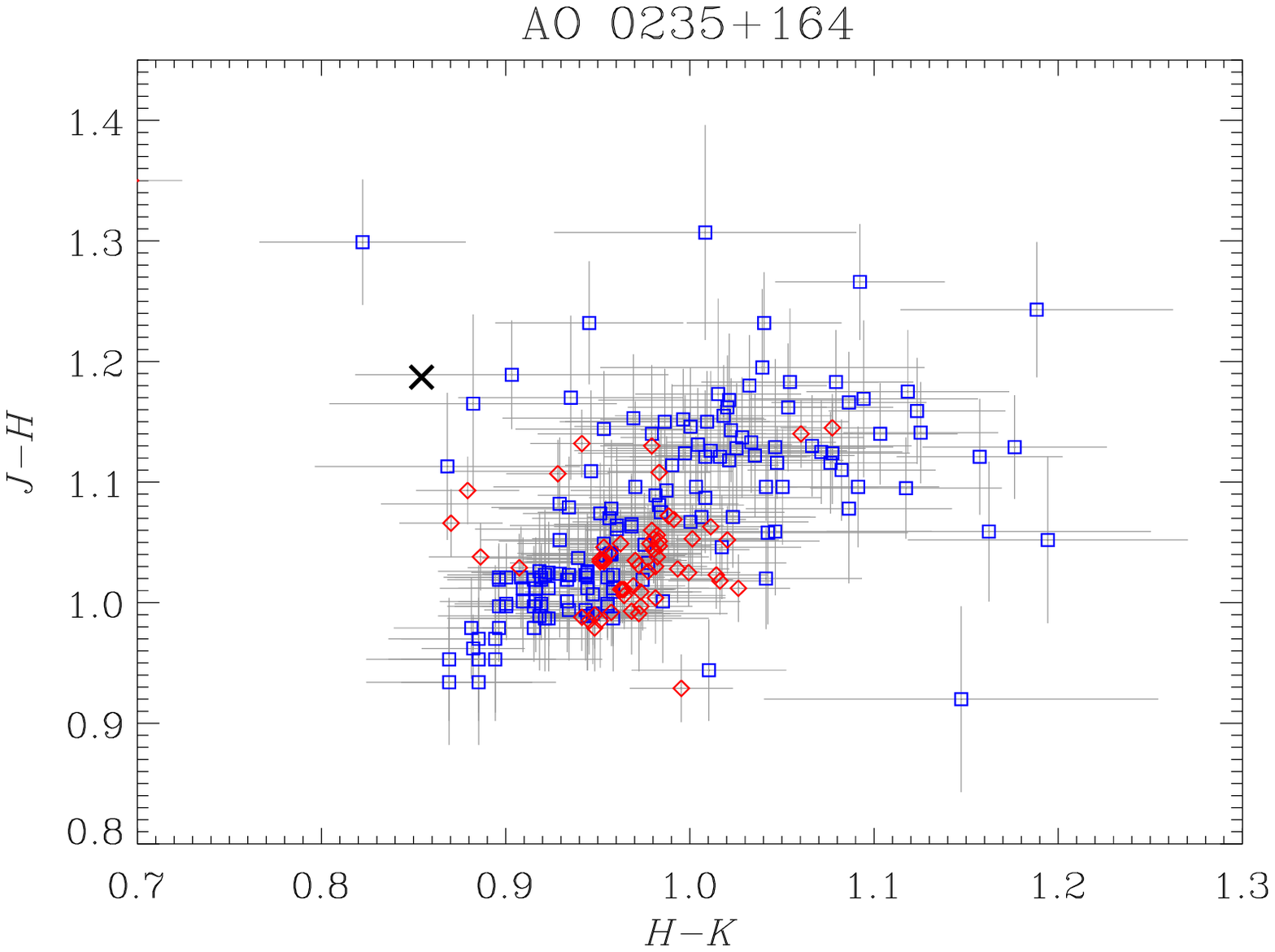,width=0.30\linewidth}
    \psfig{figure=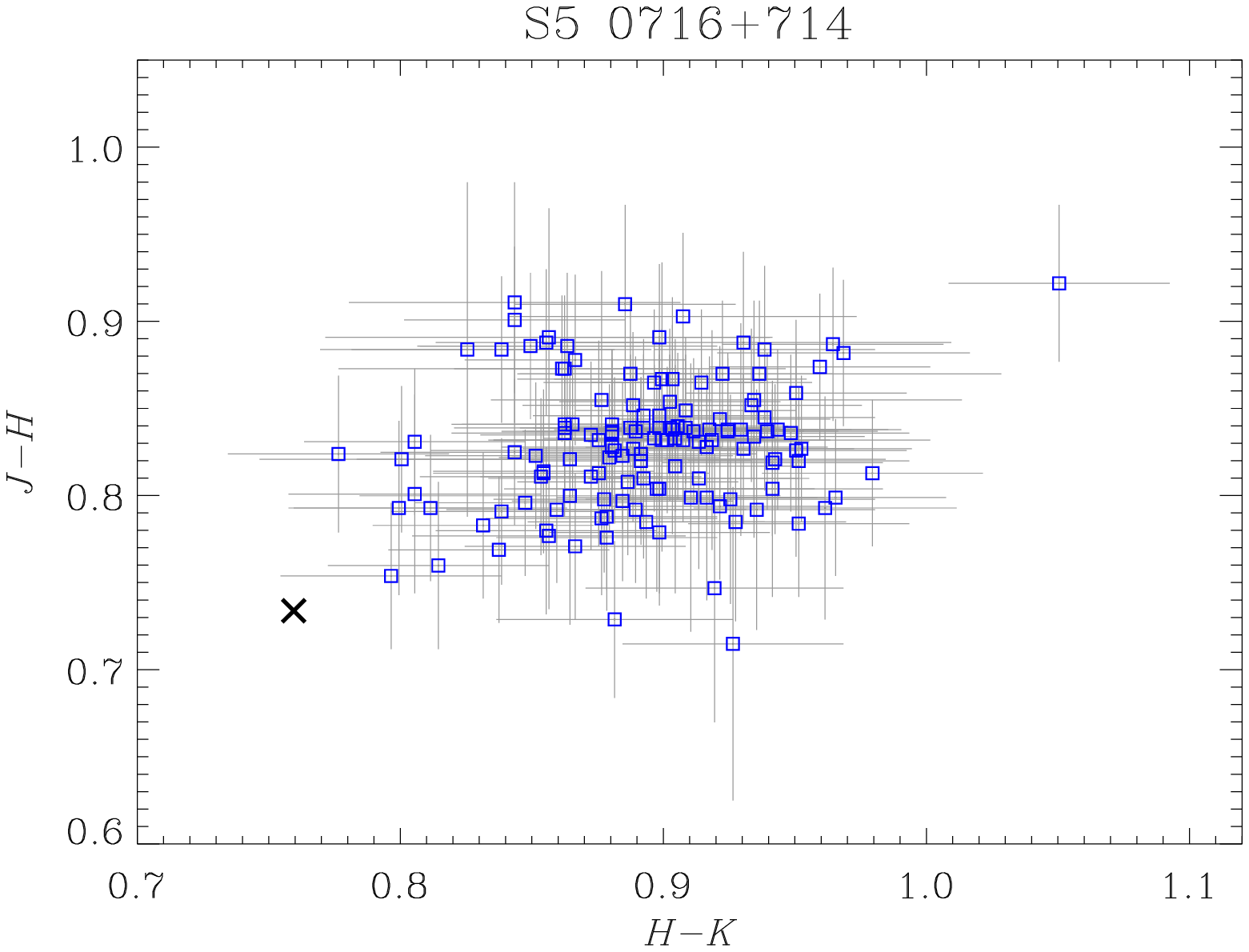,width=0.30\linewidth}}
    \vspace{0.5cm}
    \centerline{
    \psfig{figure=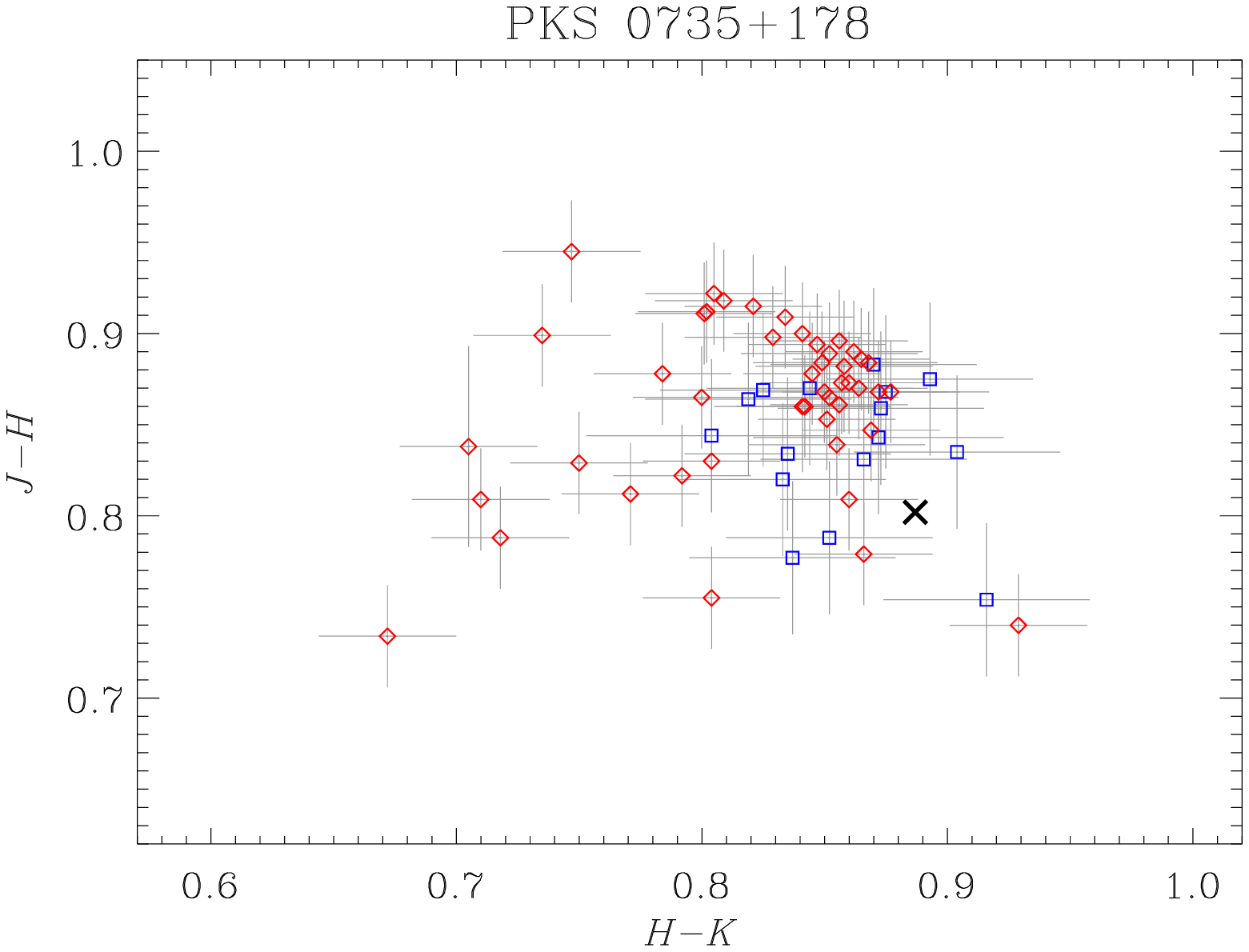,width=0.30\linewidth}
    \psfig{figure=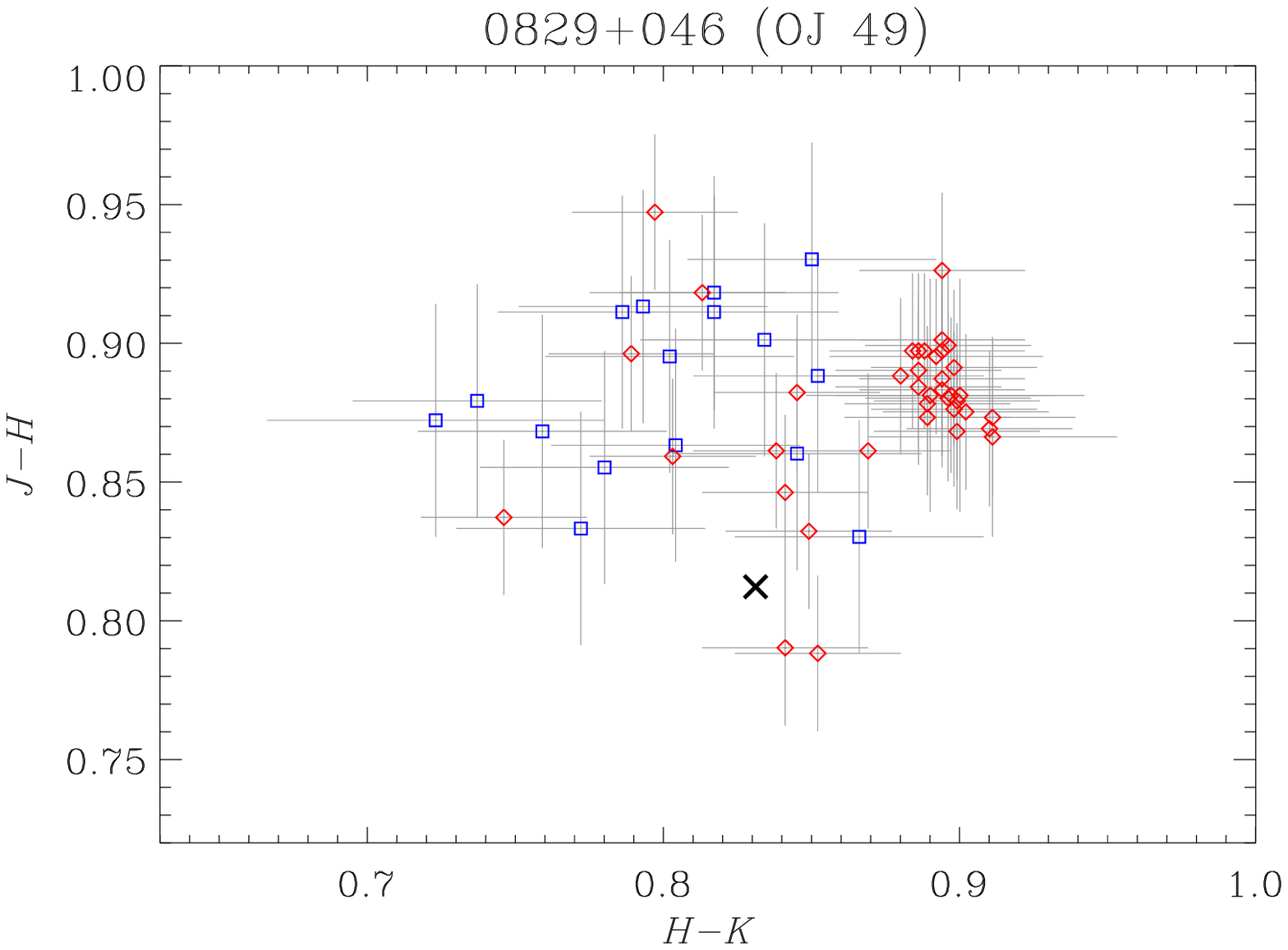,width=0.30\linewidth}
    \psfig{figure=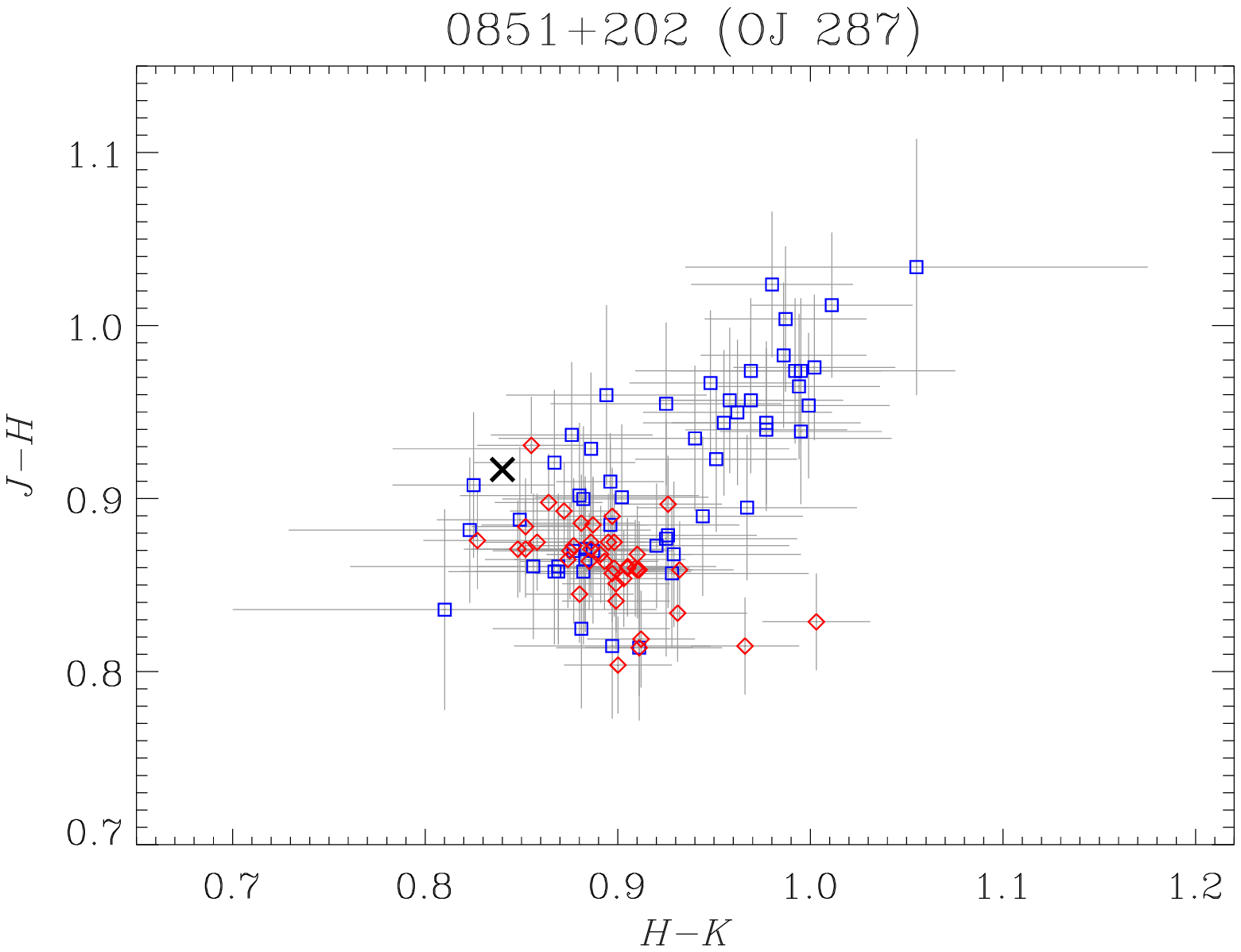,width=0.30\linewidth}}
    \vspace{0.5cm}
    \centerline{
    \psfig{figure=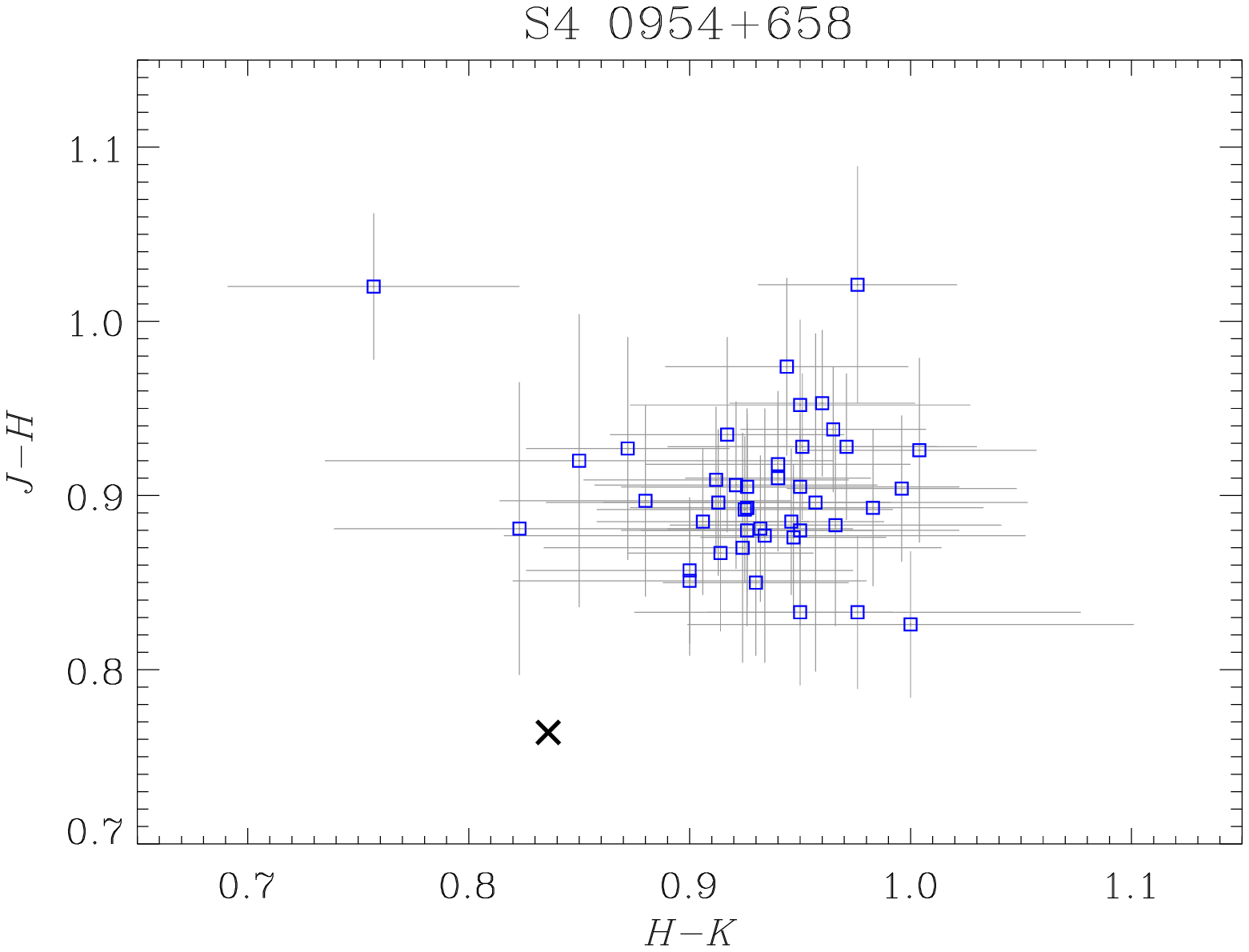,width=0.30\linewidth}
    \psfig{figure=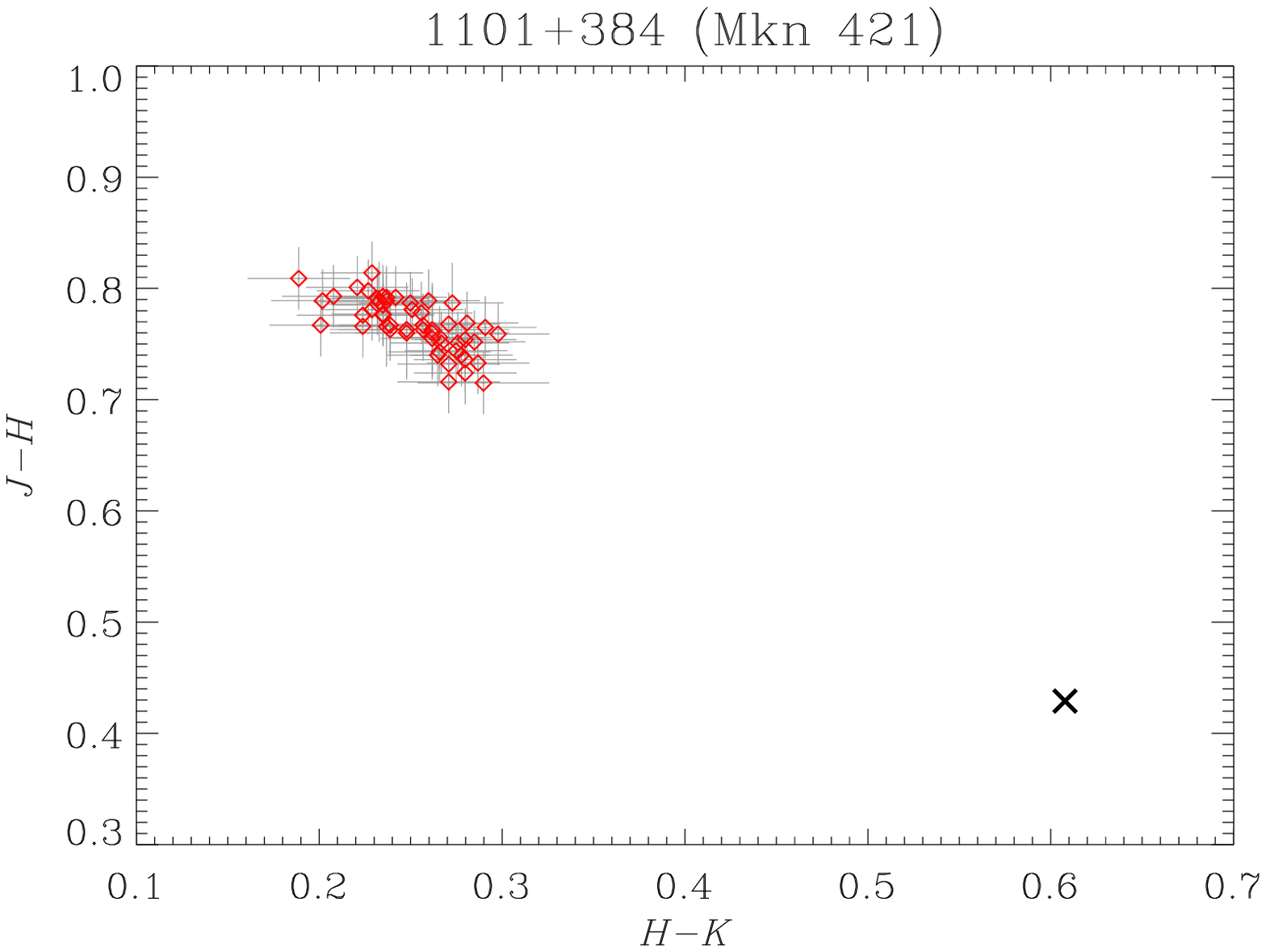,width=0.30\linewidth}
    \psfig{figure=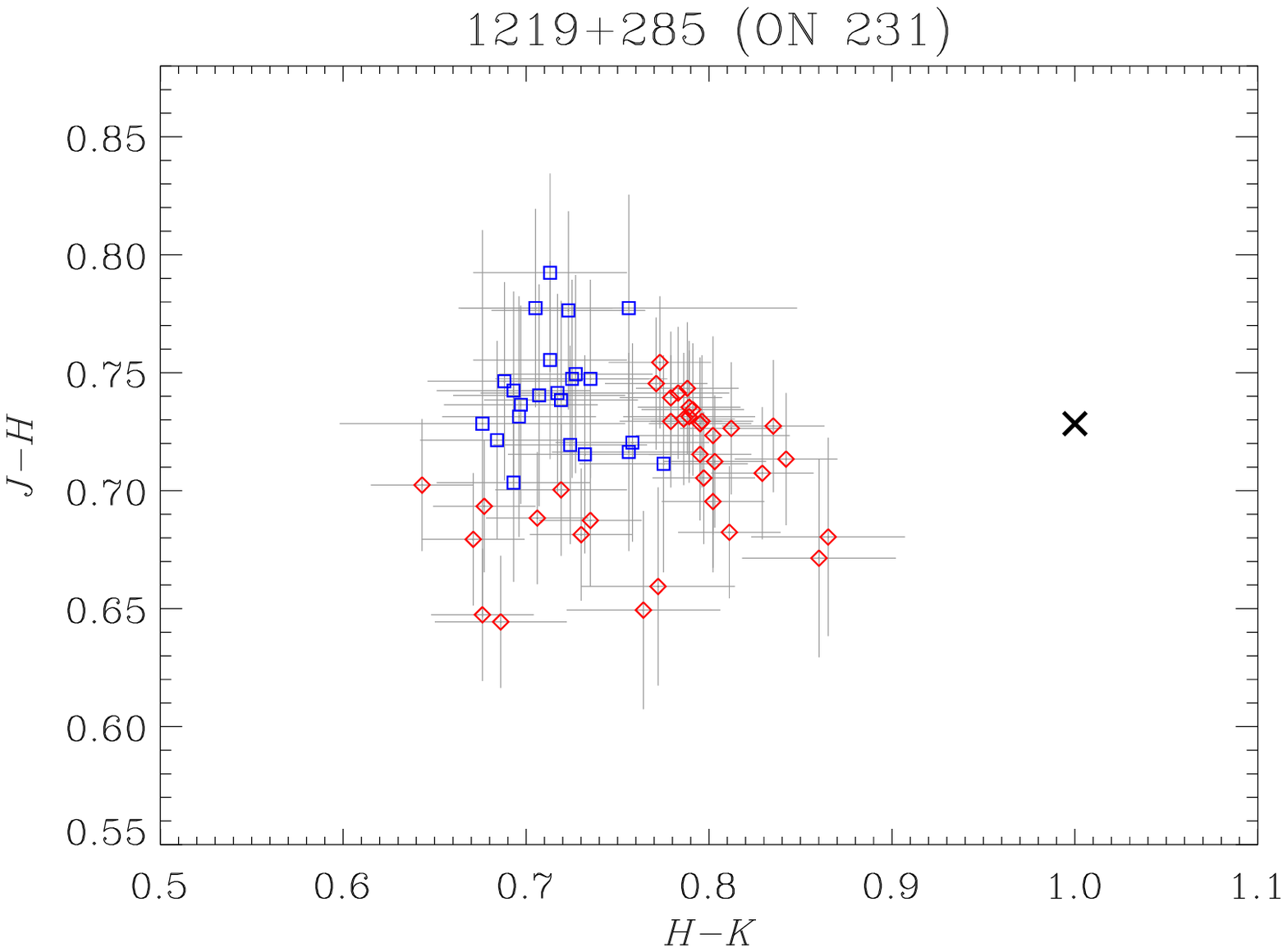,width=0.30\linewidth}}
    \vspace{0.5cm}
    \centerline{
    \psfig{figure=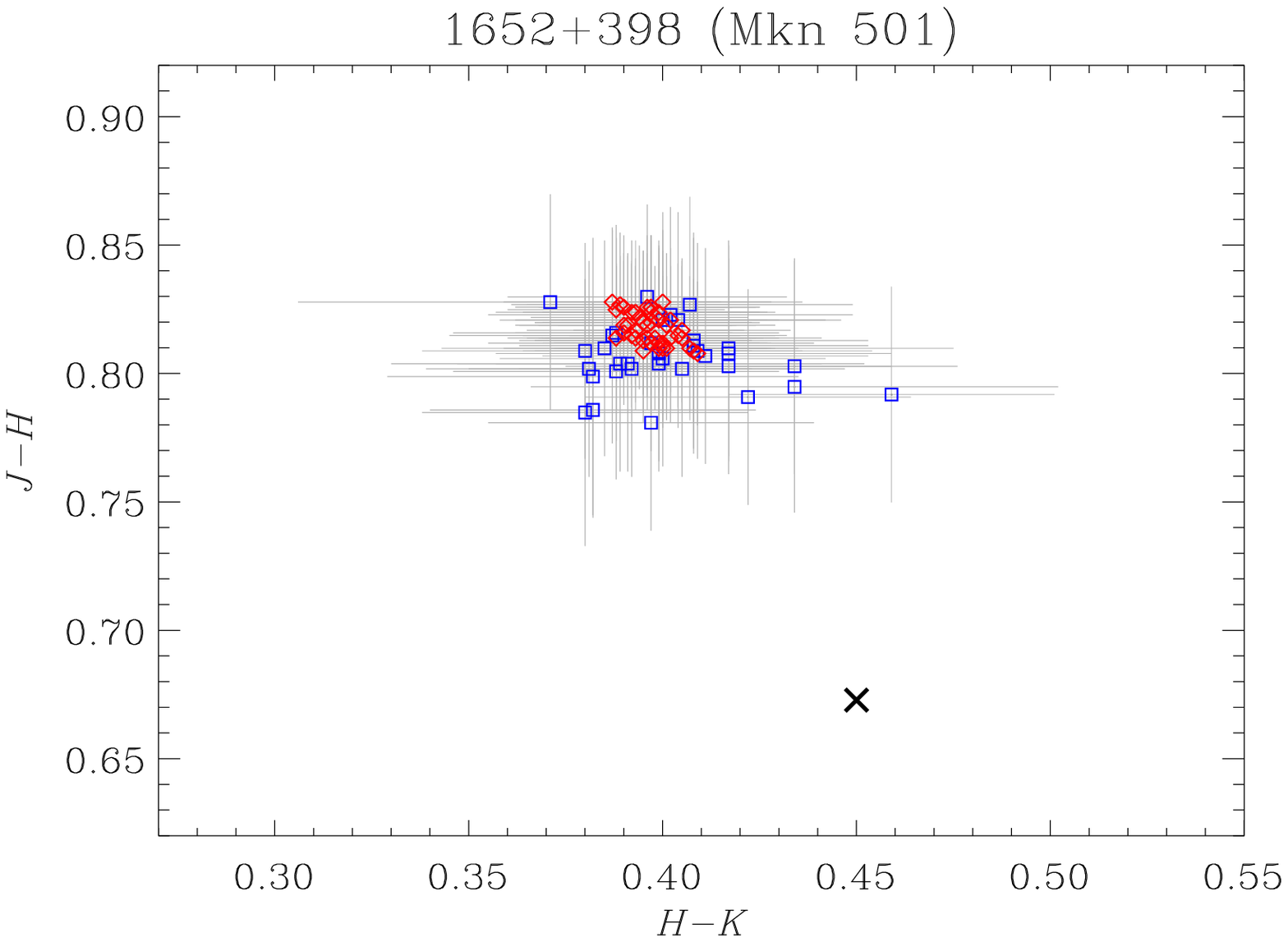,width=0.30\linewidth}
    \psfig{figure=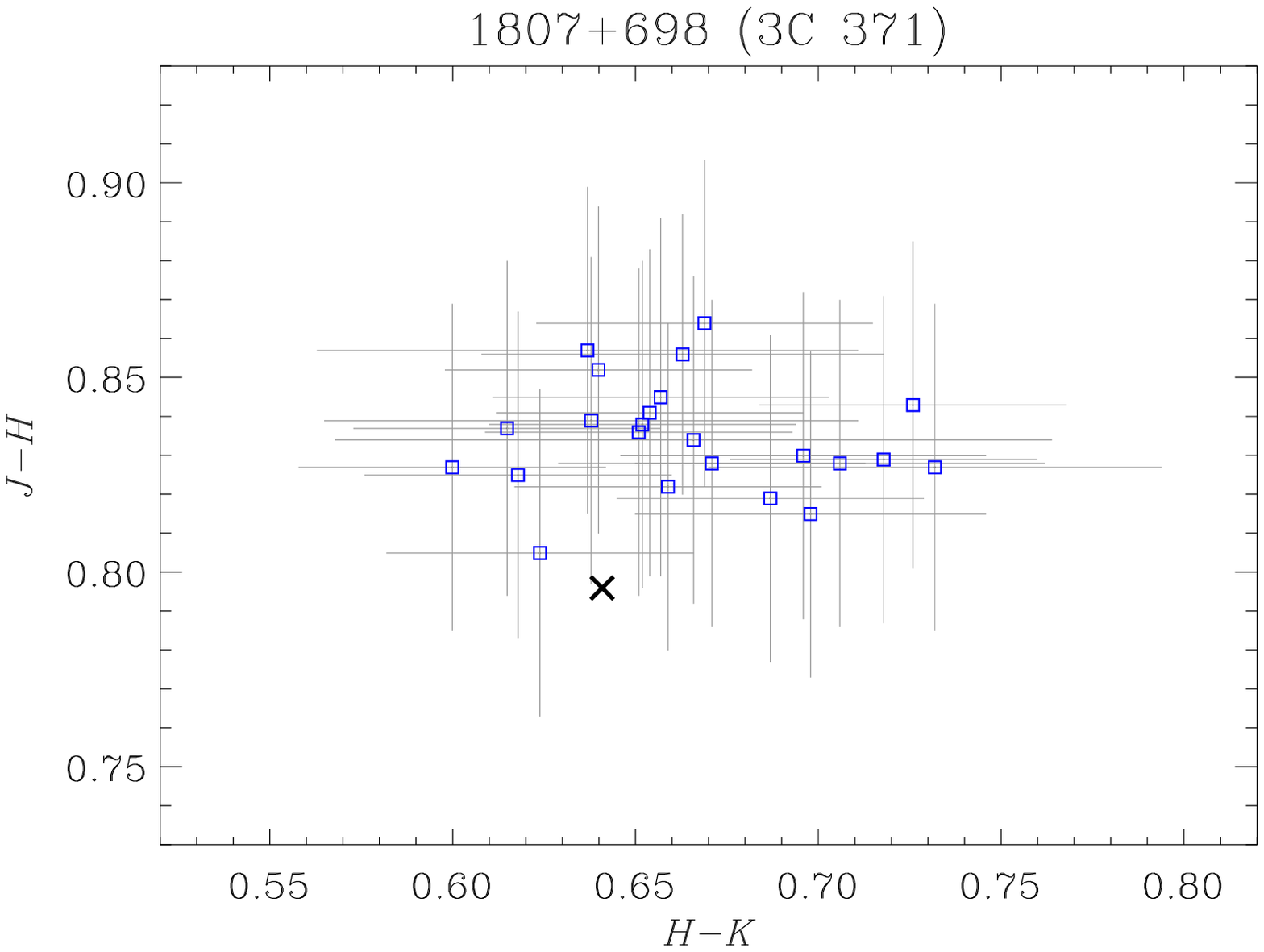,width=0.30\linewidth}
    \psfig{figure=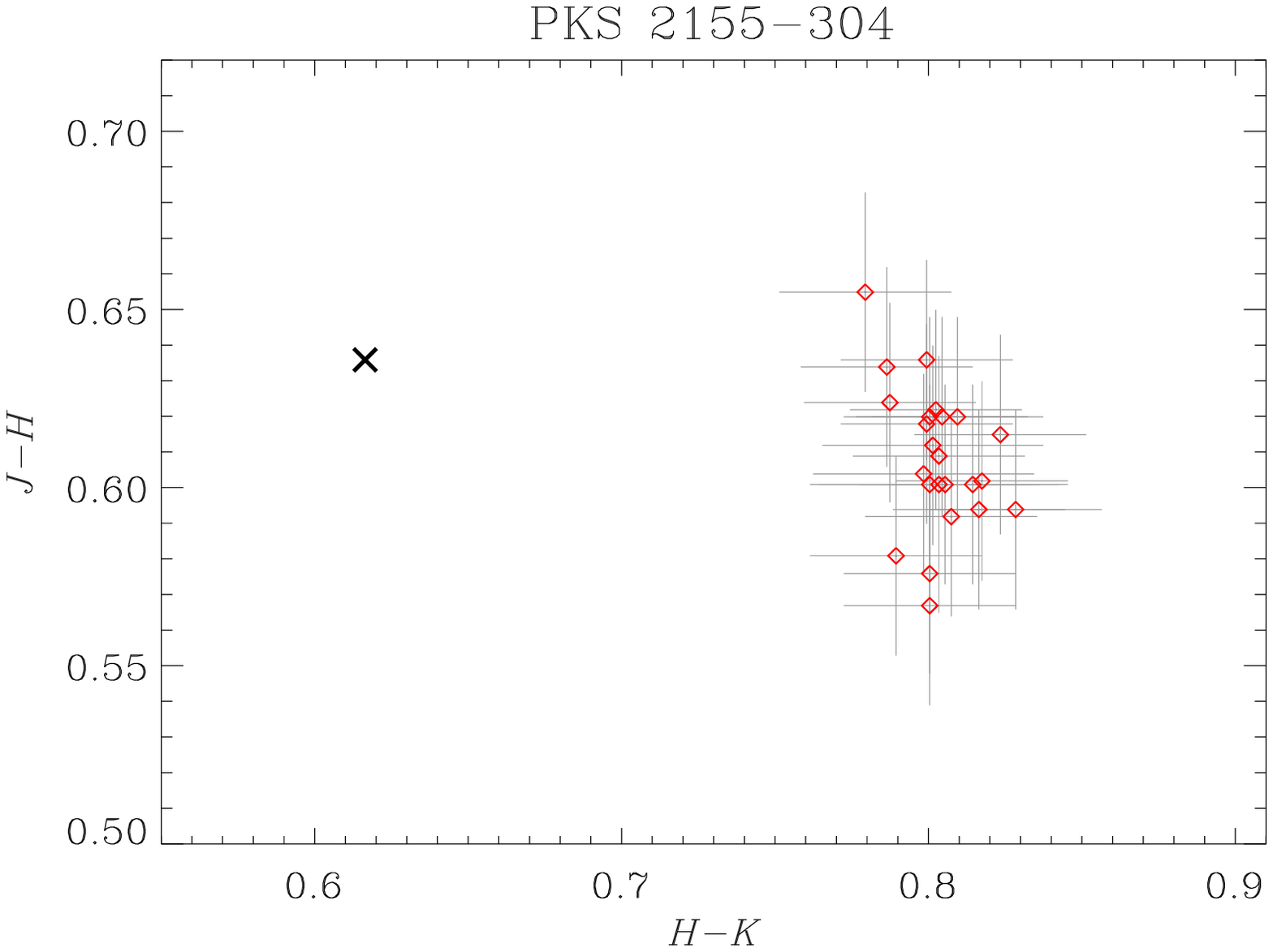,width=0.30\linewidth}}
    \vspace{0.5cm}
    \centerline{
    \psfig{figure=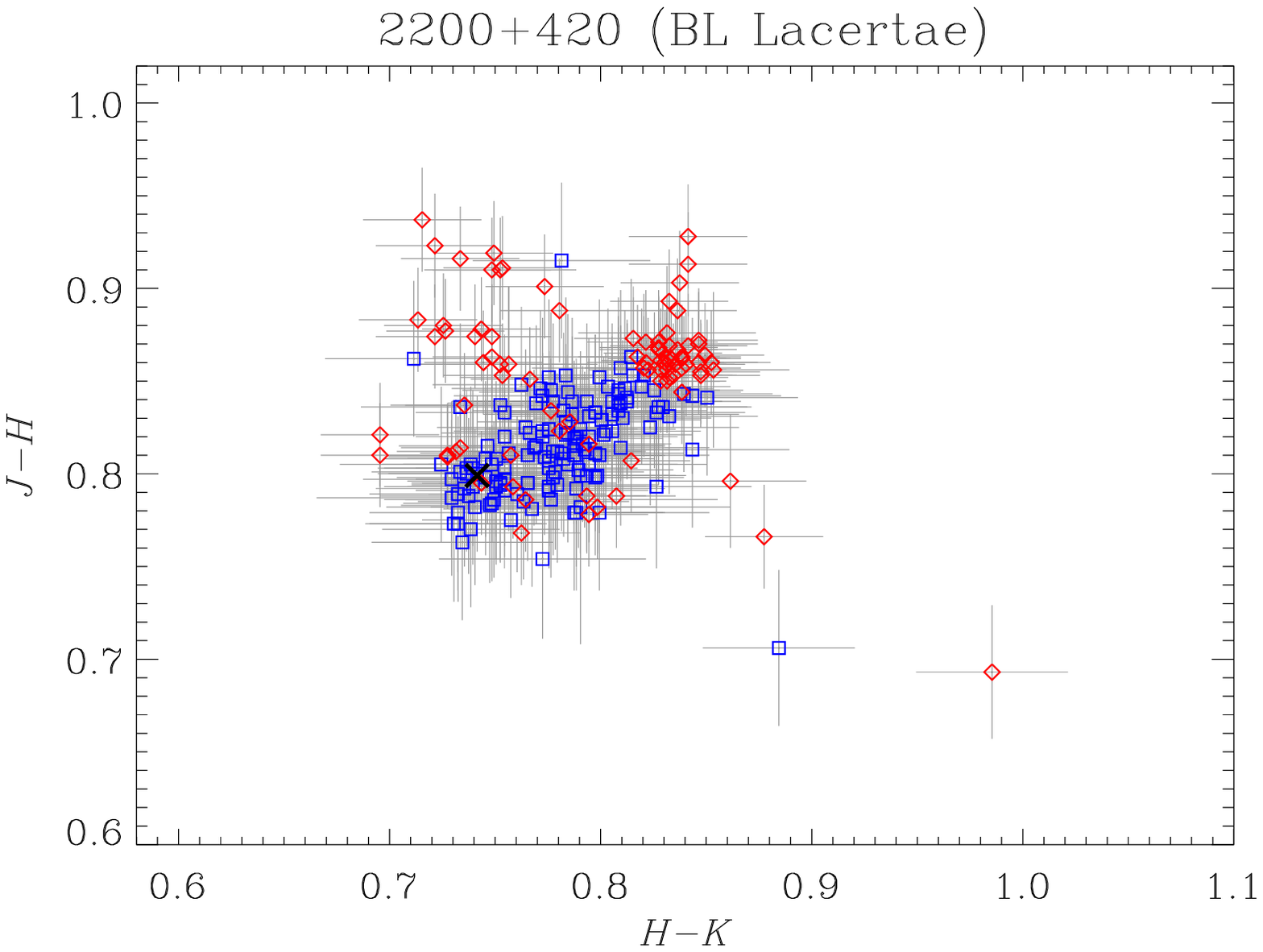,width=0.30\linewidth}
    \psfig{figure=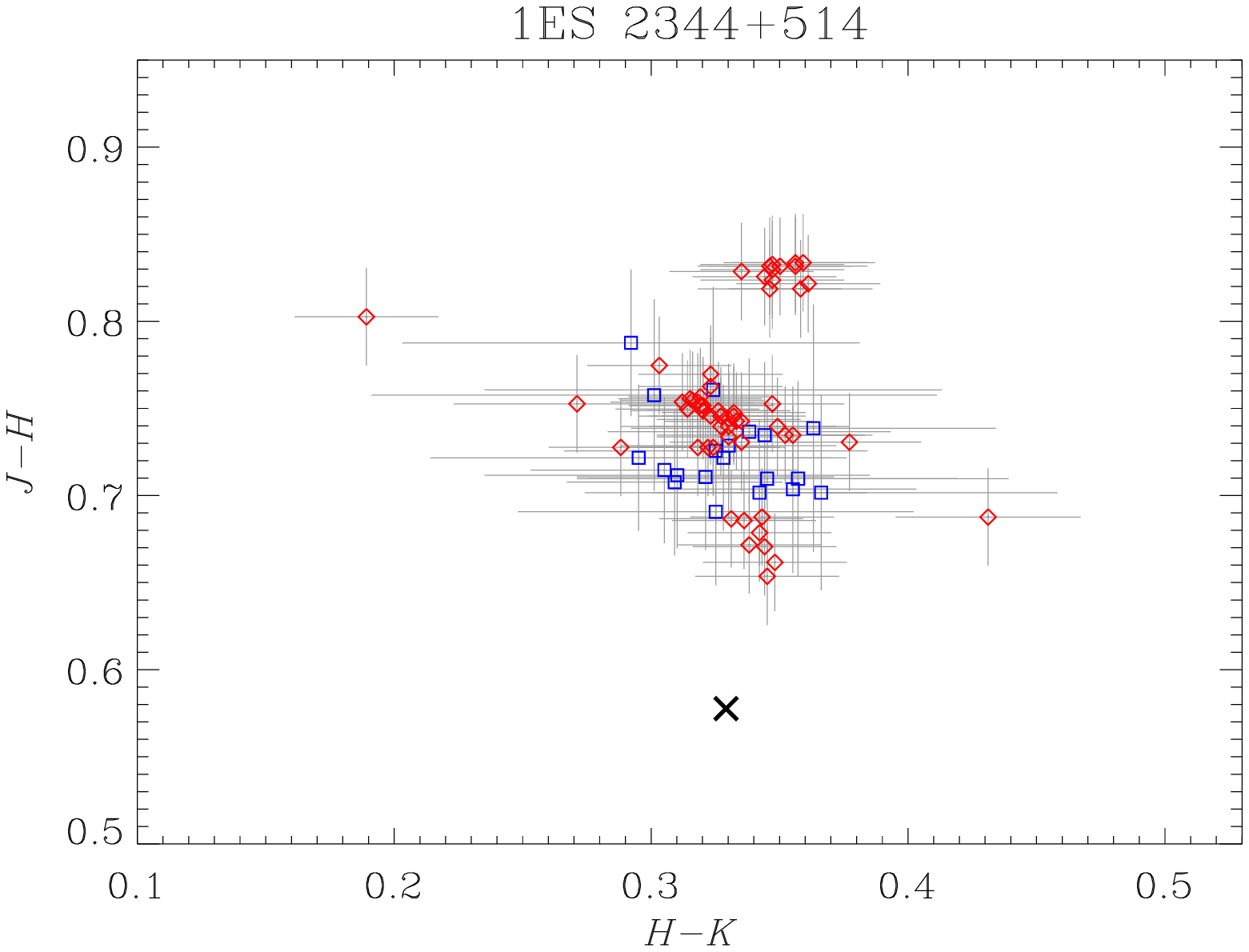,width=0.30\linewidth}}
    \caption{Near-IR colour-colour plots for the GASP-WEBT BL Lac objects. Data have been corrected for Galactic extinction. Blue squares indicate data from Campo Imperatore, red diamonds those from Teide. The black cross indicates the point obtained from the 2MASS catalogue.} 
    \label{colcol_bllacs} 
   \end{figure*}

\begin{figure*}
    \vspace{0.5cm}
    \centerline{
    \psfig{figure=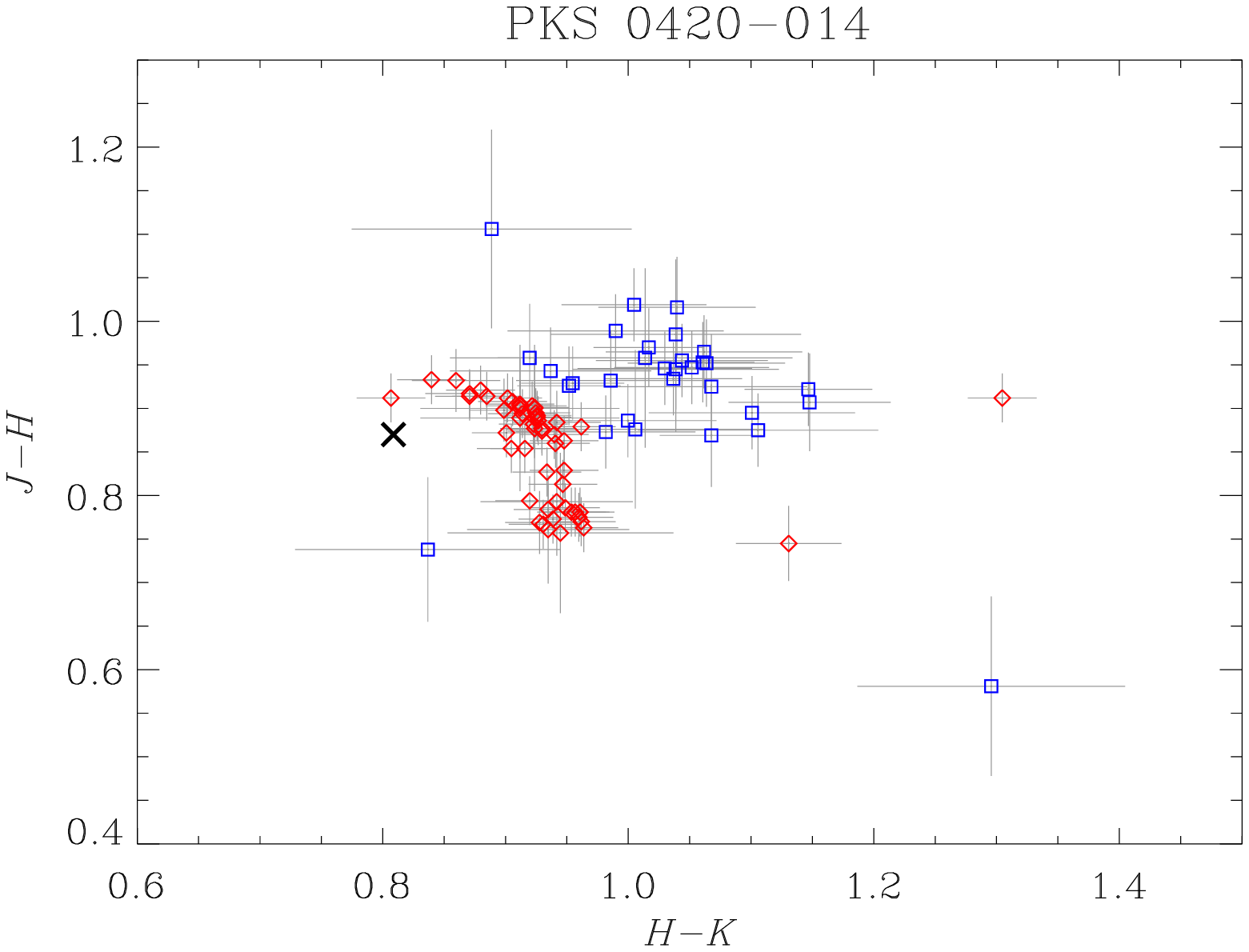,width=0.30\linewidth}
    \psfig{figure=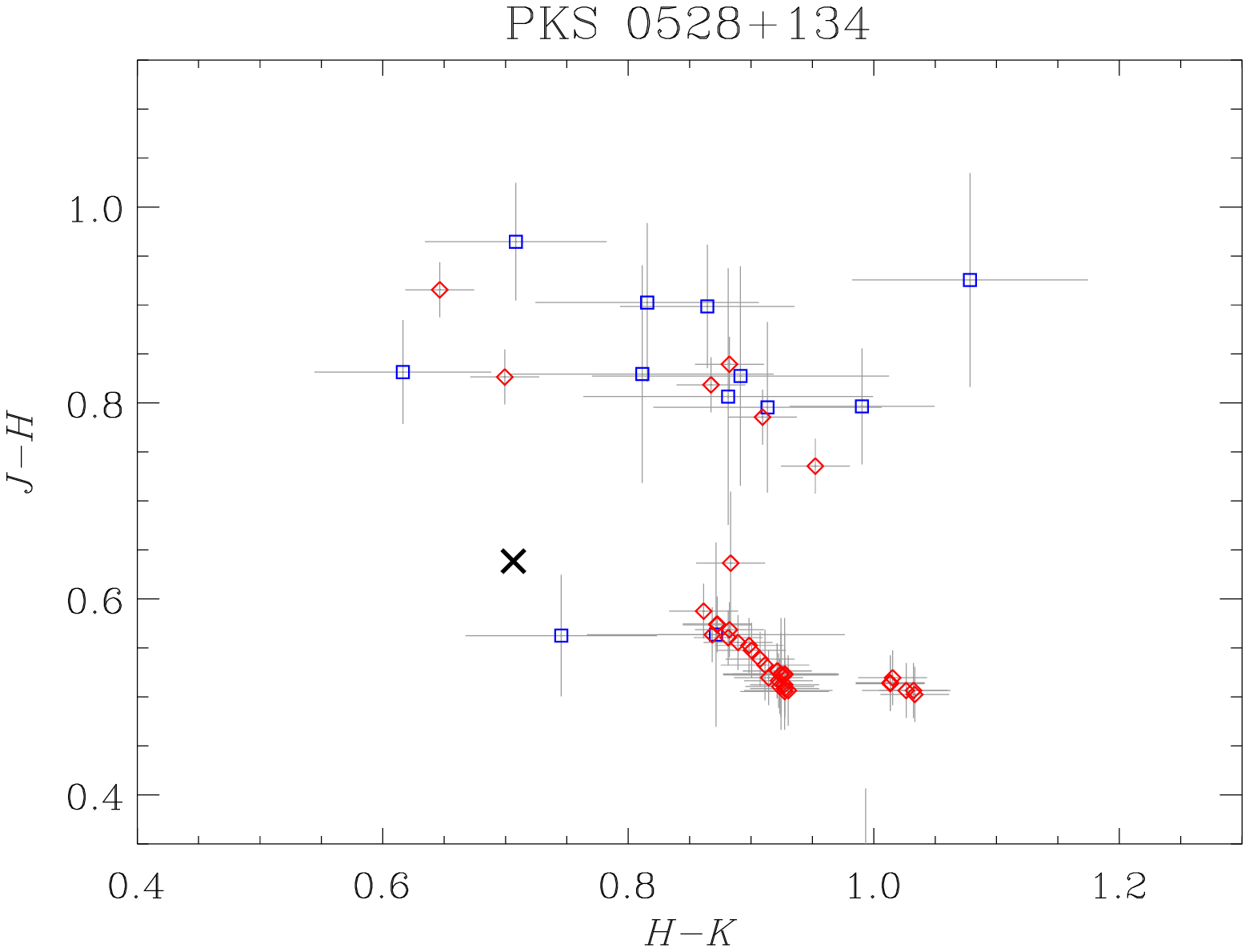,width=0.30\linewidth}
    \psfig{figure=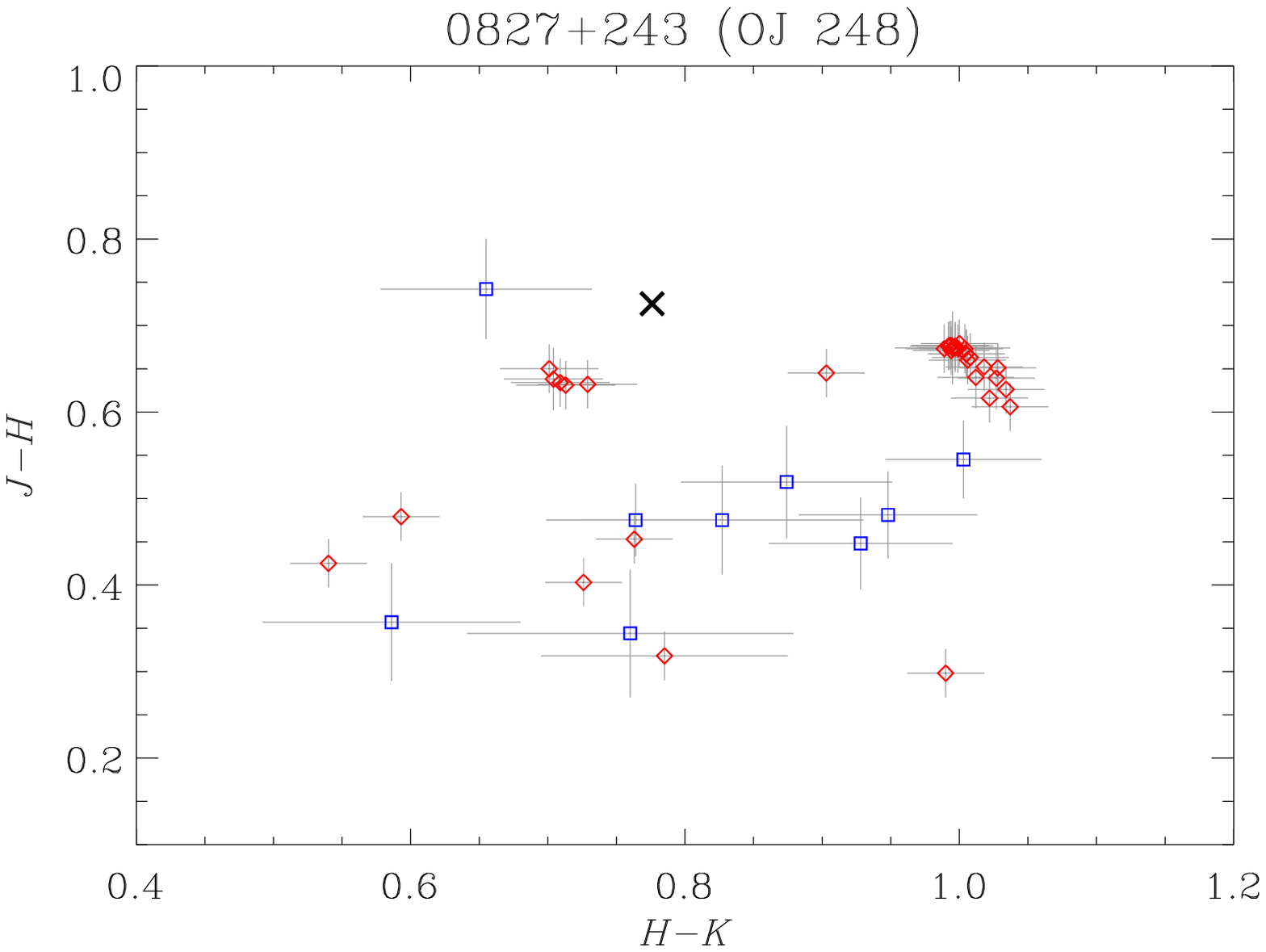,width=0.30\linewidth}}
    \centerline{
    \psfig{figure=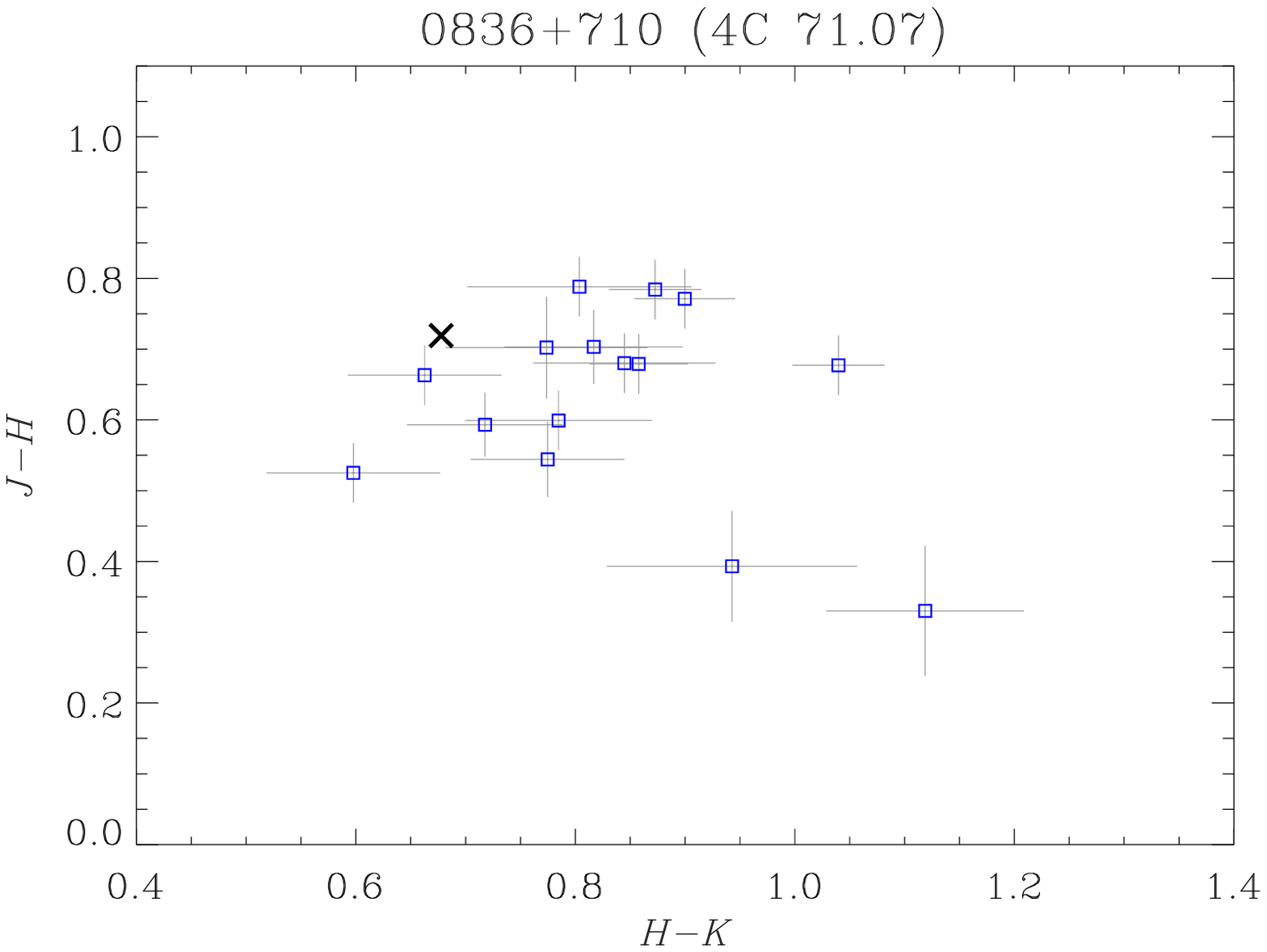,width=0.30\linewidth}
    \psfig{figure=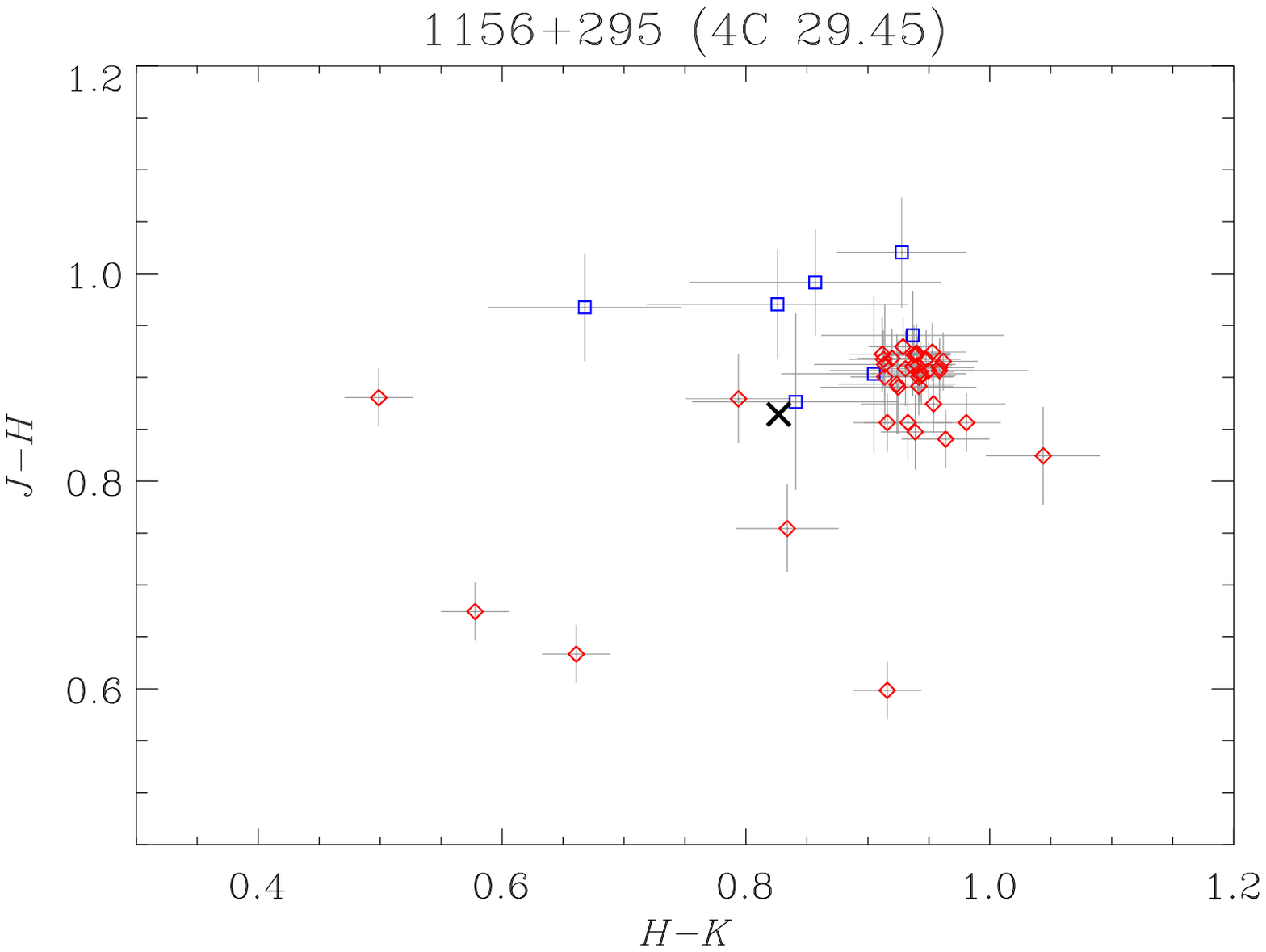,width=0.30\linewidth}
    \psfig{figure=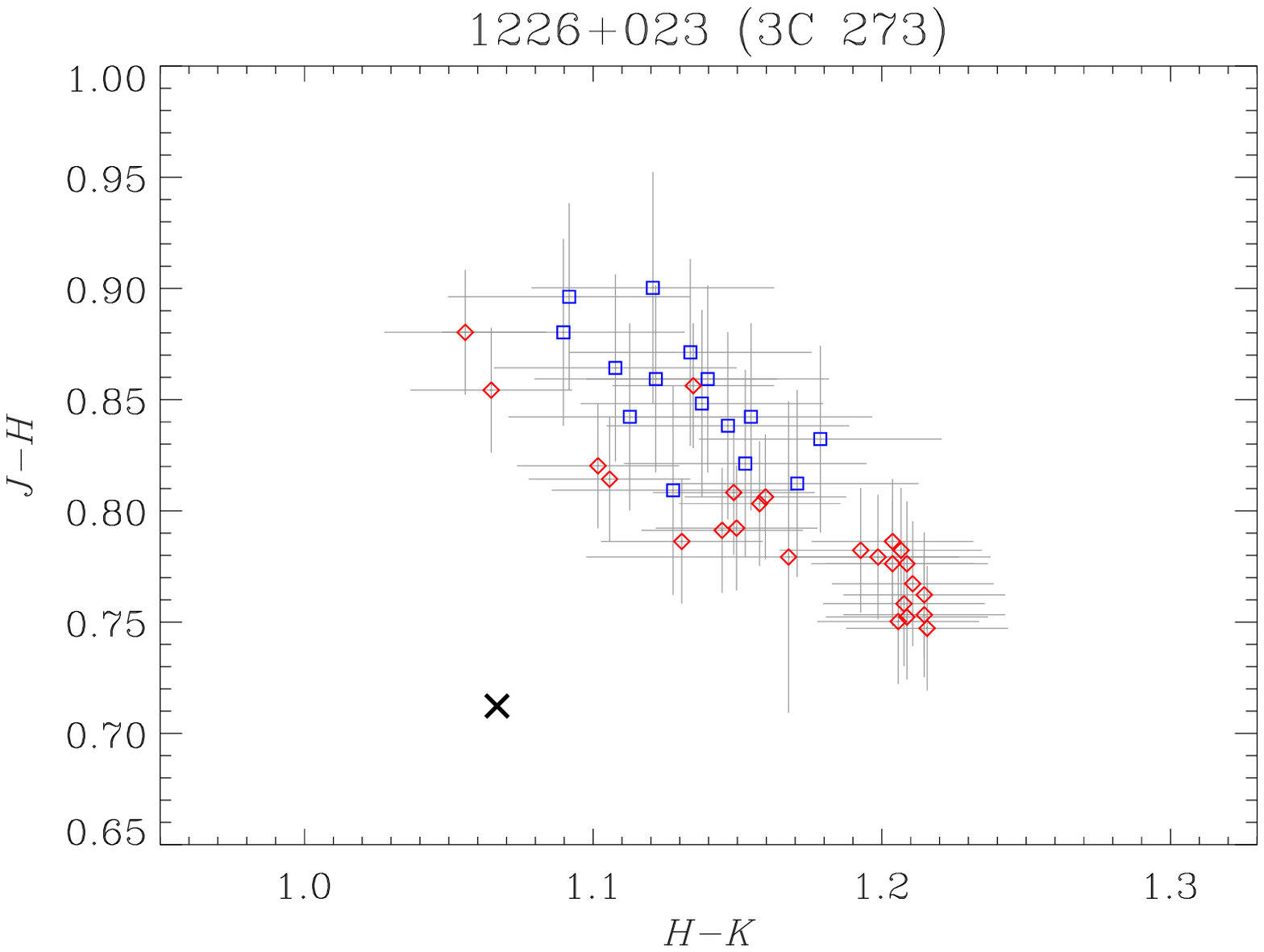,width=0.30\linewidth}}
    \centerline{
    \psfig{figure=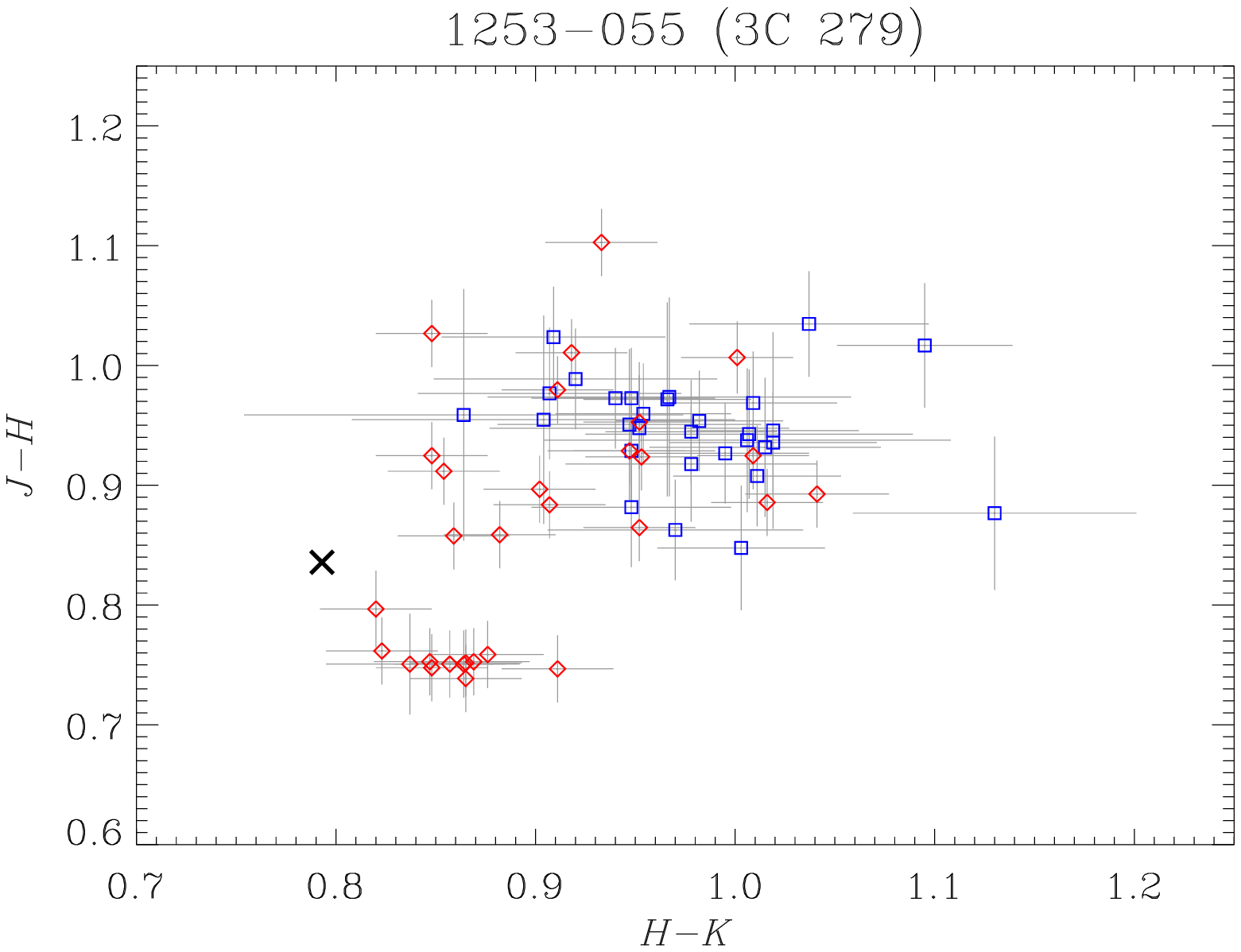,width=0.30\linewidth}
    \psfig{figure=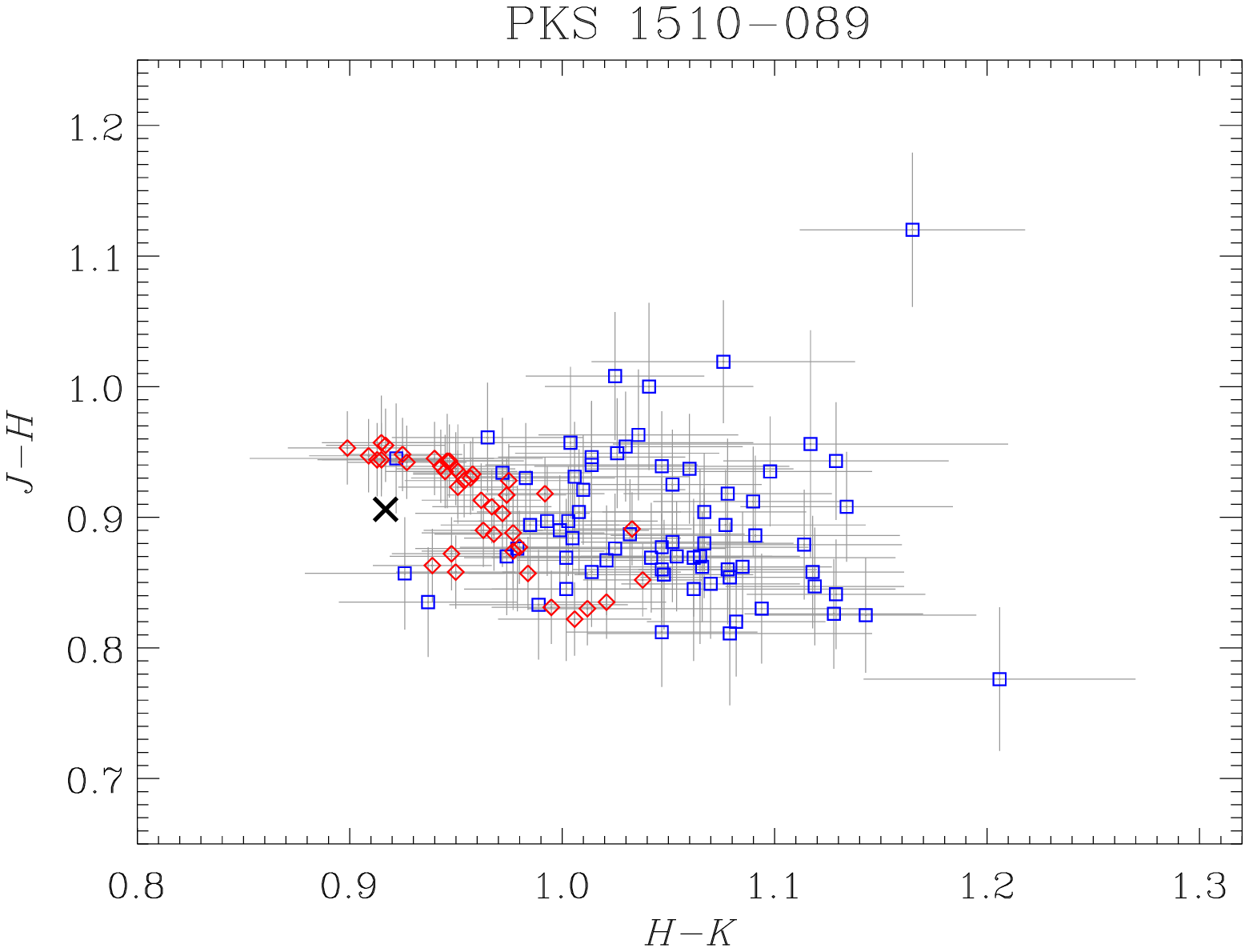,width=0.30\linewidth}
    \psfig{figure=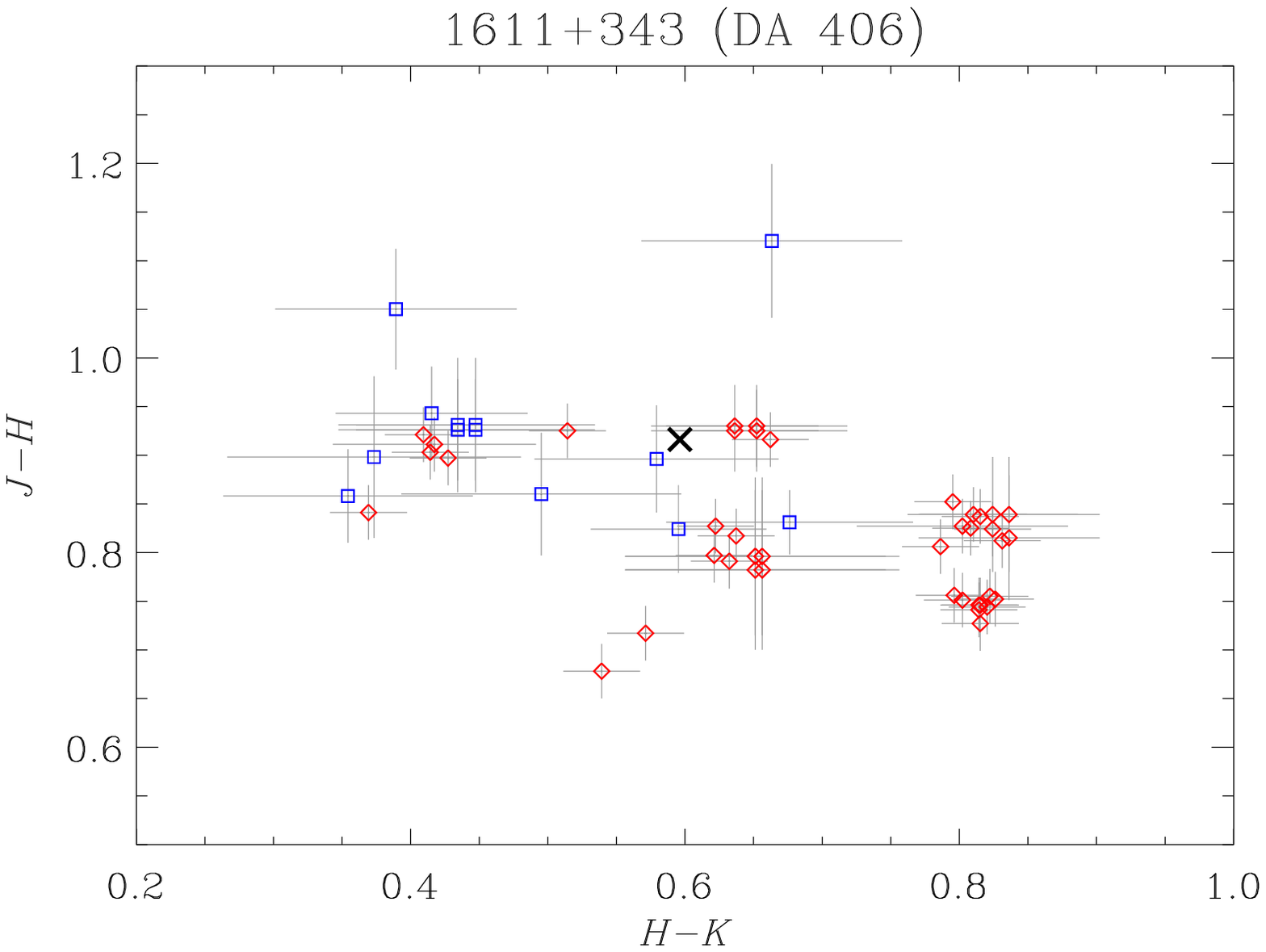,width=0.30\linewidth}}
    \centerline{
    \psfig{figure=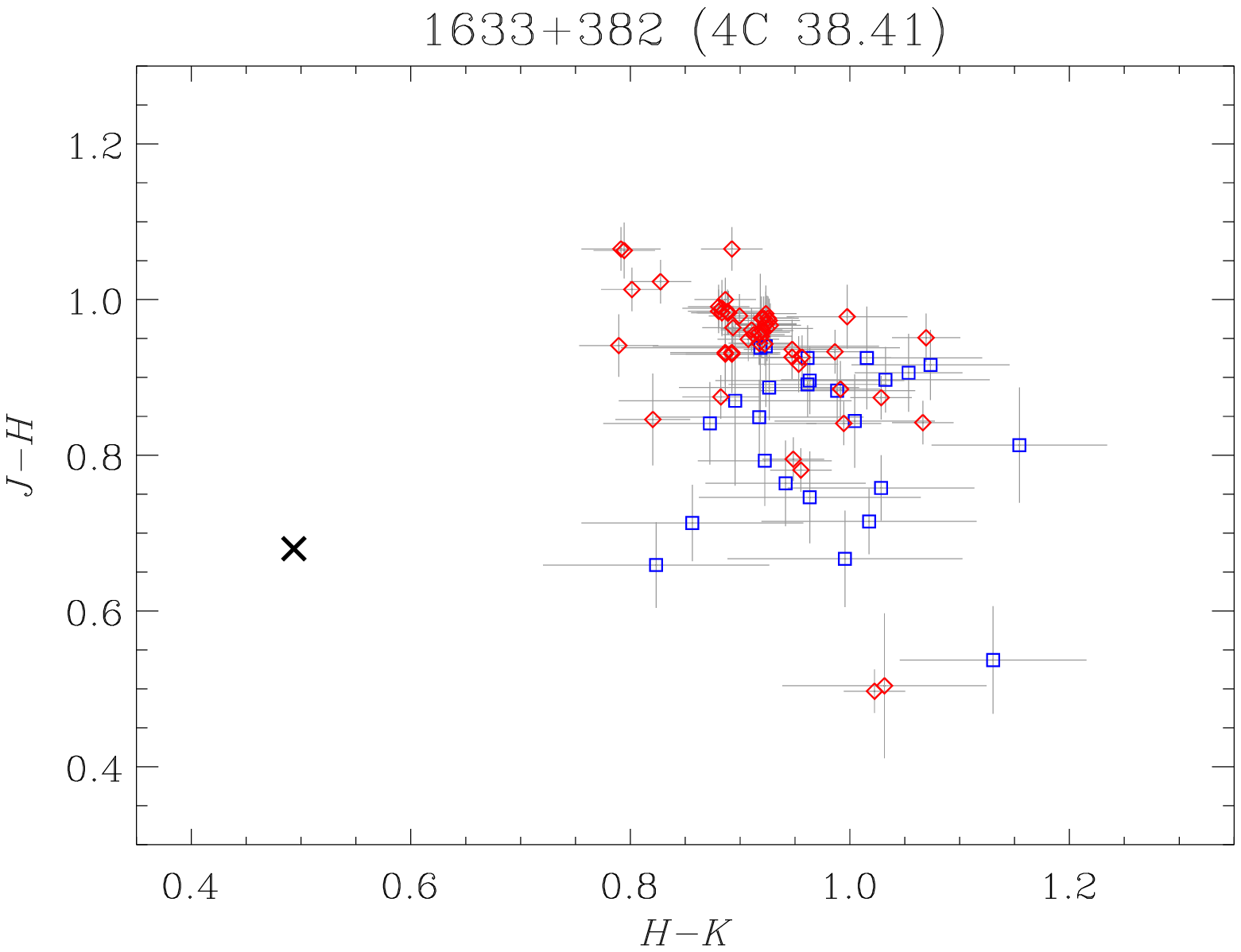,width=0.30\linewidth}
    \psfig{figure=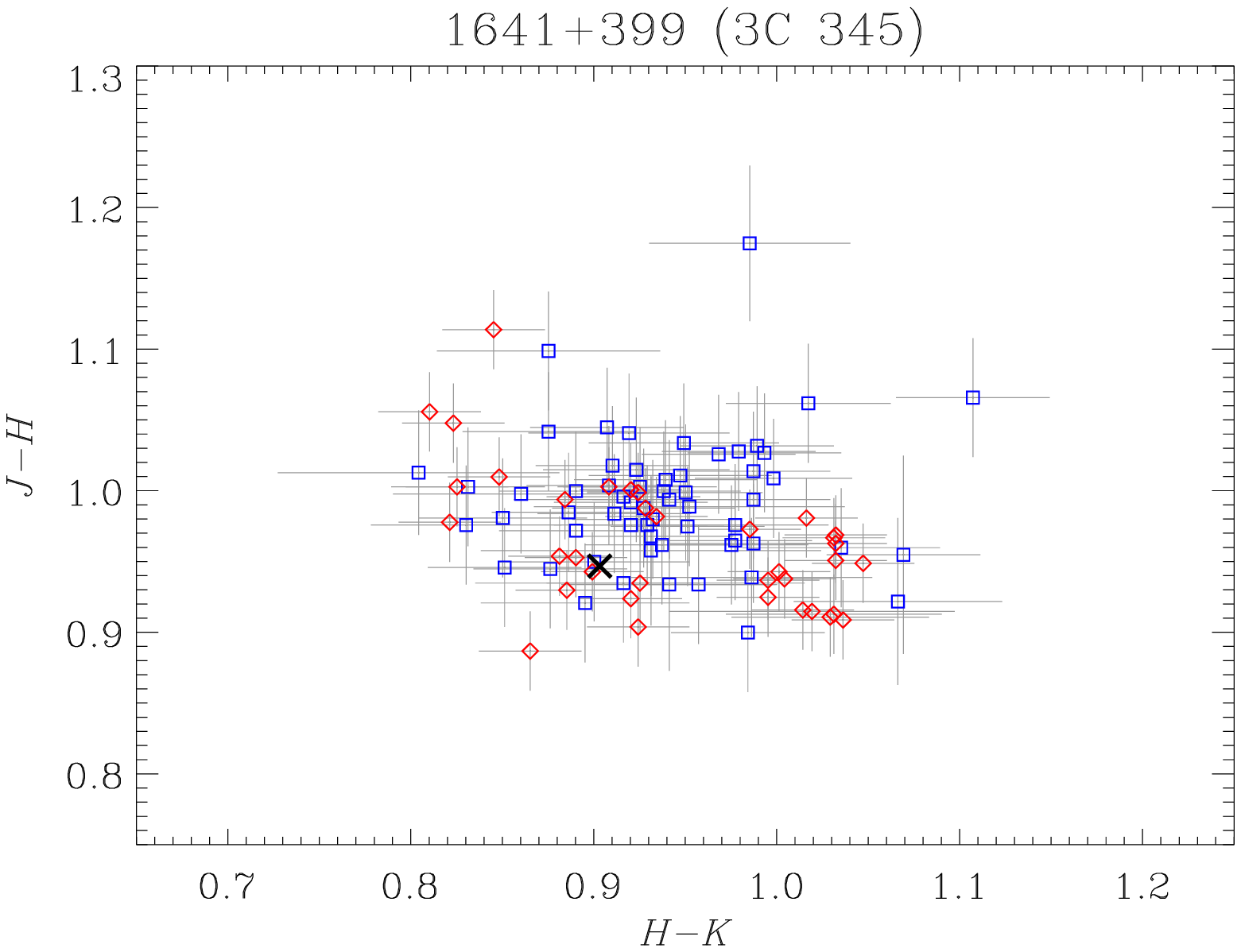,width=0.30\linewidth}
    \psfig{figure=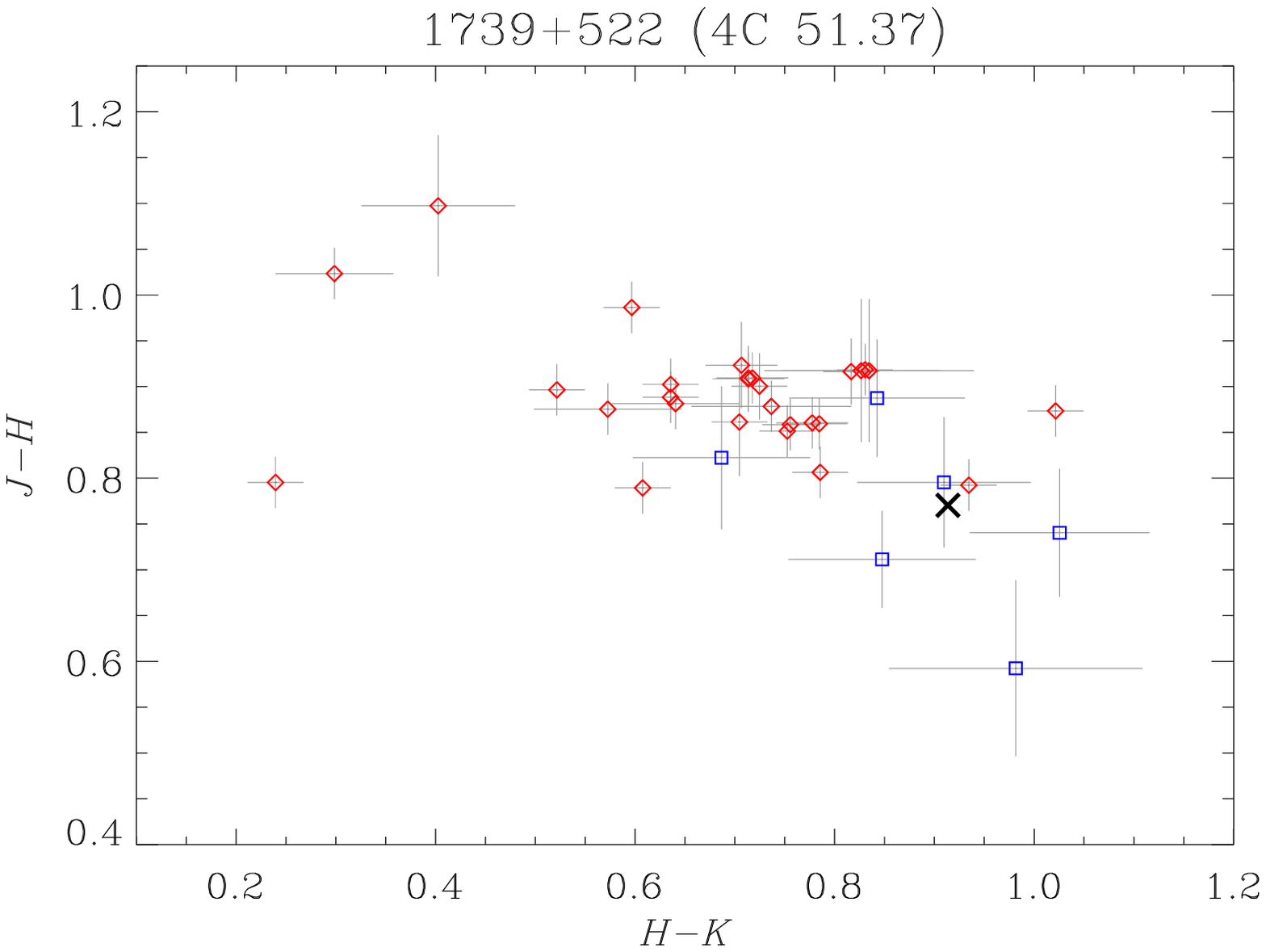,width=0.30\linewidth}}
    \centerline{
    \psfig{figure=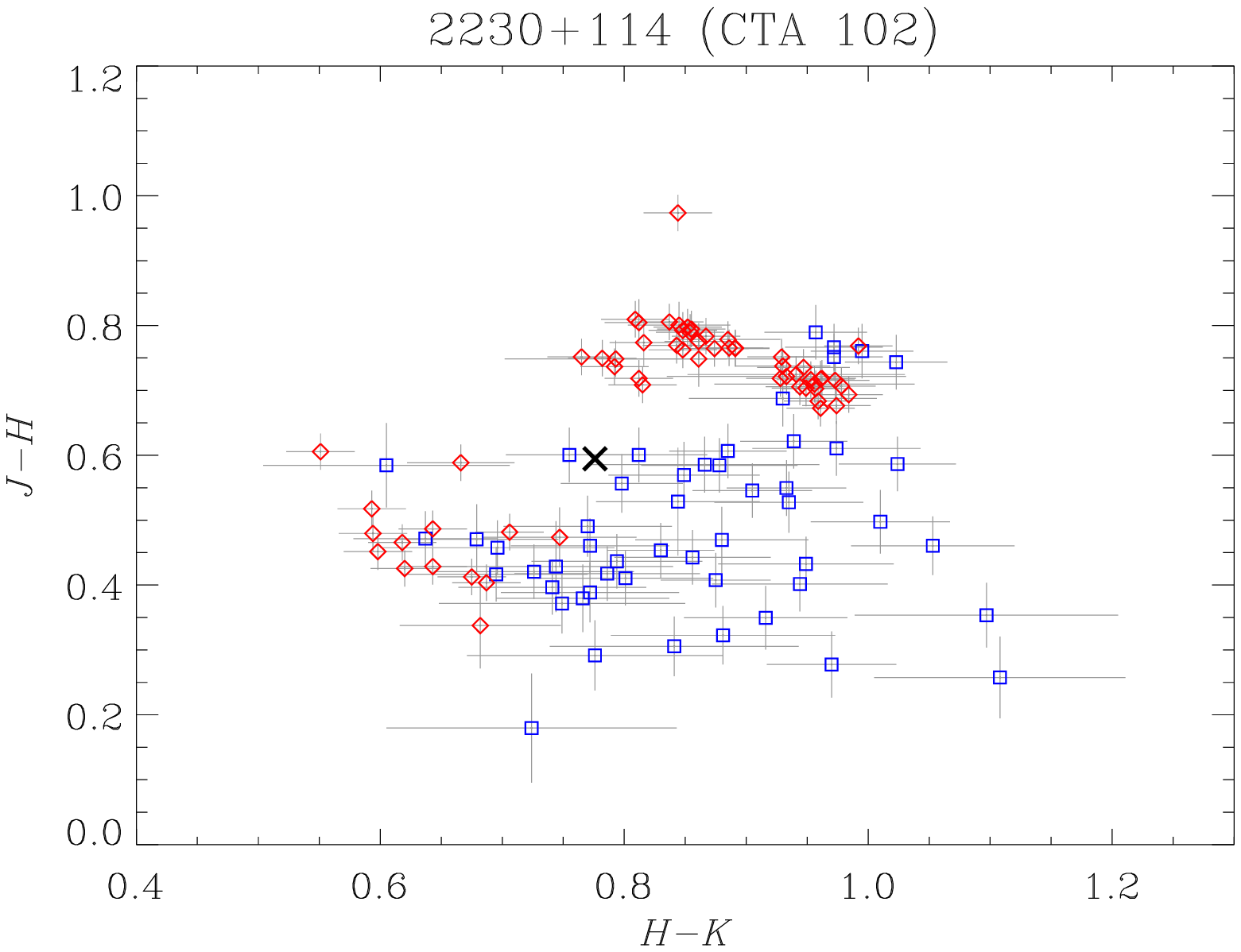,width=0.30\linewidth}
    \psfig{figure=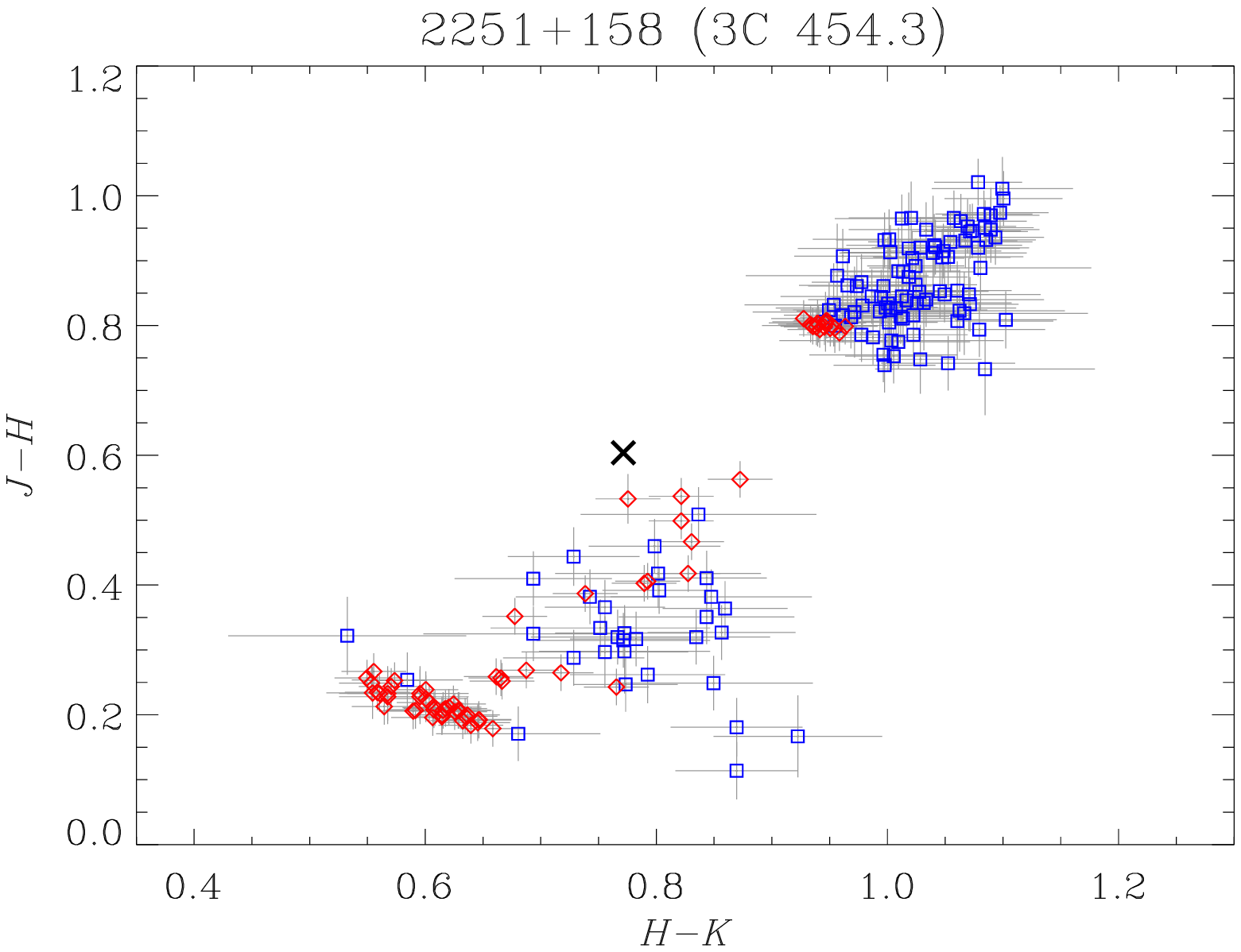,width=0.30\linewidth}}
    \caption{As in Fig.\ \ref{colcol_bllacs} but for the GASP-WEBT FSRQs.} 
    \label{colcol_fsrqs}    
   \end{figure*}

To analyse the near-infrared spectral variability, we first binned the $J$, $H$, and $K$ data of each observatory acquired within 30 minutes and then coupled the binned data with error less than 0.1 mag to build $J-H$ and $H-K$ colour indices.
Figures \ref{colcol_bllacs} and \ref{colcol_fsrqs} show the colour variability of the GASP sources in the $J-H$ versus $H-K$ plot, compared to the 2MASS value (see Fig.\ \ref{2mass_ext}).
Correction for Galactic extinction has been performed as explained in Sect.\ \ref{compa}.
In Tables 3 and 4 we report colour indices average values, their standard deviations and mean fractional variations (see Eq.\ 1). 
The latter are often imaginary numbers for BL Lac objects, which means that the corresponding colour index variability is dominated by noise. When this is not the case, the value is however low, in the range 1--7\%. This is also true for 0235+164, the source with the largest flux variability, which nonetheless is the BL Lac showing the most noticeable spectral changes. In general, larger near-IR spectral variability characterises FSRQs because of the interplay between synchrotron and quasar-like emission components, the latter including contributions from the accretion disc, broad line region, and torus (see Sect.\ \ref{sed}). Exceptions are 3C 273, PKS 1510$-$089, and 3C 345.
The sources exhibiting the largest colour indices changes are 3C 454.3 and CTA 102. 

By looking at Figs.\ \ref{colcol_bllacs} and \ref{colcol_fsrqs} one can see that sometimes the 2MASS value is outside the region where most of our points collect. This may partly reflect cross-calibration problems, but may also indicate an even larger variability.

Two sources show clear trends in the colour-colour diagram.
The anticorrelation between $J-H$ and $H-K$ in the 3C 273 plot suggests that there is a spectral break in the $H$ band.
In contrast, a correlation is visible in the colour-colour plot of 3C 454.3, where the lower-left group of points refer to the source faint state from 2011 onward, while the upper-right group of points come from the bright state before 2011. The colour-colour plot then tell us that the near-IR spectrum is bluer in faint states and redder in bright states. A redder-when-brighter trend has already been noticed when analysing the optical spectrum of this source \citep[e.g.][]{vil06,rai08c}.
These spectral behaviours will become clearer in the next section, where spectral energy distributions are presented and interpreted in terms of different emission contributions.

\section{Infrared-to-optical spectral energy distribution}
\label{sed}
In this section we present the infrared-to-optical spectral energy distributions of the GASP sources.
They are built with the survey data already presented in Sect.\ \ref{compa} ({\em WISE}, 2MASS, SDSS), literature data, and the results of our near-IR monitoring at Campo Imperatore and Teide. 
Notice that a {\em WISE} single exposure source database is available, containing multiple observations of the GASP sources. However, the use of this database is discouraged because it includes all single exposure images regardless of their quality\footnote{http://wise2.ipac.caltech.edu/docs/release/allsky/expsup}, so we did not consider it.
From the literature we extracted data from the Infrared Astronomical Satellite \citep[{\em IRAS},][]{imp88}, the Infrared Space Observatory \citep[{\em ISO},][]{haa04}, and the {\em Spitzer} satellite \citep{mal11,ogl11}.
{\em Herschel} data for 3C 454.3 are from \citet{weh12}.
For 3C 273 and 3C 371 we used the optical spectra acquired at the Telescopio Nazionale Galileo (TNG) by \citet{but09}, which are available on the NASA/IPAC Extragalactic Database (NED)\footnote{http://ned.ipac.caltech.edu}. In a few cases we reported SEDs from papers involving data from the GASP-WEBT collaboration for a more accurate modelling: \citet{dam11} for PKS 1510-089; \citet{rai12} for 4C 38.41, \citet{rai11} for 3C 454.3.

\begin{figure*}
    \vspace{0.5cm}
    \centerline{
    \psfig{figure=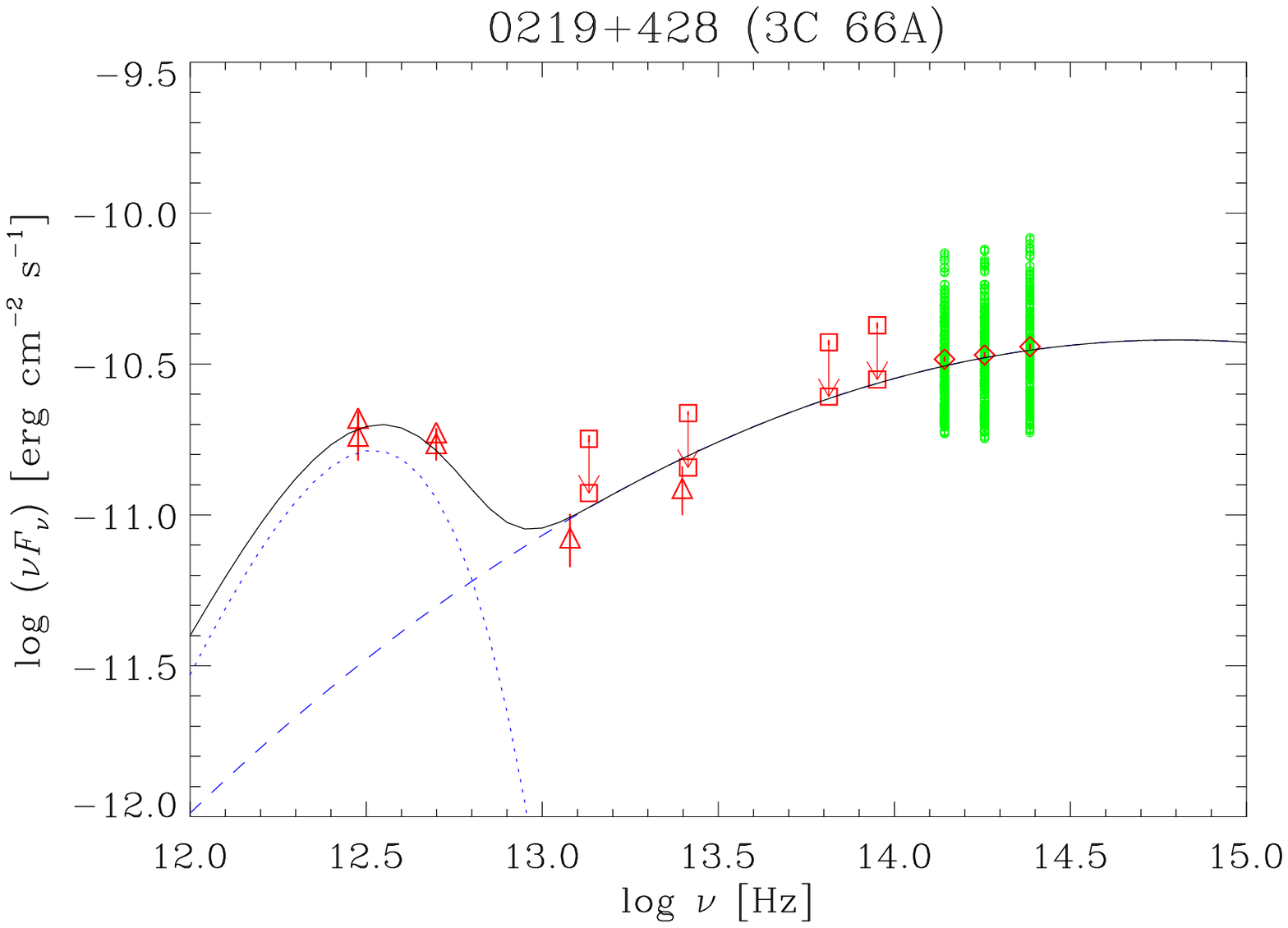,width=0.30\linewidth}
    \psfig{figure=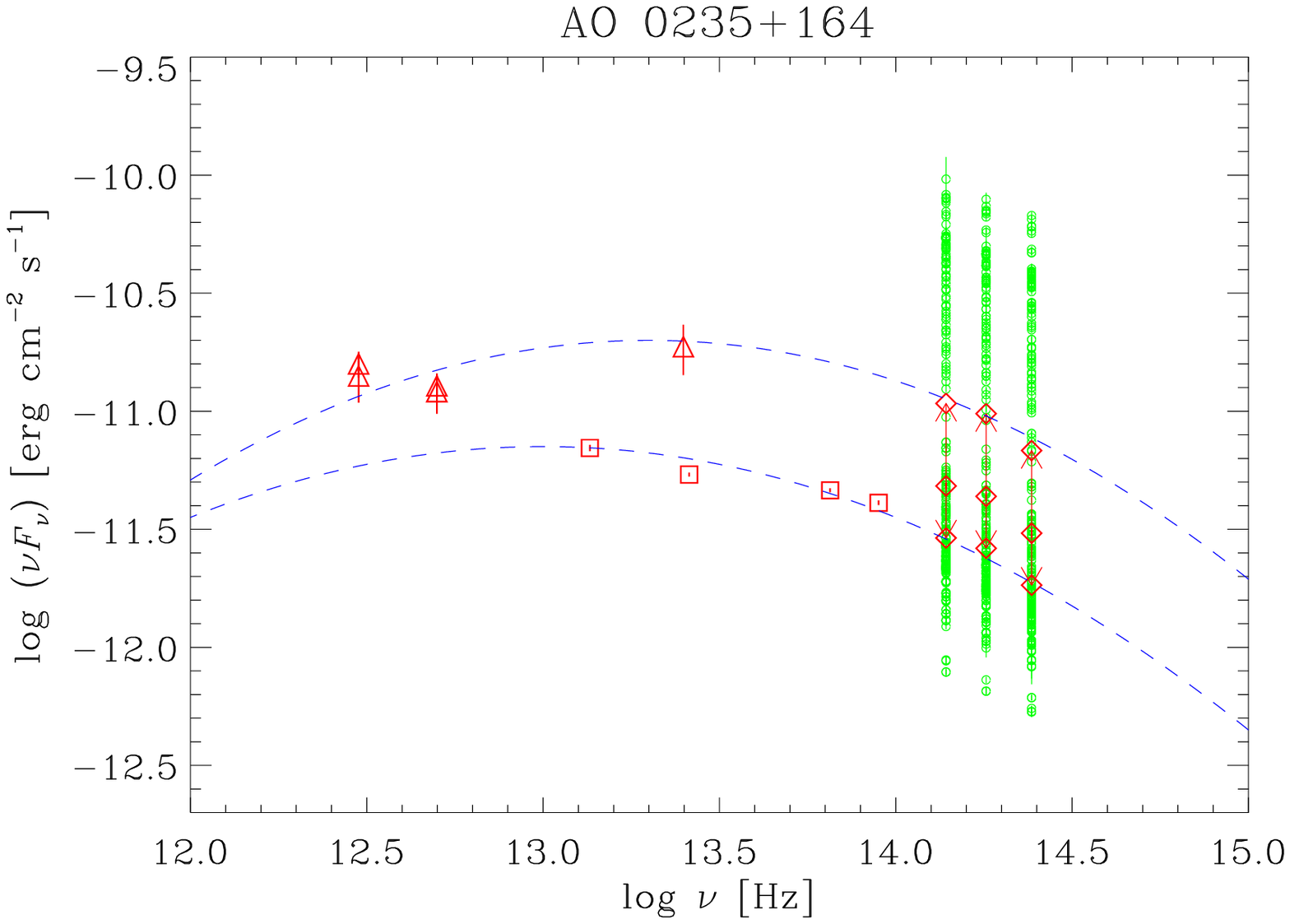,width=0.30\linewidth}
    \psfig{figure=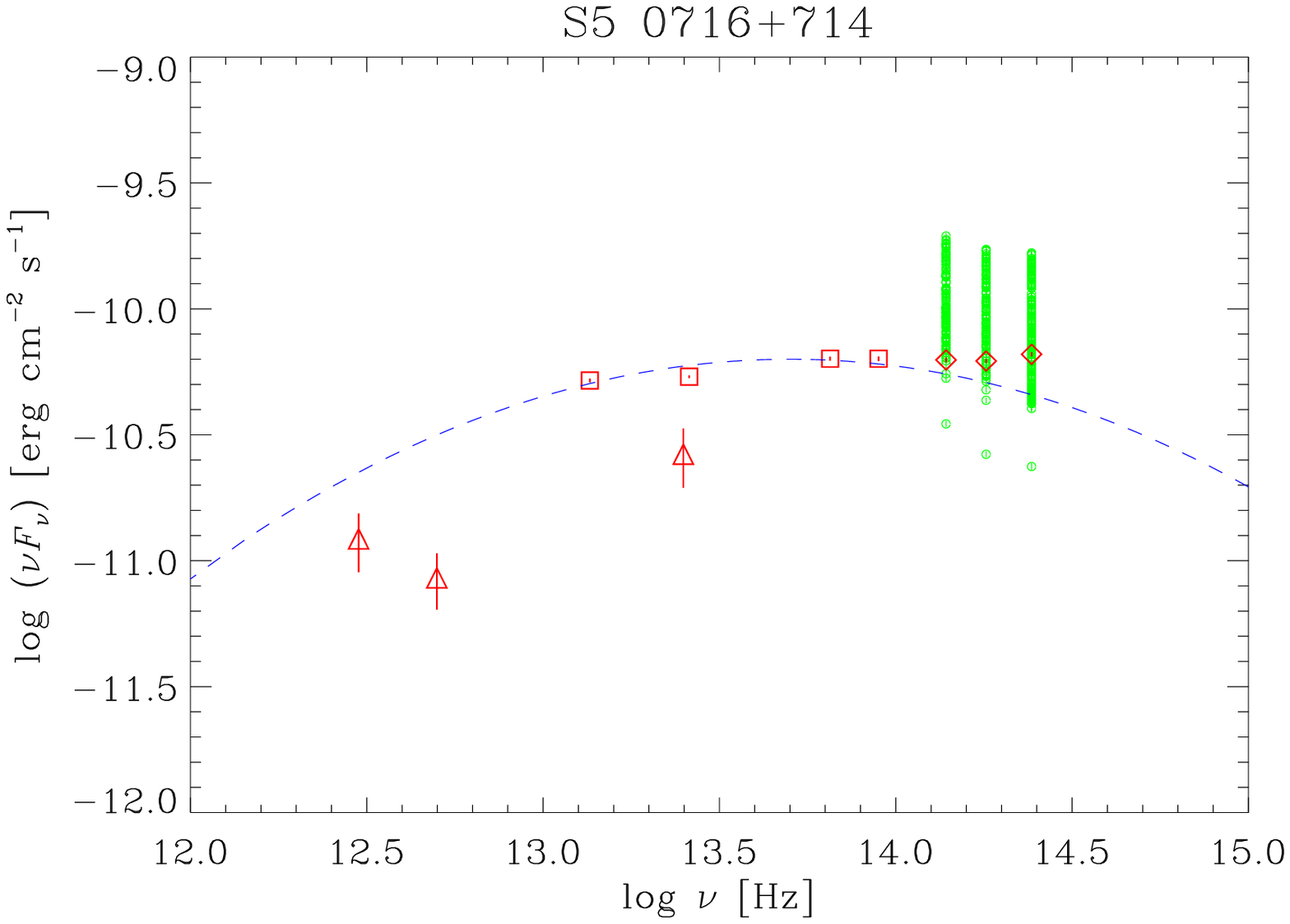,width=0.30\linewidth}}
    \vspace{0.5cm}
    \centerline{
    \psfig{figure=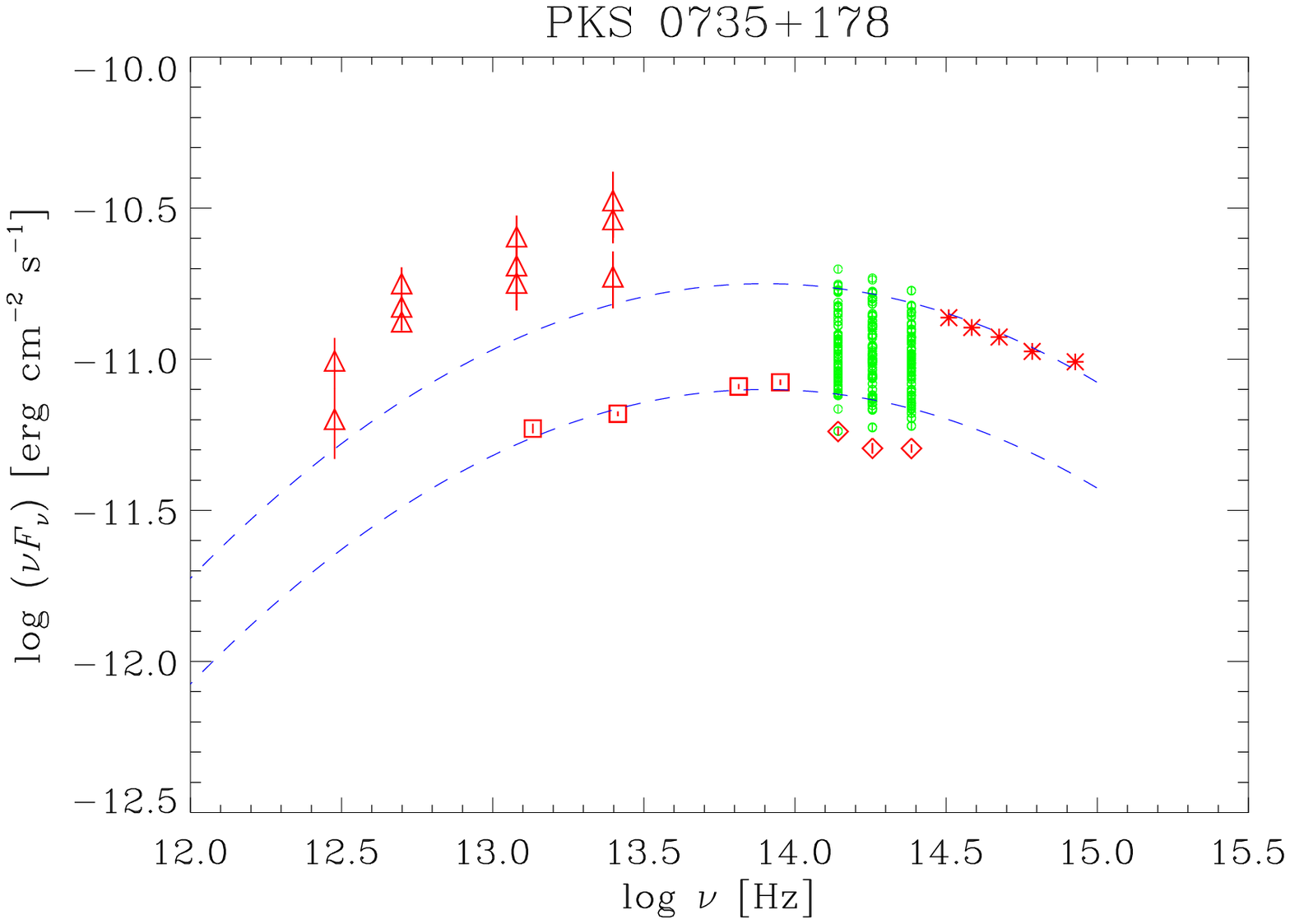,width=0.30\linewidth}
    \psfig{figure=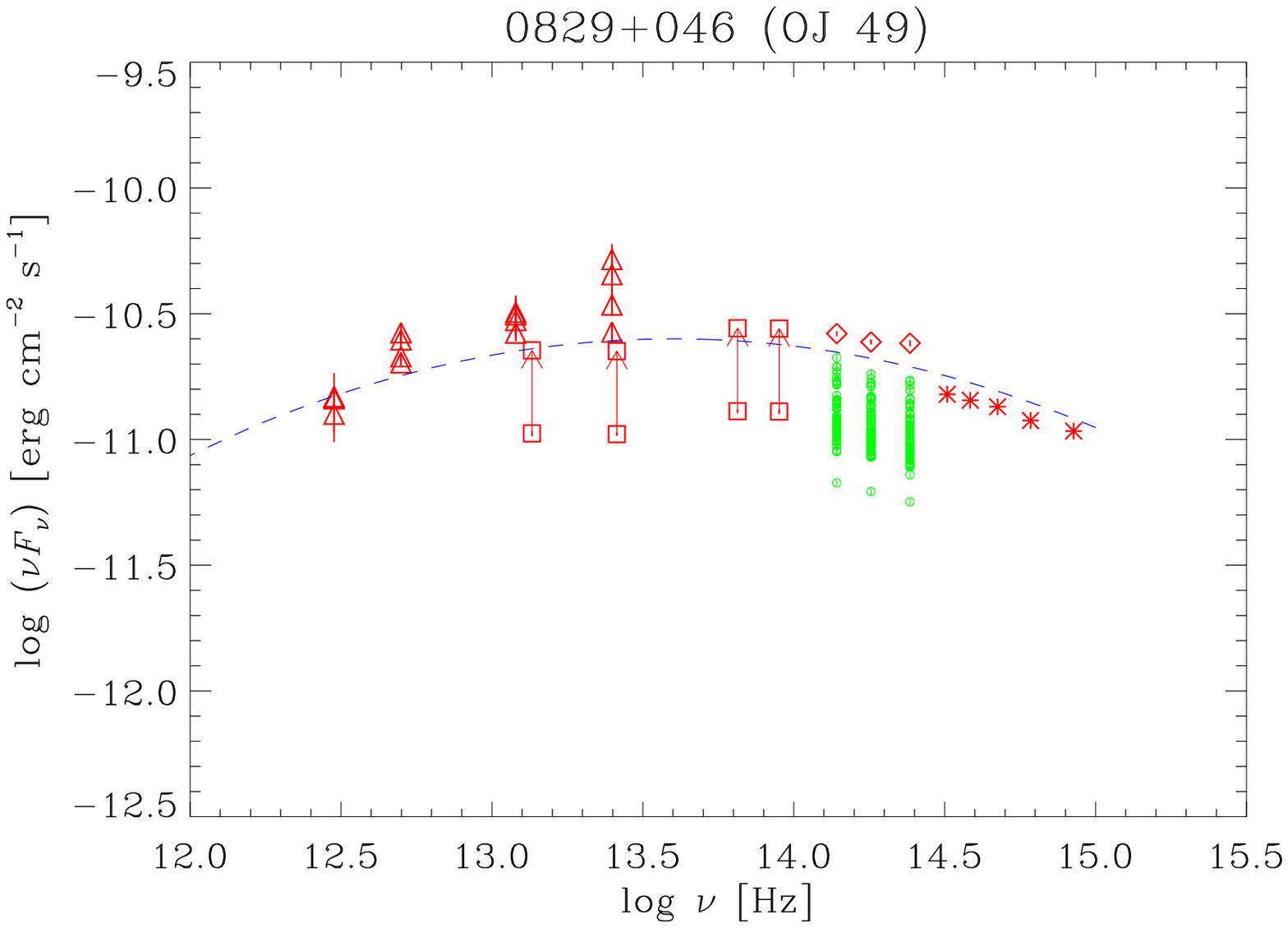,width=0.30\linewidth}
    \psfig{figure=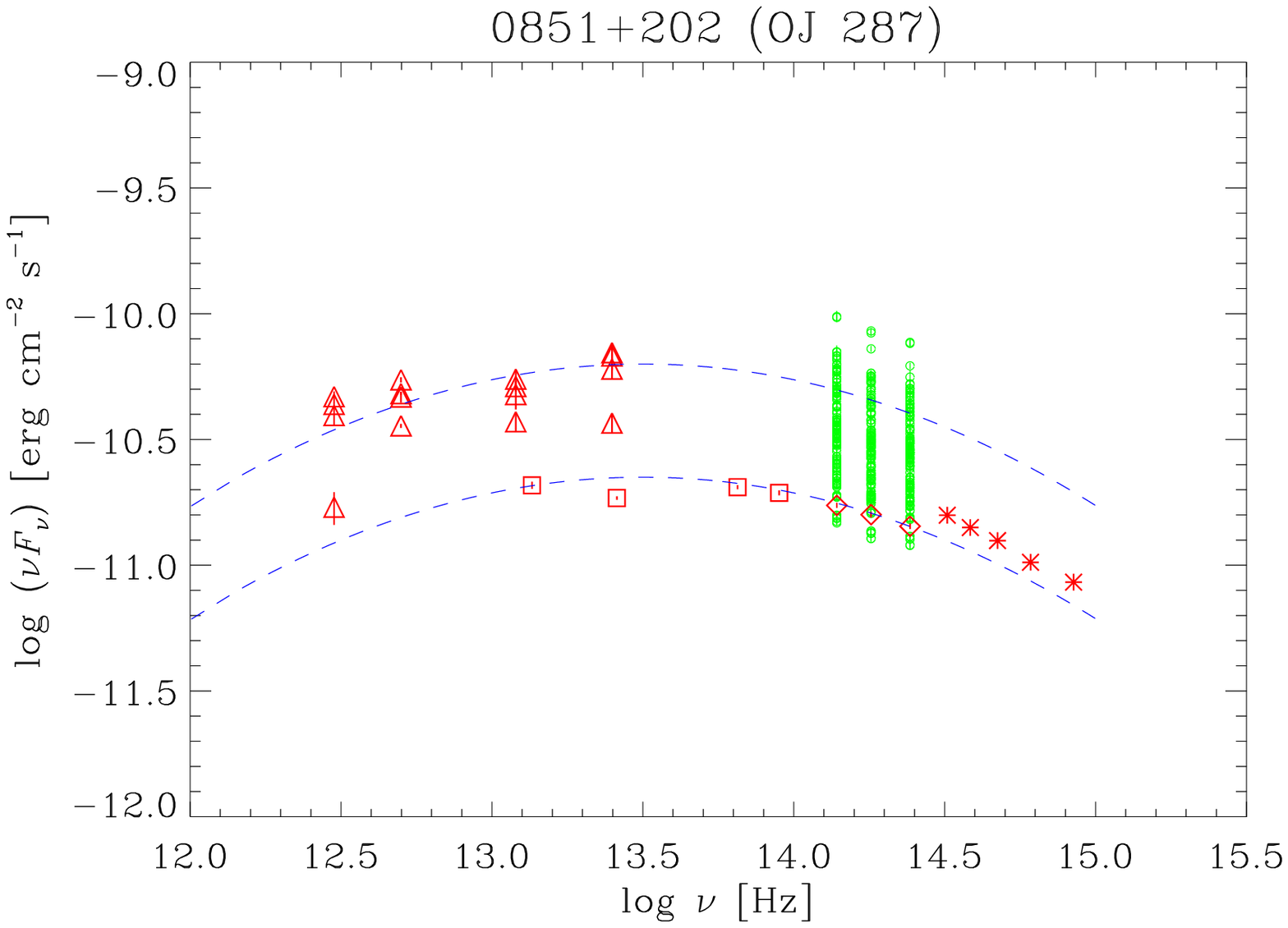,width=0.30\linewidth}}
    \vspace{0.5cm}
    \centerline{
    \psfig{figure=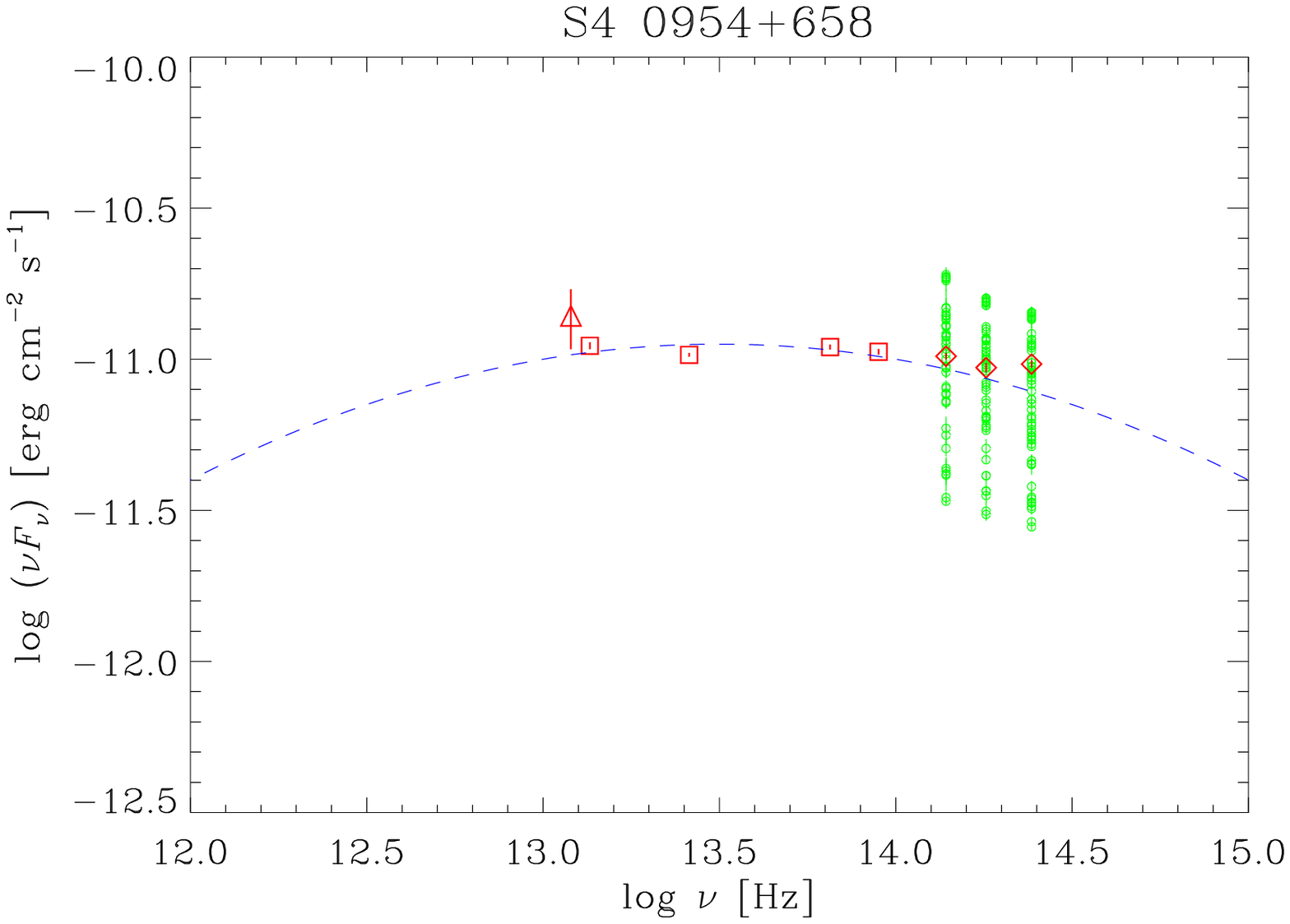,width=0.30\linewidth}
    \psfig{figure=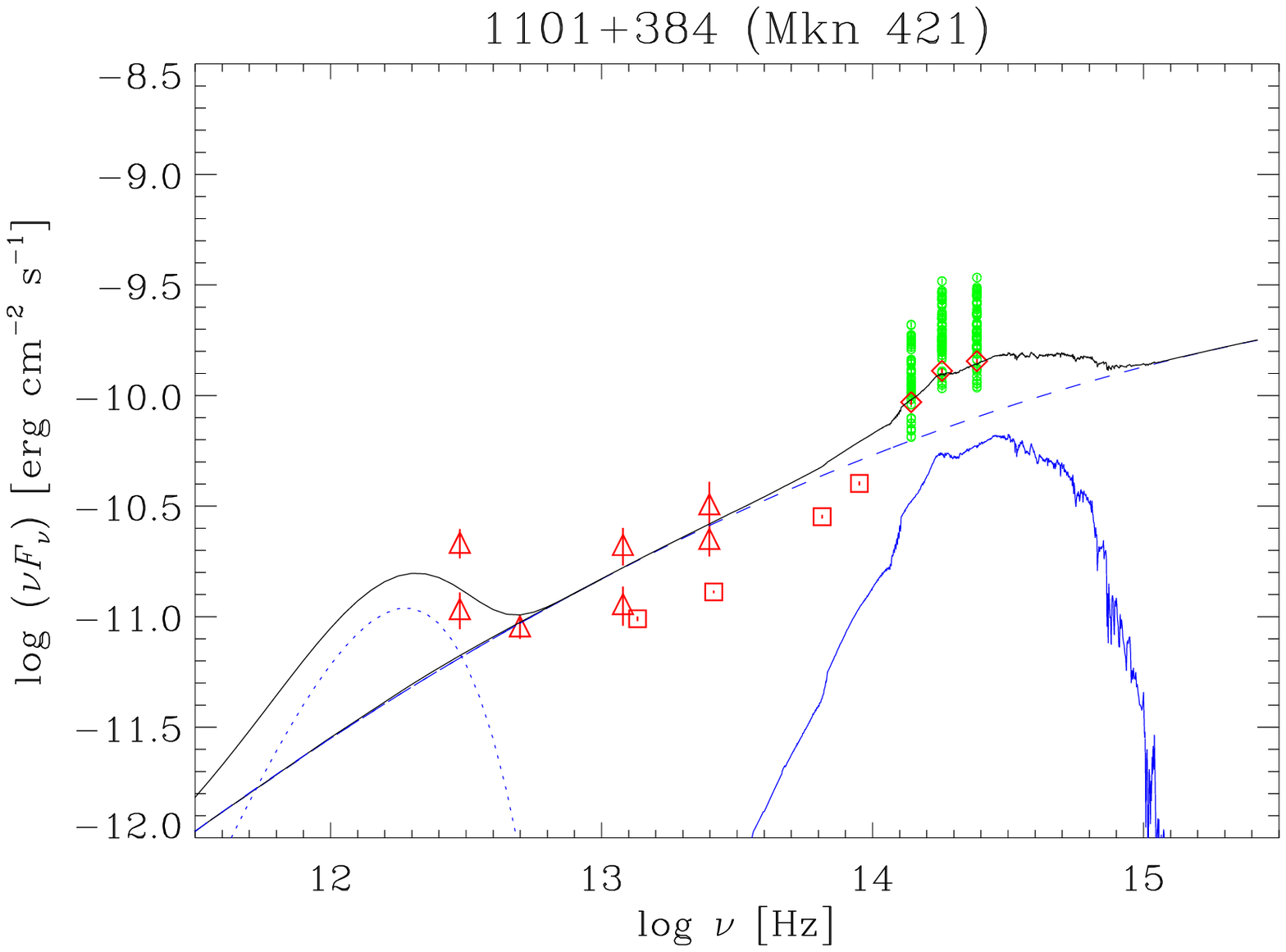,width=0.30\linewidth}
    \psfig{figure=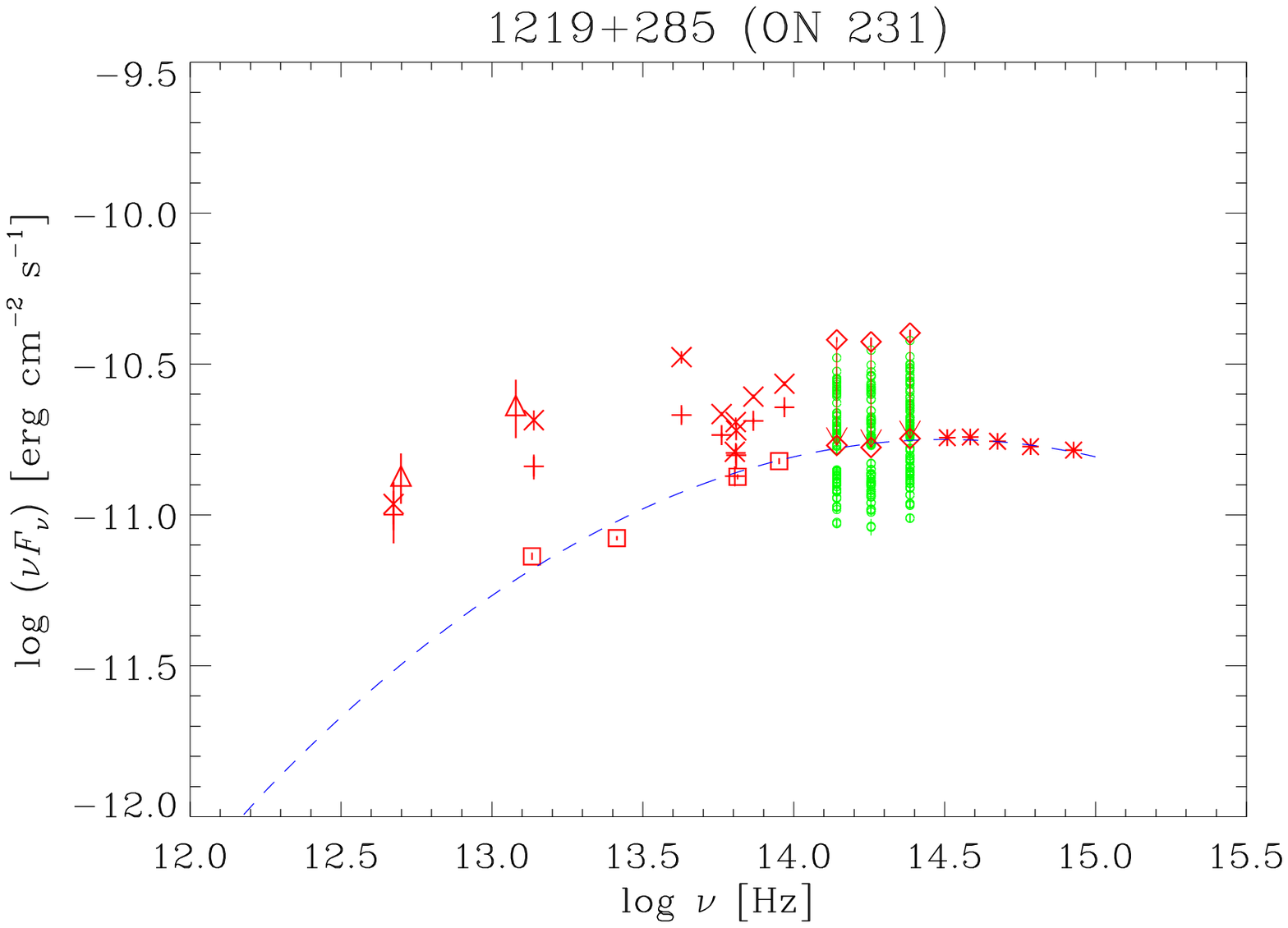,width=0.30\linewidth}}
    \vspace{0.5cm}
    \centerline{
    \psfig{figure=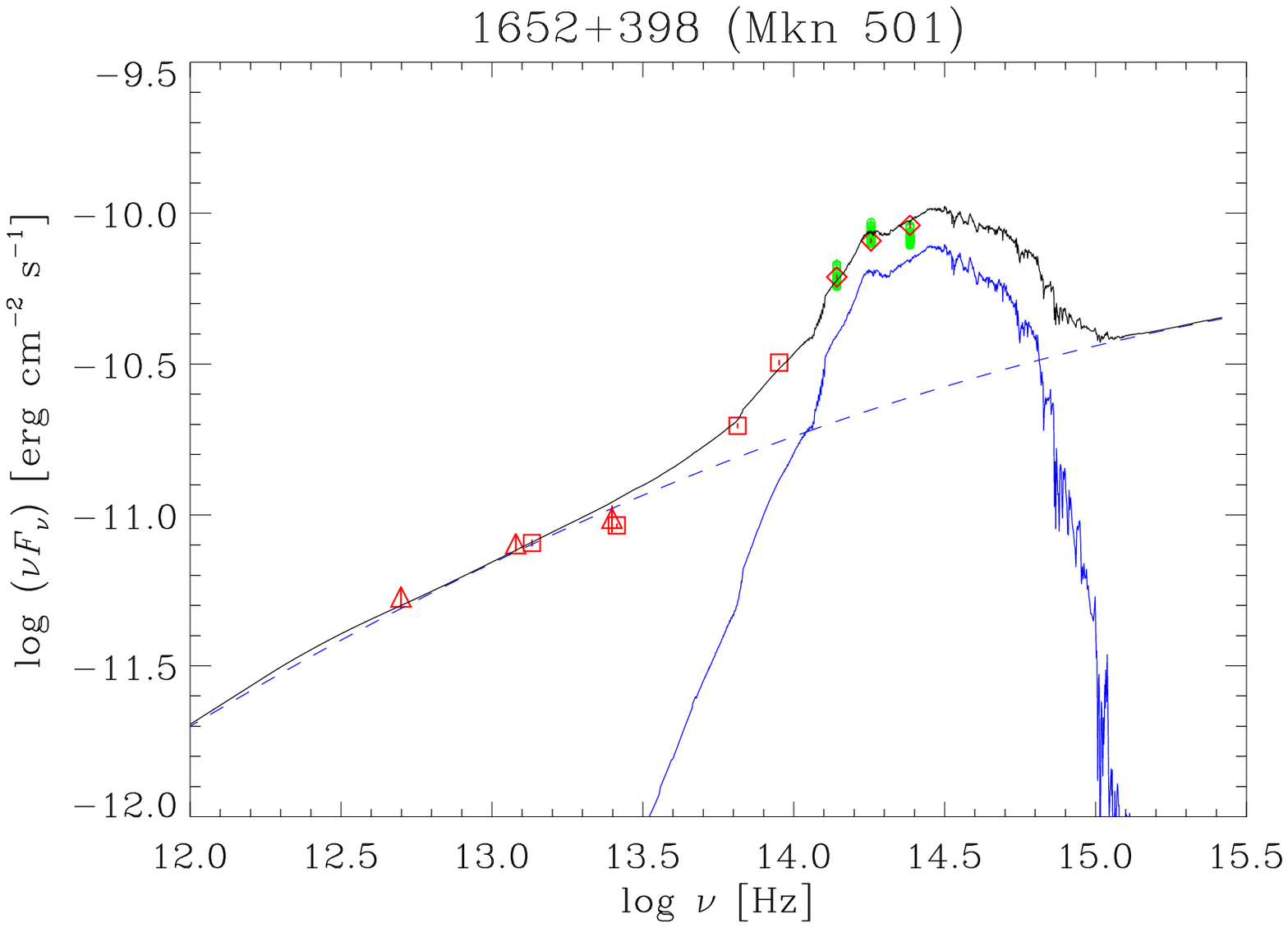,width=0.30\linewidth}
    \psfig{figure=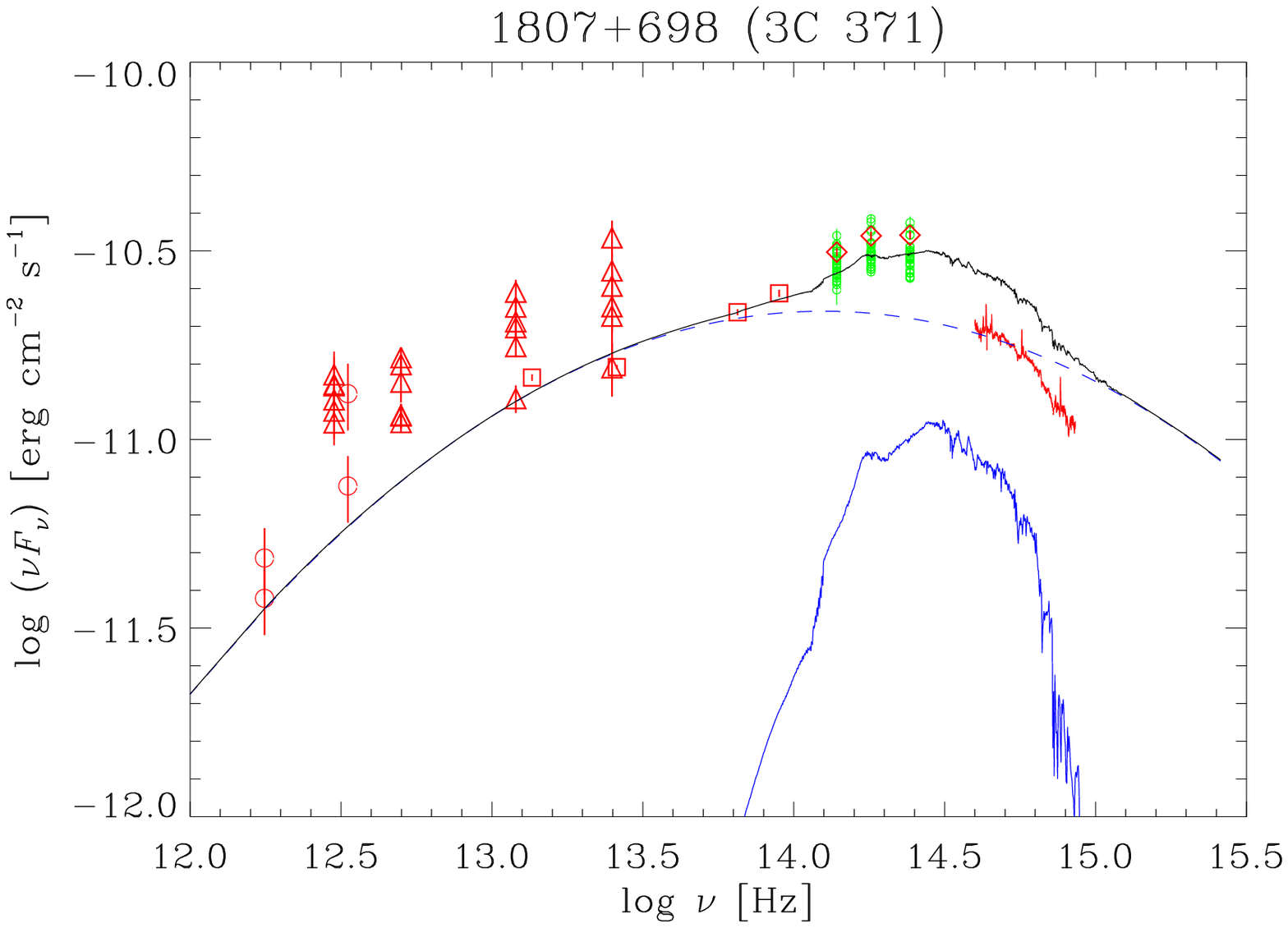,width=0.30\linewidth}
    \psfig{figure=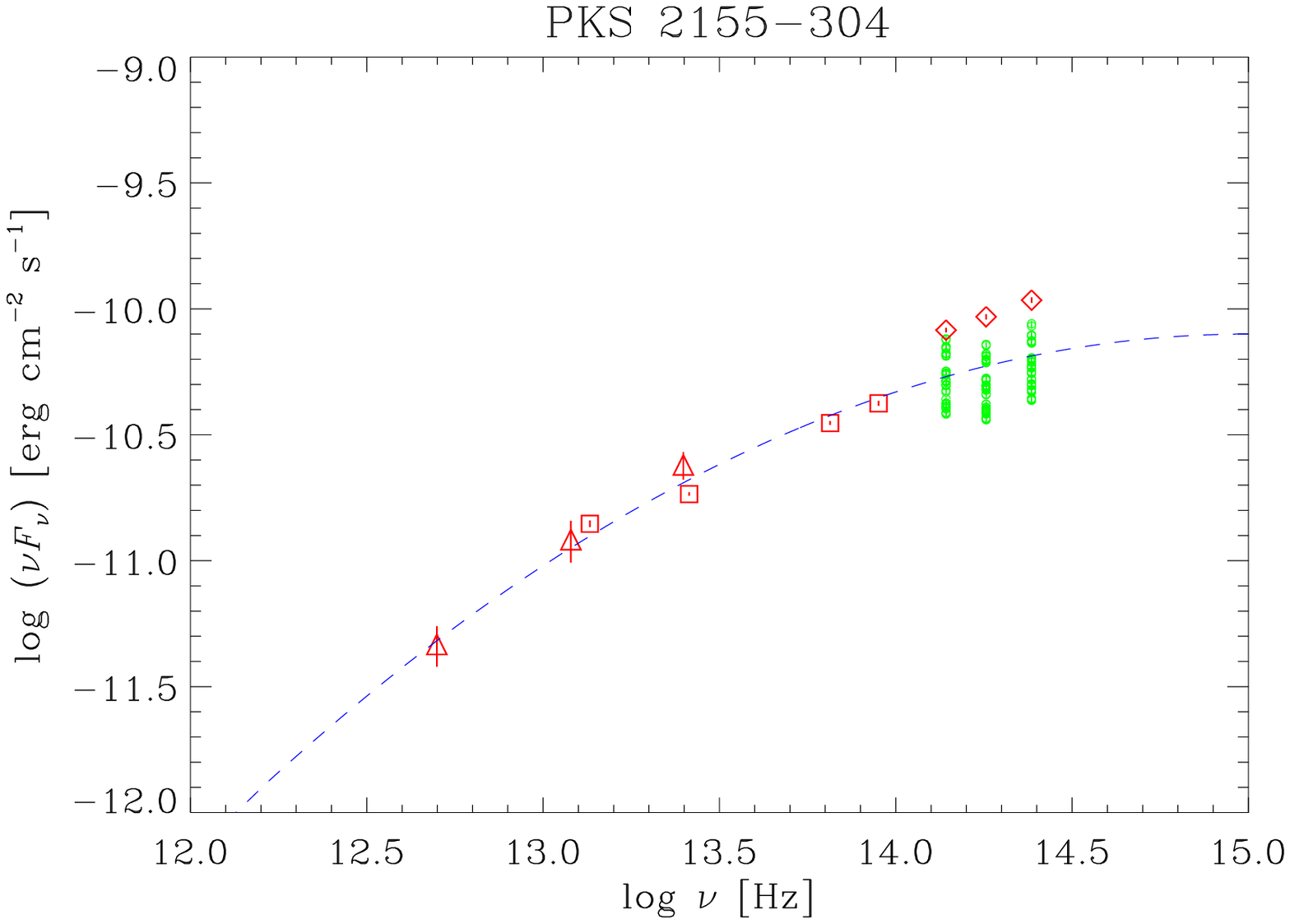,width=0.30\linewidth}}
    \vspace{0.5cm}
    \centerline{
    \psfig{figure=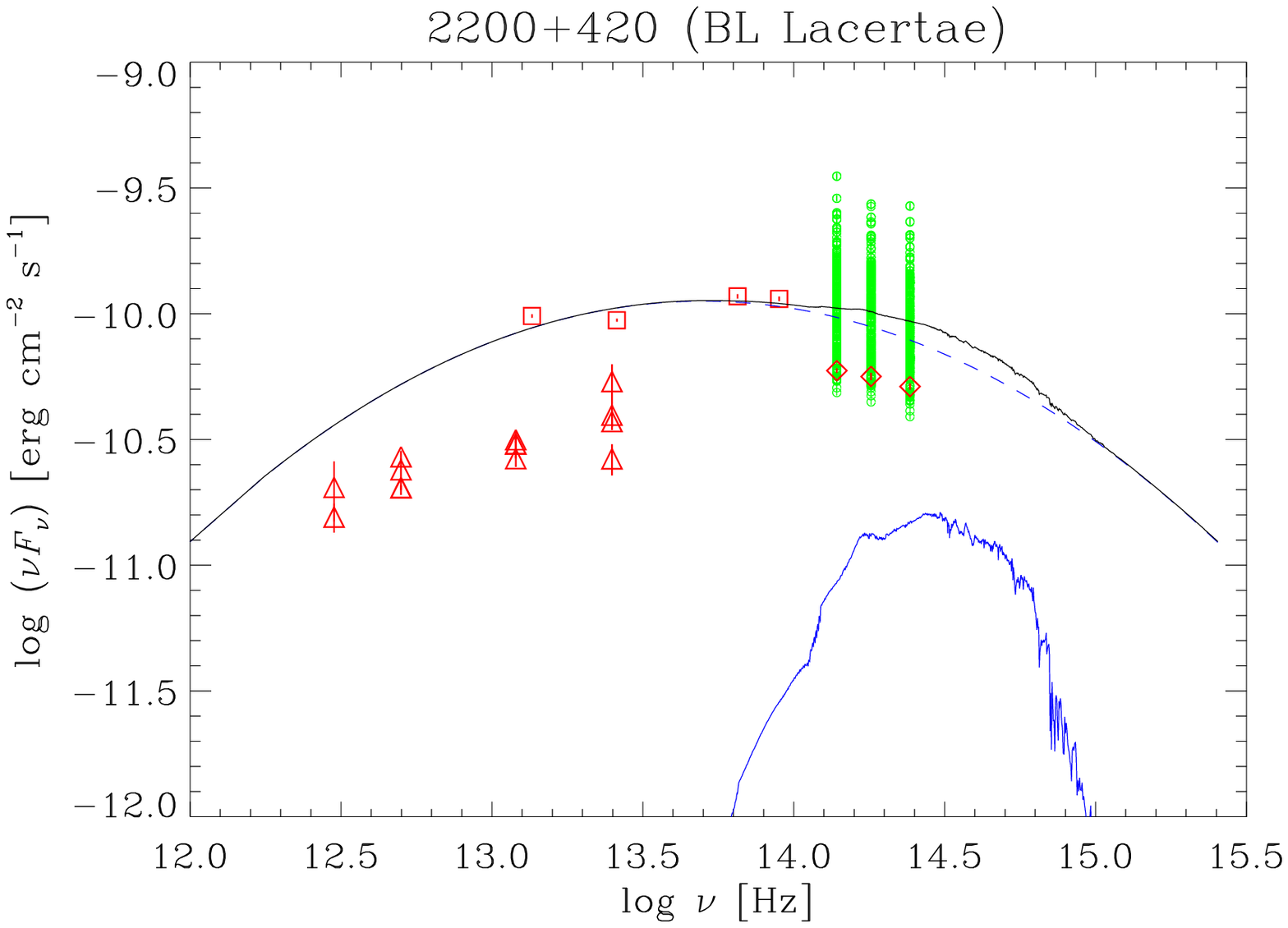,width=0.30\linewidth}
    \psfig{figure=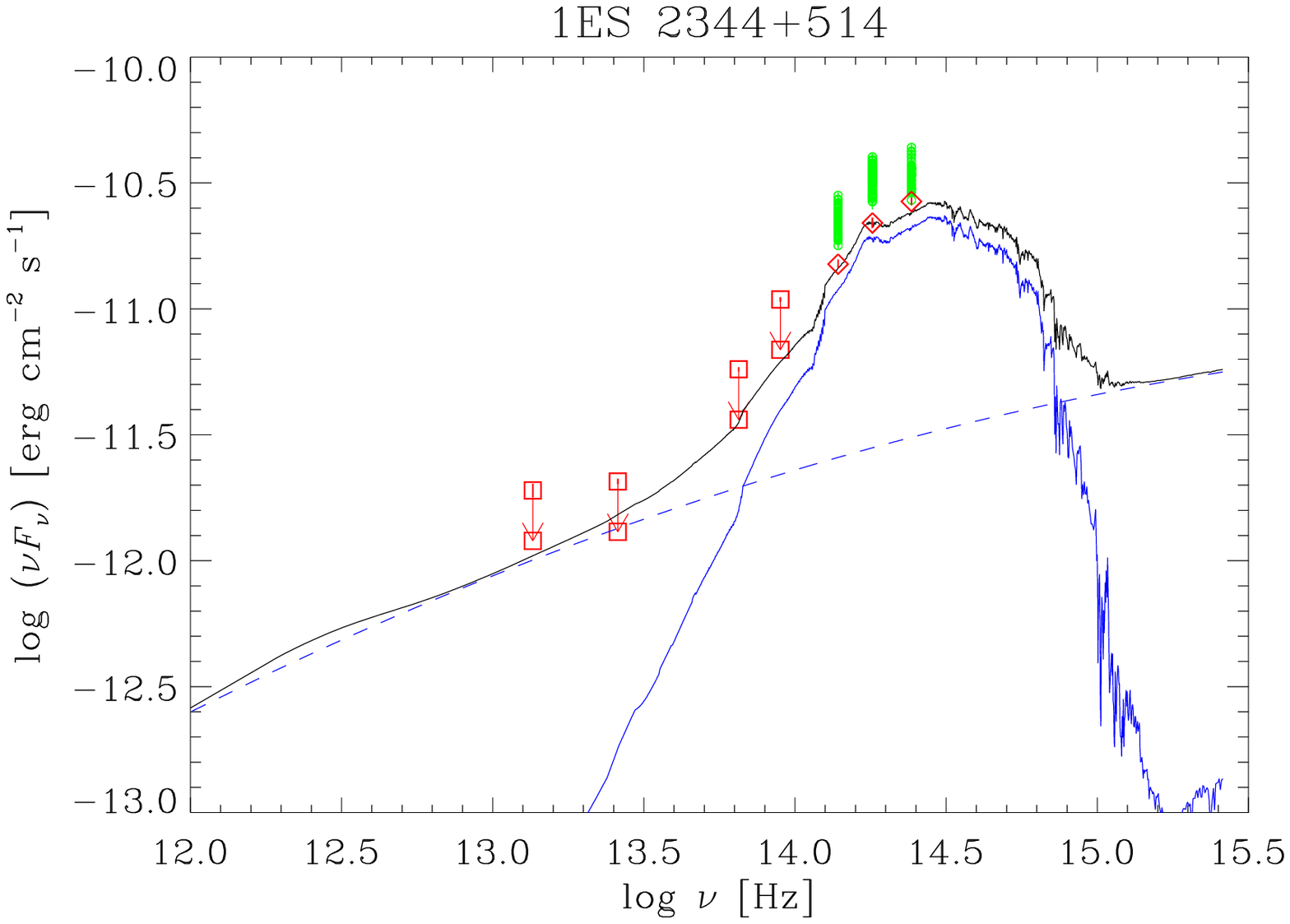,width=0.30\linewidth}}
    \caption{Spectral energy distribution of the GASP-WEBT BL Lac objects. Data from {\em ISO} (red circles), {\em IRAS} (triangles), {\em Spitzer} (plus signs and crosses), {\em WISE} (squares), 2MASS (diamonds), and SDSS (asterisks) are shown. The green circles represent the results of the GASP near-IR monitoring at the Campo Imperatore and Teide observatories. The optical spectrum of 1807+698 by \citet{but09} is also plotted.
Contributions from the synchrotron (dashed line) and host galaxy (solid line) emission are displayed in blue. In the cases of 3C 66A and Mkn 421, thermal dust emission is added, which is modelled as a black-body (blue dot-dashed line). The black line represents the sum of all components.} 
    \label{sed_bllacs} 
   \end{figure*}

Although in general the SEDs do not contain simultaneous data, they can help us understand what are the photons sources intervening in this frequency range.
In Figs.\ \ref{sed_bllacs} and \ref{sed_fsrqs} we show the SEDs together with model fits that take into account the possible emission contributions: 
\begin{itemize}
\item synchrotron radiation from the jet, which is modelled as a log-parabola following \citet{mas04},
\begin{equation}
\log [\nu F(\nu)] = \log [\nu_{\rm p} F(\nu_{\rm p})] - b \, (\log \nu - \log \nu_{\rm p})^2 , 
\label{logpa}
\end{equation}
where $\nu_{\rm p}$ is the frequency of the synchrotron peak $\nu_{\rm p} F(\nu_{\rm p})$;
\item host galaxy emission, for which we consider the SWIRE template of a 13 Gyr old elliptical galaxy;  
\item QSO-like nuclear contribution from accretion disc, broad line region, and dust torus, all included in the QSO1 SWIRE templates, i.e.\ templates of QSO with broad emission lines;
\item thermal radiation from a black body to simulate dust signatures in the infrared or to enhance the accretion disc flux in FSRQs. 
\end{itemize}

In the plots presented for BL Lac objects in Fig.\ \ref{sed_bllacs}, sometimes catalogue data are shifted in $\nu F_\nu$ (the shift is marked by arrows) in order to simulate simultaneous SEDs that can ease the modelling task. Although this is not strictly correct because the spectral shape may change with flux, it is justified by the small spectral variability that we noticed in Sect.\ \ref{colour} for BL Lac objects.

Fits to the SEDs of BL Lac objects require a synchrotron emission, plus the contribution of a host galaxy for the nearest objects. This is dominant in the near-IR in the cases of Mkn 501 and 1ES 2344+514, and it is very important for Mkn 421 and 3C 371.
For only two BL Lacs (3C 66A and Mkn 421) the far-infrared data from {\em IRAS} suggest the presence of dust, as already noticed by \citet{imp88}.

In general, the FSRQs SEDs need a synchrotron plus a QSO-like emission contribution due to accretion disc, broad line region, and torus. 
There are three QSO1 templates in the SWIRE database. We first used TQSO1 because it shows a prominent H$\alpha$ line, which is known to be present in the spectra of our objects.
However, its high IR/optical flux ratio (rest frame) appears inadequate to model some FSRQs SEDs, so we built a composite QSO1 template, combining the higher-frequency part of TQSO1, i.e.\ its disc plus BLR emission, with the lower-frequency part of BQSO1, representing a fainter torus emission.
Moreover, in some FSRQs a brighter disc is required, which we obtained by adding a black-body component to the QSO1 template.
In the following we comment on single source features.

\subsection{BL Lac objects}

\subsubsection{0219+428}
The model fit indicates that the synchrotron peak falls at optical frequencies.
As already noticed by \citet{imp88}, there is good evidence for thermal dust emission, which we fitted with a blackbody spectrum with temperature of $\sim 41 \, \rm K$. 

\subsubsection{0235+164}
This is the farthest among the GASP-WEBT BL Lac objects and has often revealed an FSRQ-like behaviour \citep[e.g.][]{rai07a,ghi11}. 
As discussed in Sect.\ \ref{colour}, notwithstanding the strong variability in the near-IR, the spectral slope does not change much.
There is no hint of a $J$-band flux excess due to the contribution of the H$\alpha$ emission line even in the faintest states. 
This means that BLR emission is undetected in our observations. 
Data are compatible with a shift of the synchrotron peak toward higher energies when the flux increases, as shown by the model fits.

\subsubsection{0716+714}
The near-IR data from the GASP indicate a steep spectrum, which appears steeper in faint states \citep[see also][]{vil08}. 
In contrast, the 2MASS spectrum is flat. 
{\em IRAS} data likely belong to a fainter state.

\subsubsection{0735+178}
By shifting in $\nu F_\nu$ the same log-parabola model, we obtained reasonable fits of two different brightness states traced by SDSS and {\em WISE} data, respectively.

\subsubsection{0829+046}
The model fit accounts for a bright state marked by SDSS, 2MASS, and the lowest {\em IRAS} data points.

\subsubsection{0851+202}
The lower model fit reproduces the {\em WISE} and 2MASS SED and its shape also agrees with the SDSS spectrum. The upper model is obtained by shifting vertically the previous one and goes through the scattered {\em IRAS} data.

\subsubsection{0954+658}
As in the case of 0716+714, the near-IR data from the GASP show a steep spectrum, in contrast with the 2MASS spectral slope.

\subsubsection{1101+384}
This is an HBL, whose synchrotron peak lies in the X-ray band, with impressive frequency shifts \citep{pia98}. According to \citet{nil07}, the host galaxy has a magnitude $R=13.18 \pm 0.05$ and contributes up to $\sim 10 \, \rm mJy$ (for a 10\arcsec\ aperture radius and good seeing conditions) to the source photometry.
The 2MASS data are from the All-Sky Extended Source Image Server and show a faint state of the source. 
An infrared excess is visible in the {\em IRAS} data suggesting thermal emission from dust, as pointed out by \citet{imp88}. We then interpreted the SED as a superposition of a jet spectrum, host galaxy contribution, and a black-body dust component with temperature of about 23 K.
The model fit presented in Fig.\ \ref{sed_bllacs} goes through the 2MASS and {\em IRAS} data.

\subsubsection{1219+285}
The {\em Spitzer} data, from \citet{mal11}, were acquired in June--July 2007 (plus signs) and January 2008 (crosses); observations with the different instruments (IRS, IRAC, and MIPS) are not simultaneous and this explains the discontinuities of the spectra. The model fit peaks in the $z$ band.

\subsubsection{1652+398}
This is another HBL, whose near-IR--optical spectrum shows a strong host galaxy signature. The host brightness is $R=11.92 \pm 0.06$ \citep{nil07}; its contribution in Fig.\ \ref{sed_bllacs} is $\sim 12 \, \rm mJy$, in agreement with the contaminating flux estimate within an aperture radius of 7.5 arcsec.

\subsubsection{1807+698}
In Fig.\ \ref{sed_bllacs} data from the Infrared Space Observatory ({\em ISO}) by \citet{haa04} and the spectrum from \citet{but09} are also included. The model fit accounts for the {\em WISE} and faintest {\em IRAS} and {\em ISO} data. It requires an important contribution from the host galaxy. 

\subsubsection{2155$-$304}
This is the most southern among the GASP sources.
The synchrotron peak of the model fit falls in the UV. 

\subsubsection{2200+420}
This is one of the most studied sources by the GASP-WEBT \citep[see e.g.][and references therein]{vil02,rai13}. 
The host galaxy has $R=15.5 \pm 0.02$ \citep{sca00}, and affects the source photometry in the near-IR and optical bands, especially in faint states.

\subsubsection{2344+514}
This is the third HBL in the GASP sample. Its near-IR--optical spectrum is dominated by the host galaxy, whose brightness is $R=13.90 \pm 0.06$ \citep{nil07}.
In Fig.\ \ref{sed_bllacs} we show a fit to the faint state traced by the 2MASS data.
The jet contribution is superposed to a host galaxy contribution corresponding to the contaminating flux entering an aperture radius of 3 arcsec when the FWHM is 3 arcsec.

\subsection{Flat spectrum radio quasars}

\subsubsection{0420$-$014}
The near-IR data from the GASP indicate a steep spectrum, with a flux excess in $J$ band during faint states. Indeed, the source redshift implies a contribution from the H$\alpha$ emission line in this band, as shown by the SWIRE TQSO1 template. 
The lack of medium or far-infrared data in a low state prevents us to derive information on the torus emission.

\subsubsection{0528+134}
The near-IR spectra traced by both 2MASS and GASP data in faint states reveal a concave shape, suggesting the transition from a spectrum dominated by jet emission to a spectrum dominated by QSO-like emission.
The H$\alpha$ and H$\beta$ emission lines affect the $K$ and $H$ band fluxes, respectively.
Modelling the near-IR spectrum with a log-parabola plus the SWIRE TQSO1 template would overproduce the mid-infrared flux detected by {\em WISE} because of the prominent infrared contribution from the dust torus. 
This implies that the IR/optical flux ratio of the QSO-like contribution in this source must be smaller.
We thus combined the disc+BLR component of the TQSO1 template at $\lambda < 11000$ \AA\ (rest frame) with the torus component of the BQSO1 template at $\lambda \ge 11000$ \AA. 
The result is shown in Fig.\ \ref{sed_fsrqs} as a grey solid line.
However, this composite template cannot explain the curvature of the near-IR spectrum at intermediate flux levels, which requires a brighter disc.
We thus added a black-body component (grey dashed line in Fig.\ \ref{sed_fsrqs}) to the composite template to obtain a final QSO-like contribution (blue solid line) that, once added to the jet emission (dot-dashed blue line), can produce a reasonable fit (black solid line) to the observing data. This implies a softening of the optical spectrum with increasing flux, confirming the guess by \citet{pal11}.

\subsubsection{0827+243}
As in the 0528+134 case, the signature of a QSO-like emission contribution is already evident in the near-IR spectrum. In particular, the $J$-band excess is due to the contribution of the broad H$\alpha$ emission line, while the H$\beta$ line enters the SDSS $z$ band, and the Mg II and Fe lines, giving rise to the little blue bump, contribute to the flux between the $r$ and $g$ bands.
The composite QSO1 template explains the spectral slope of the faintest NIR data points. 
The model presented in Fig.\ \ref{sed_fsrqs} reproduces a brightness state traced by the 2MASS and SDSS data; it requires, besides the non-thermal contribution, an enhanced disc contribution, which was obtained, as in the case of 0528+134, by adding a black-body component to the QSO1 composite template. 

\subsubsection{0836+710}
This is the farthest FSRQs in the GASP sample. The H$\alpha$ line contributes to the $K$ band, and the H$\beta$ line enters the $H$ band. Also in this case the near-infrared data suggest that the accretion disc may be brighter than predicted by the QSO1 template.

\subsubsection{1156+295}
Emission from the QSO-like component is not prominent: it appears in the SDSS spectrum, which shows the clear contribution of the Mg II line in the $g$ band.
Figure \ref{sed_fsrqs} displays a possible fit to the source SED at the SDSS and {\em IRAS} brightness level.

\subsubsection{1226+023}
The SED of 3C 273 in the infrared--optical band is known to be complex, and various authors have proposed interpretations in terms of synchrotron plus various dust components \citep[see e.g.][]{tur06} or dust plus various synchrotron contributions \citep[see e.g.][]{sol08}.
The model fit we show in Fig.\ \ref{sed_fsrqs} includes a log-parabola synchrotron component (dot-dashed blue line) superimposed to a QSO-like emission (solid blue line) obtained from the composite QSO1 template (solid grey line), with enhanced disc (obtained by adding the dashed grey black body to the template) and enhanced dust around the W1 {\em WISE} band (obtained by adding the dotted grey black body). 
This latter component is responsible for the spectral break that causes the anticorrelation between $J-H$ and $H-K$ noticed in the previous section.
The brightest state traced by the {\em IRAS} data can be reproduced by increasing the jet flux. 
Notice that the 2MASS $J$ band flux is underproduced by the model, but the spectral shape of the GASP data applied to the 2MASS points would suggest a lower value.
We plotted the optical spectrum (red continuous line) from \citet{but09} to help us model the disc emission. 

\subsubsection{1253$-$055}
The near-IR spectrum of 3C 279 remains steep even in faint states, so that the presence of a QSO-like emission contribution cannot be inferred. However, evidence for 
thermal emission was found by \citet{pia99} when analysing UV data of the source at a minimum brightness level.

\subsubsection{1510$-$089}
This source is know to present noticeable spectral variations in the near-IR--optical band \citep[e.g.][]{dam11}. Spectral variability is also suggested from the data points in Fig.\ \ref{sed_fsrqs}, including observations from {\em Spitzer} \citep[plus and cross symbols,][]{mal11}.
To help SED modelling, in the figure we reported the faintest SED by \citet{dam11}, built with near-IR and optical data from the GASP (red dots), and optical--UV data from {\em Swift} (red empty circles) acquired in March 2008. 
Two model fits are then presented in the figure: one passing through the March 2008 spectrum, requiring a synchrotron peak at about $\log \nu =13$, while the other one reproduces a higher state involving 2MASS data, with the synchrotron peak shifted at a higher energy, in agreement with \citet{dam11}.

\subsubsection{1611+343}
The shape of the near-IR spectrum is explained by the contributions from the H$\alpha$ and H$\beta$ lines to the $H$ and $J$ fluxes, respectively.
The model fit displayed in Fig.\ \ref{sed_fsrqs} reproduces reasonably well the shape of the SDSS optical spectrum. The peak of the disc emission falls in the $u$ band, which receives a contribution from the C IV] line. Mg II and Fe lines affect the $r$ flux.

\subsubsection{1633+382}
A big observing effort on this source has recently been carried out by the GASP-WEBT; the results have been published by \citet{rai12}. In Fig.\ \ref{sed_fsrqs} we display their faintest near-IR--UV SED, which was obtained in 2010 April ($\rm JD=2455296$). Red dots represent ground-based data, red empty circles are from the UVOT instrument onboard {\em Swift}\footnote{We enhanced the UVOT $v$-band flux by 10\% to account for calibration problems \citep[see][]{rai12}.}. The optical spectrum shows a brightness level similar to that of the SDSS. The observed flux in $u$, $b$, and $i$ bands is enhanced by the contribution from the Ly$\alpha$, C IV], and Mg II lines, respectively. 
The very hard near-IR spectrum traced by the 2MASS data points is not confirmed by the GASP data (neither those presented in this paper, nor those published in \citealt{rai12}), which indicate a higher flux in the $K$ band. The model fit shown in Fig.\ \ref{sed_fsrqs} satisfactorily reproduces the {\em WISE} spectrum as well as the near-IR--optical spectrum traced by the SDSS and 2010 April data, with a QSO-like component in agreement with the $R=17.85$ estimate given in \citet{rai12}.

\subsubsection{1641+399}
The near-IR spectrum always maintains a steep slope, showing no hints of QSO-like emission. The SDSS data points suggest a spectral flattening with increasing optical flux, which is not easy to explain.

\subsubsection{1739+522}
The GASP observations were done during a low brightness state and draw a near-IR spectrum that reveals flux contributions from the H$\alpha$ and H$\beta$ lines in the $H$ and $J$ bands, respectively. The lack of a flux excess in the $H$ band in the 2MASS data implies that the QSO-like component becomes negligible at that flux level.

\subsubsection{2230+114}
Marginal evidence for thermal emission was found by \citet{imp88} in {\em IRAS} data and tentative detection of dust was reported by \citet{mal11} analysing {\em Spitzer} data.
{\em Spitzer} observed the source in two epochs: 2007 June--July (crosses in Fig.\ \ref{sed_fsrqs}), and 2007 December -- 2008 January (plus signs). Some of these data are affected by large errors.
The model fit presented in Fig.\ \ref{sed_fsrqs} aims at reproducing a high brightness state as traced by {\em IRAS}, {\em Spitzer}, and SDSS data points. The excess flux in $J$ band can be accounted for by the H$\alpha$ line flux contribution.

\subsubsection{2251+158}
\begin{figure*}
    \vspace{0.5cm}
    \centerline{
    \psfig{figure=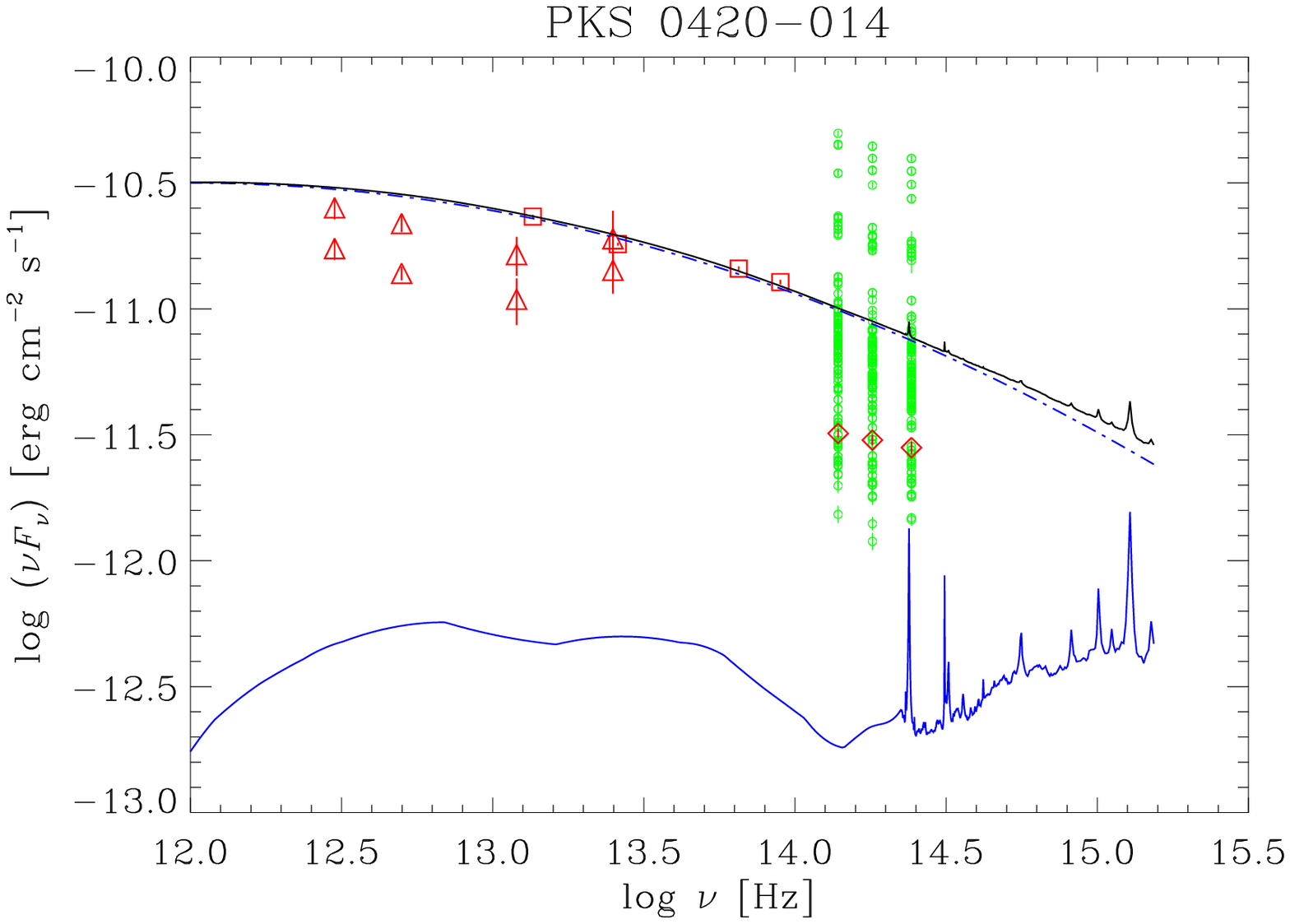,width=0.30\linewidth}
    \psfig{figure=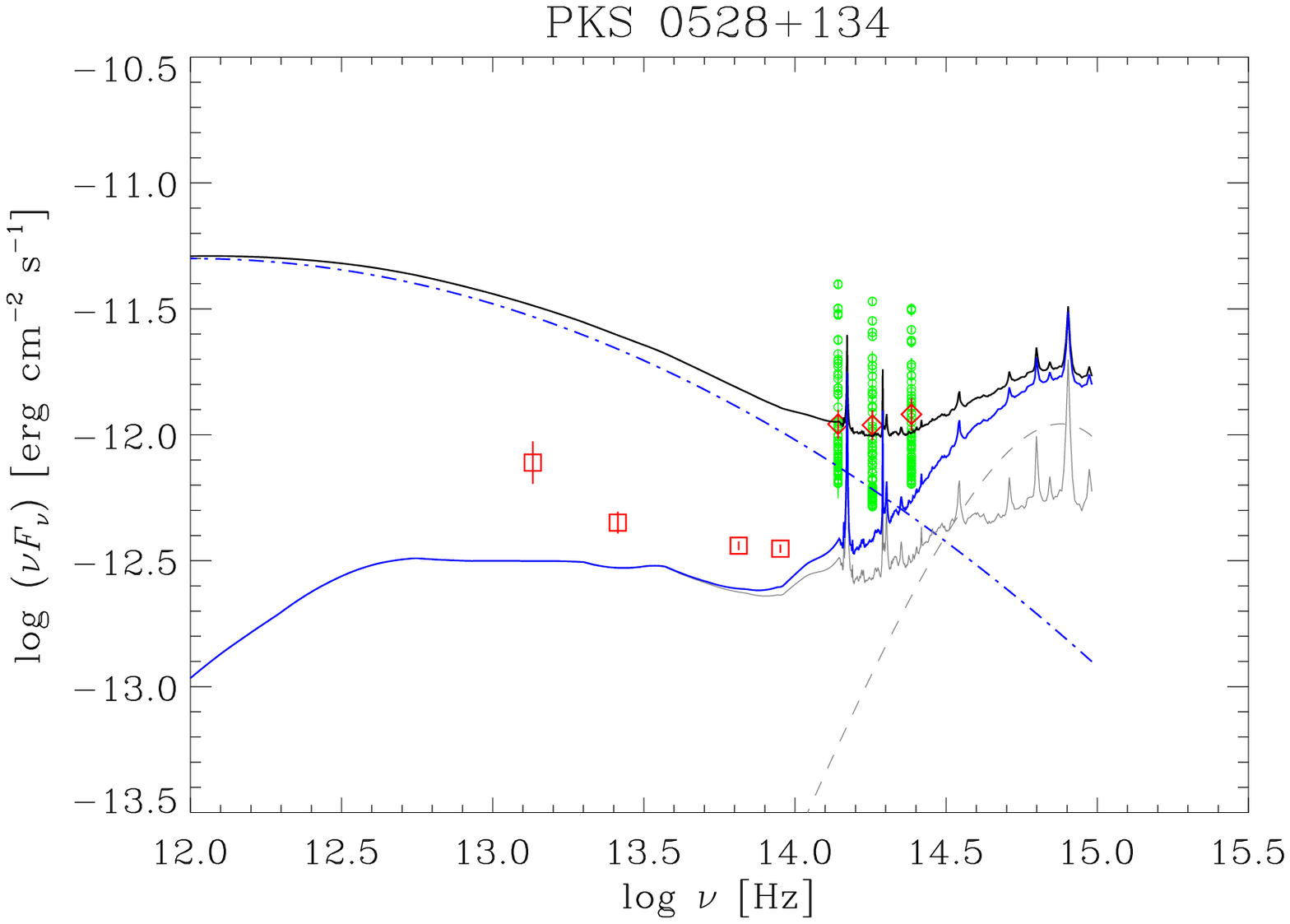,width=0.30\linewidth}
    \psfig{figure=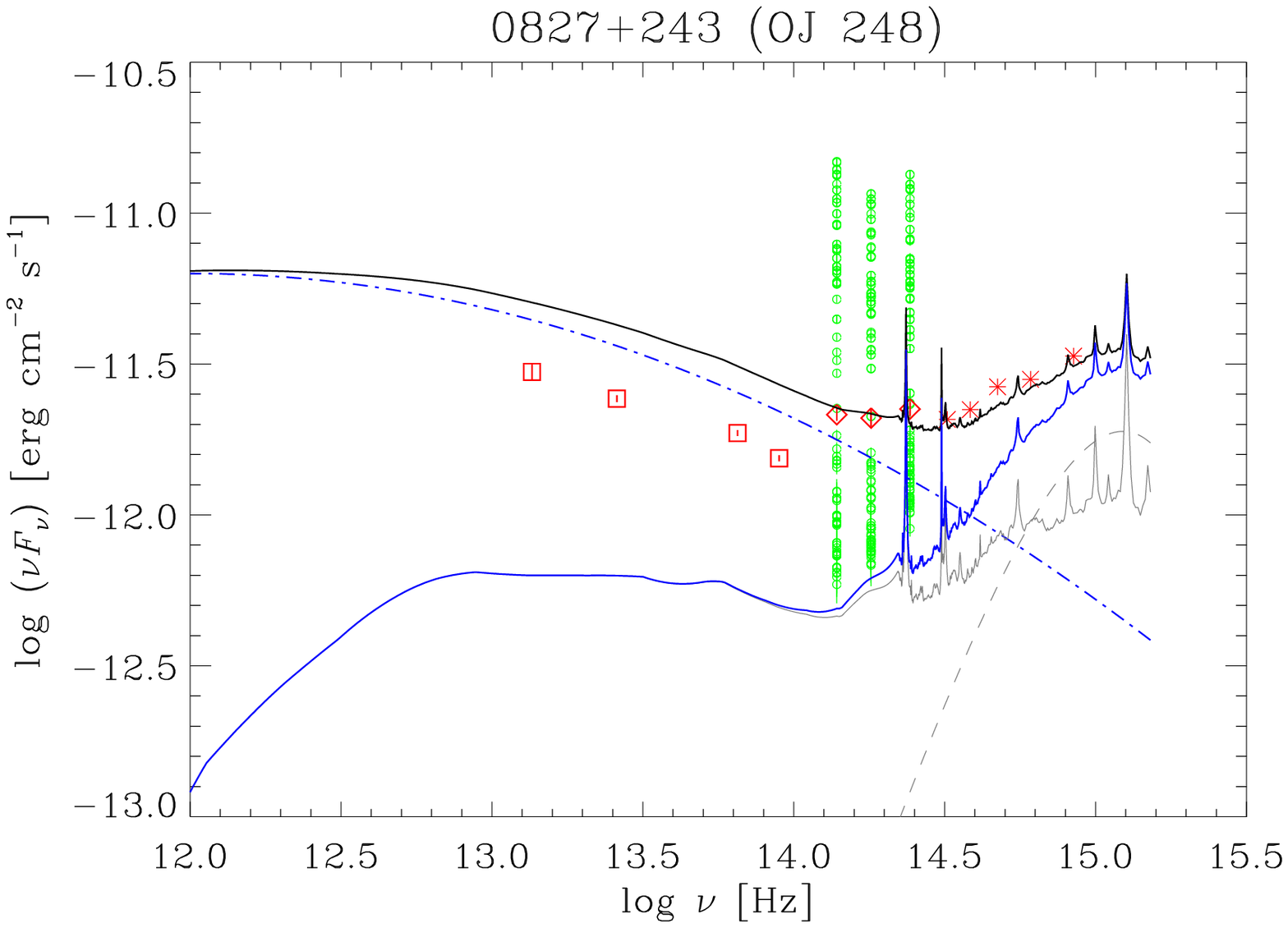,width=0.30\linewidth}}
    \centerline{
    \psfig{figure=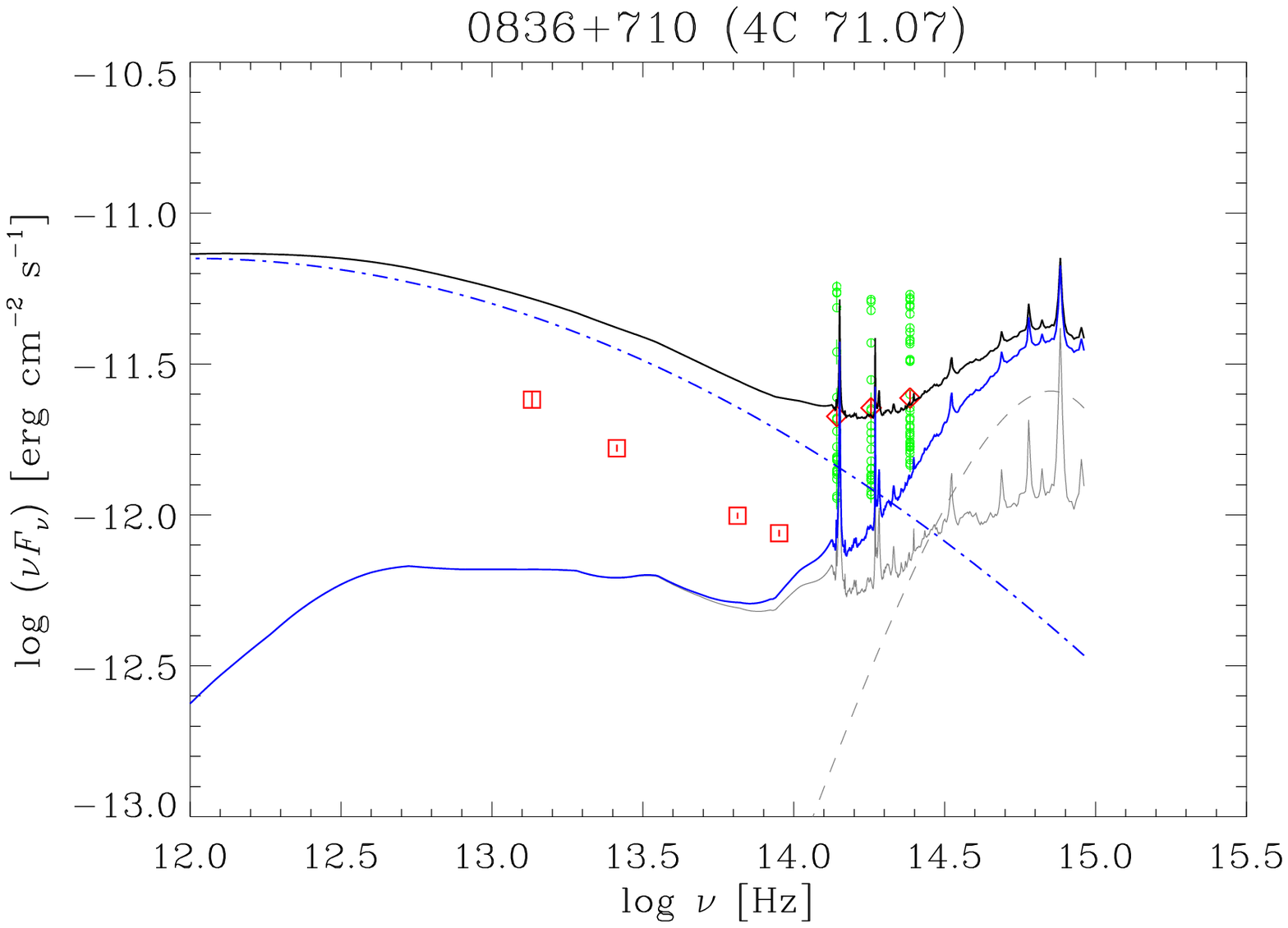,width=0.30\linewidth}
    \psfig{figure=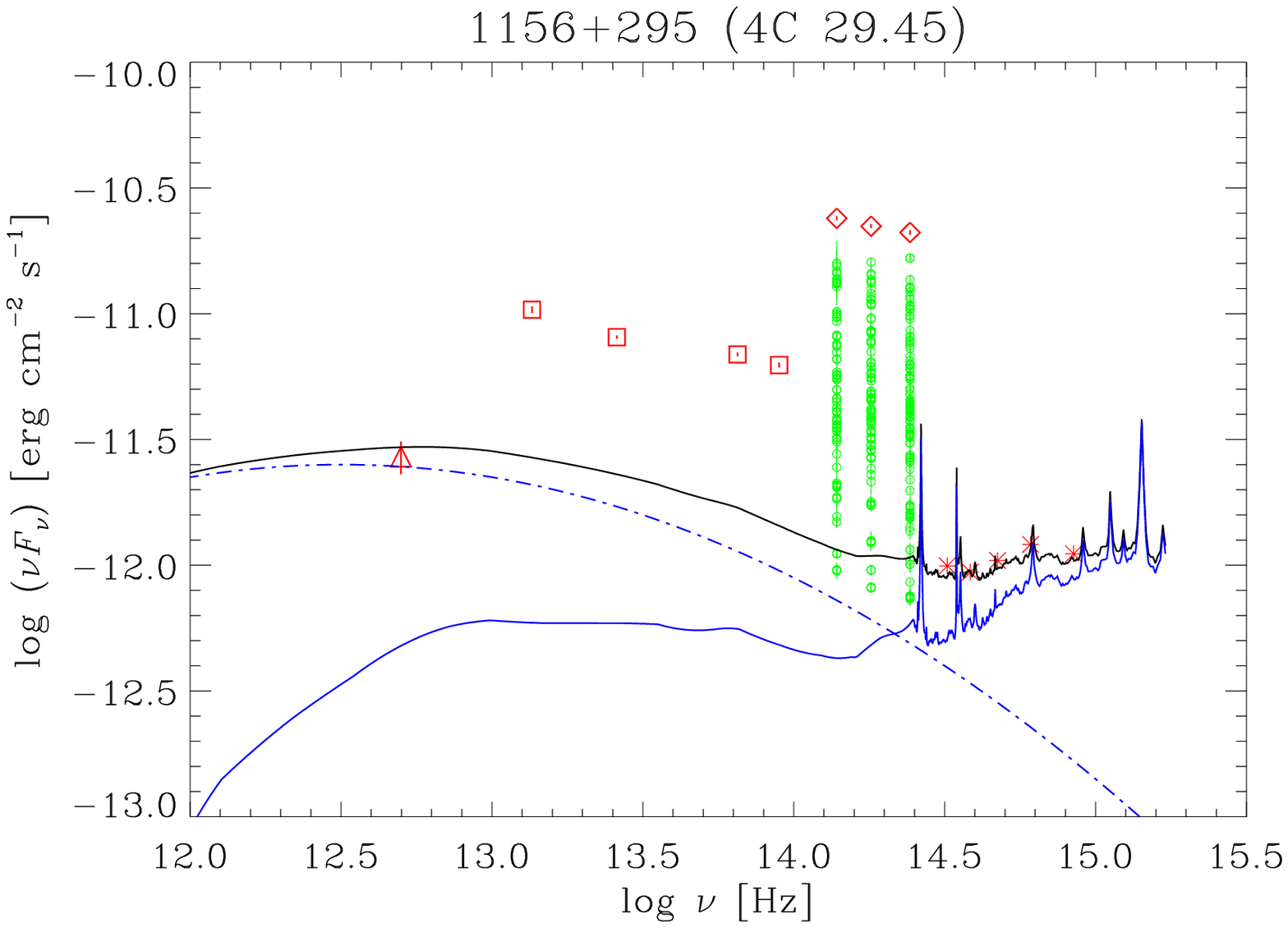,width=0.30\linewidth}
    \psfig{figure=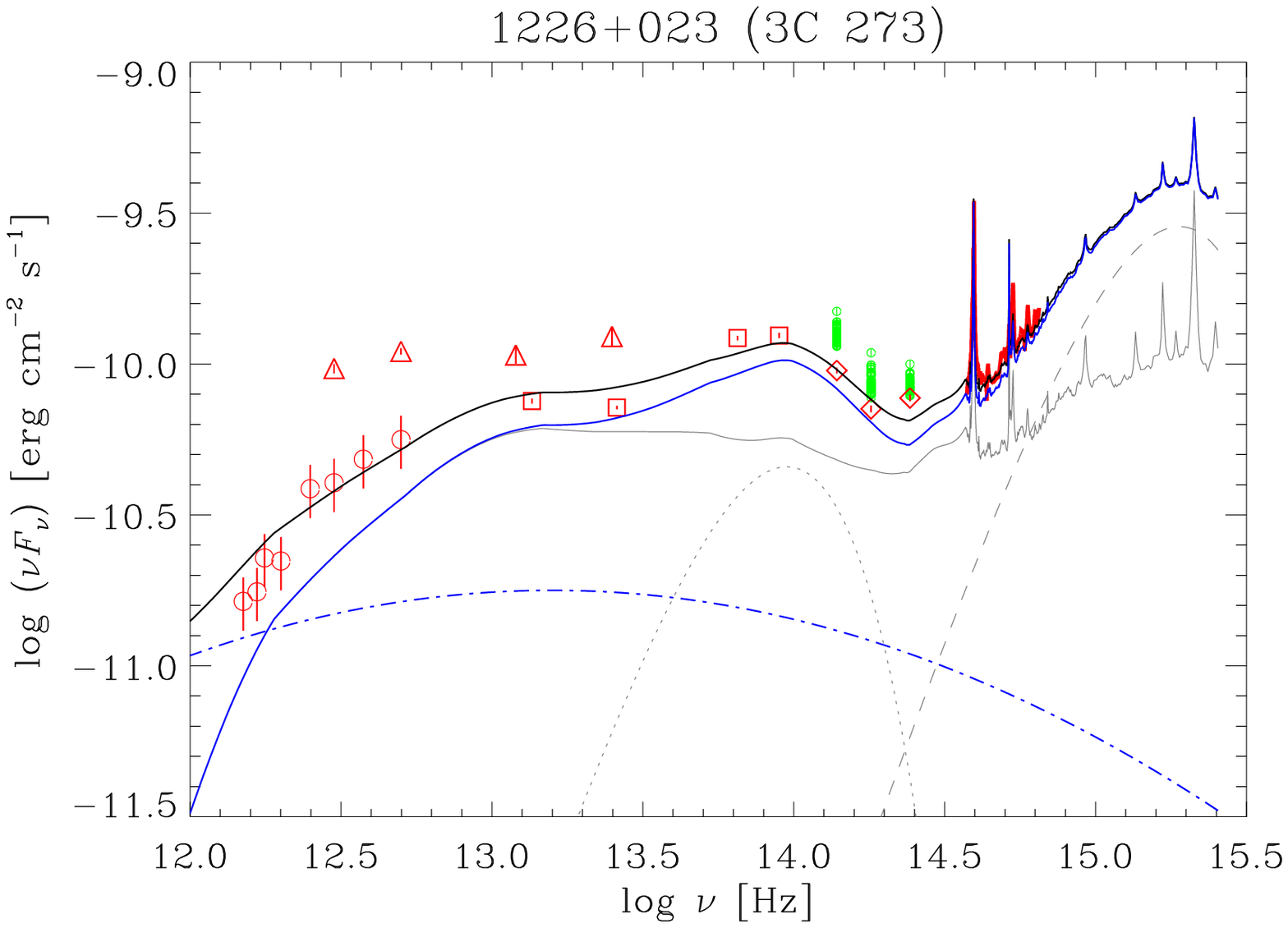,width=0.30\linewidth}}
    \centerline{
    \psfig{figure=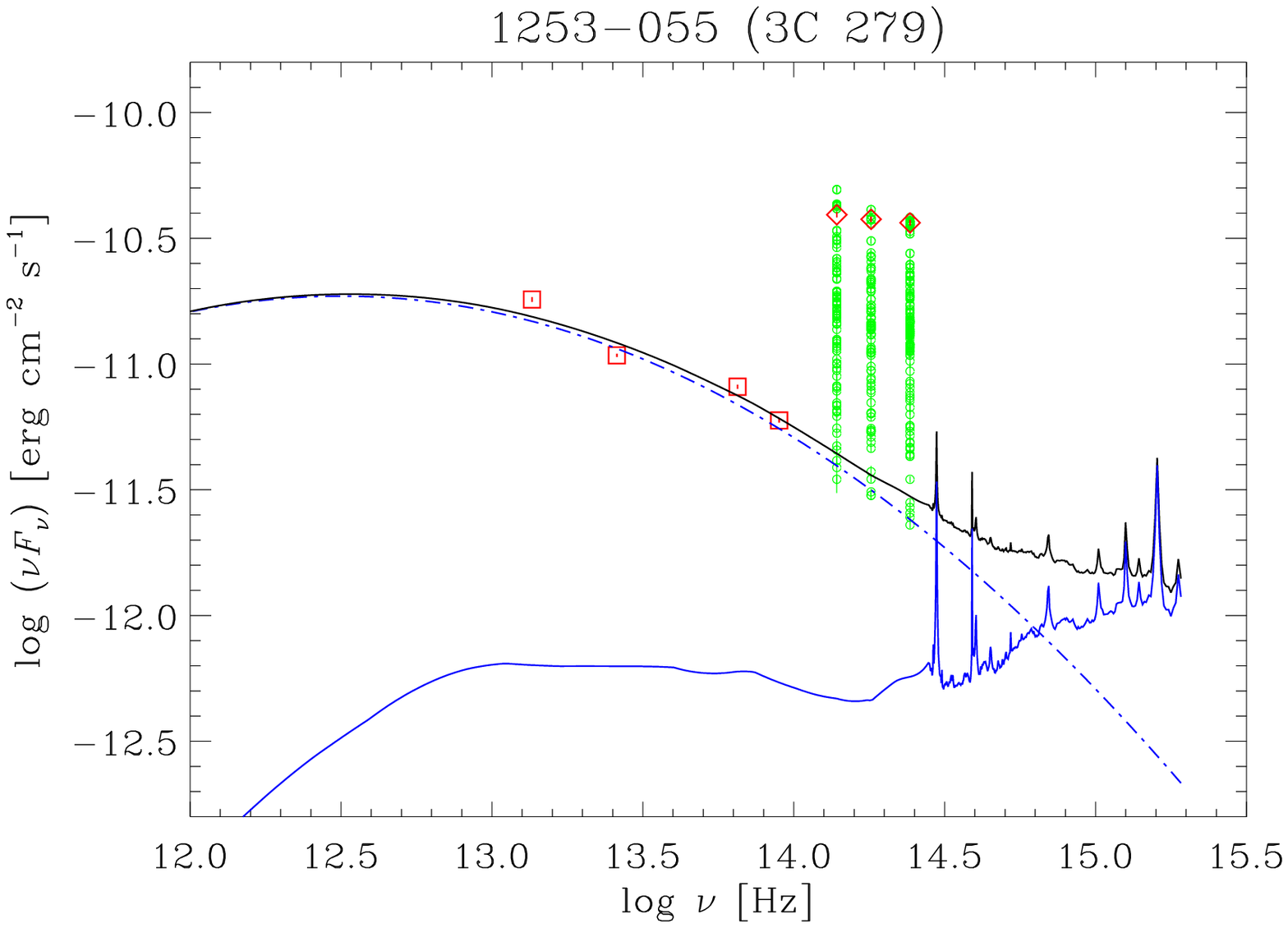,width=0.30\linewidth}
    \psfig{figure=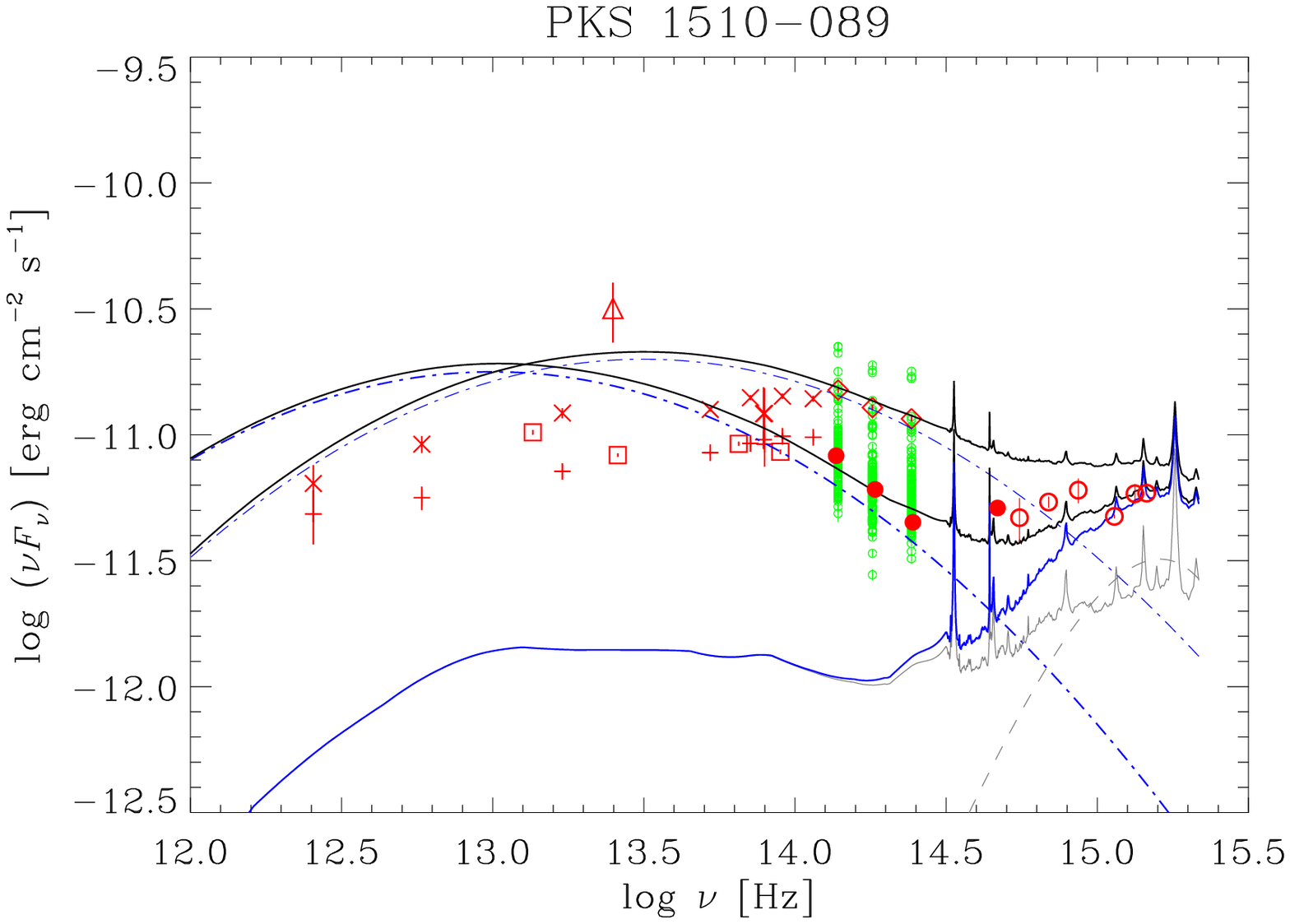,width=0.30\linewidth}
    \psfig{figure=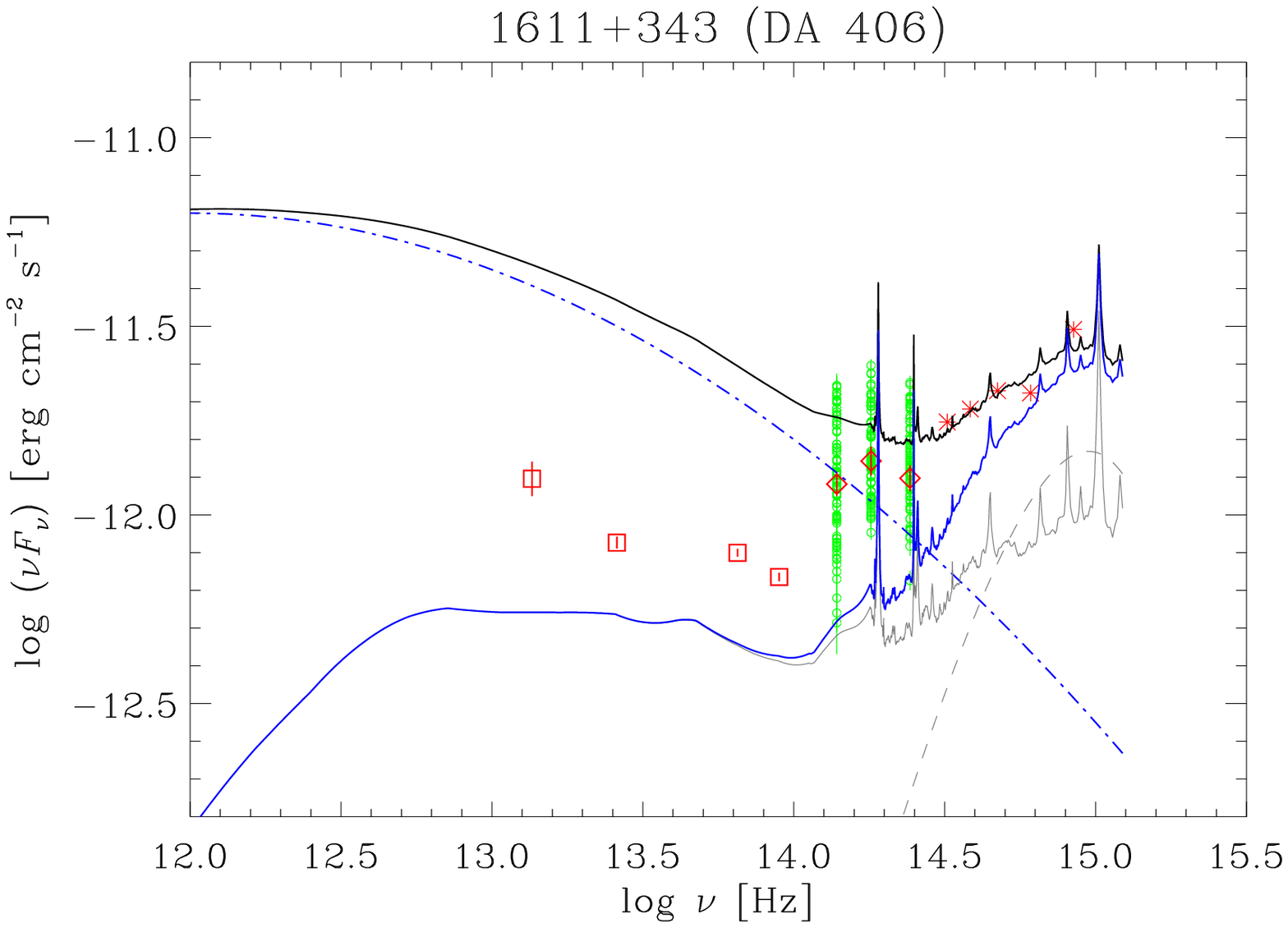,width=0.30\linewidth}}
    \centerline{
    \psfig{figure=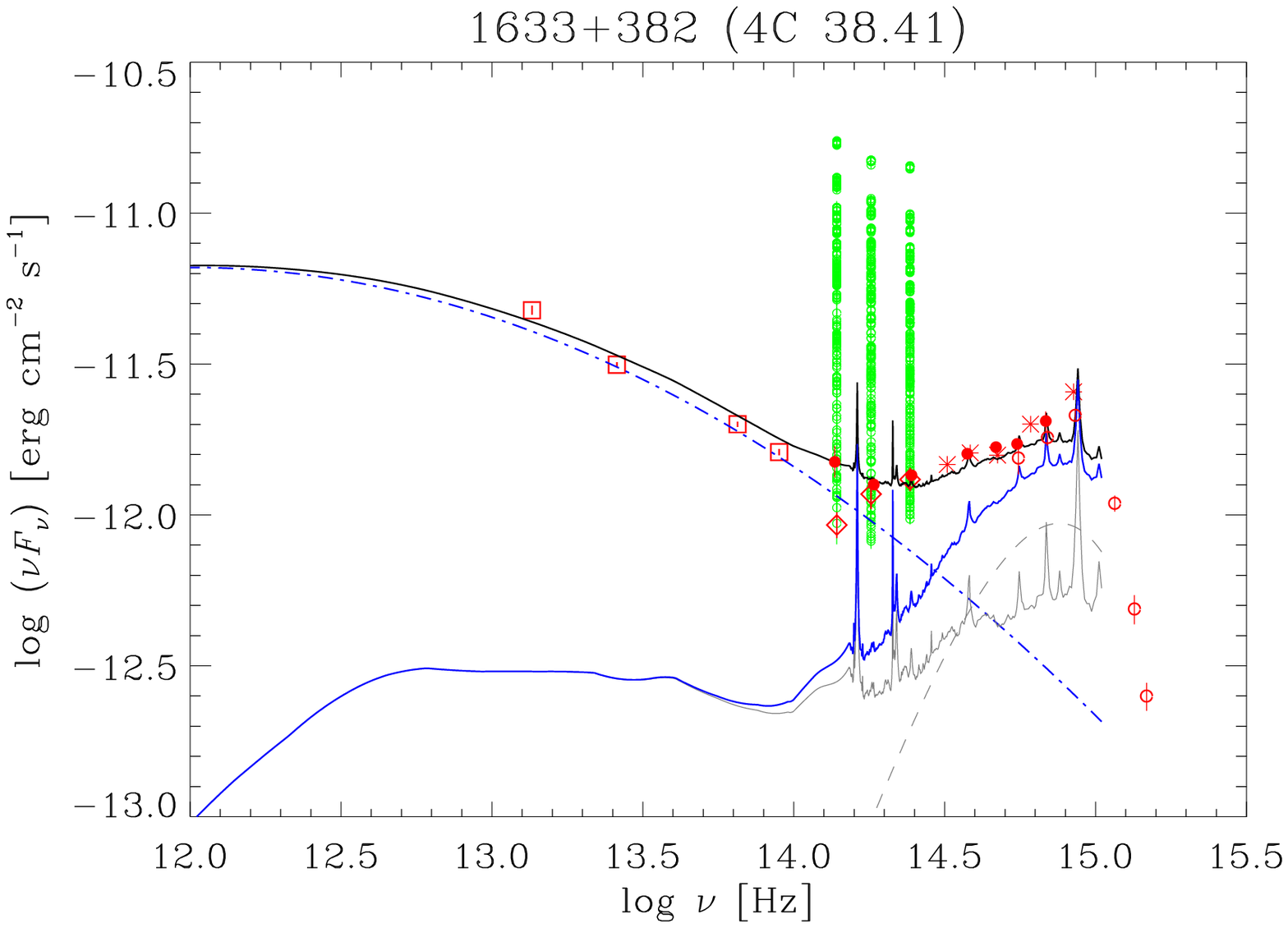,width=0.30\linewidth}
    \psfig{figure=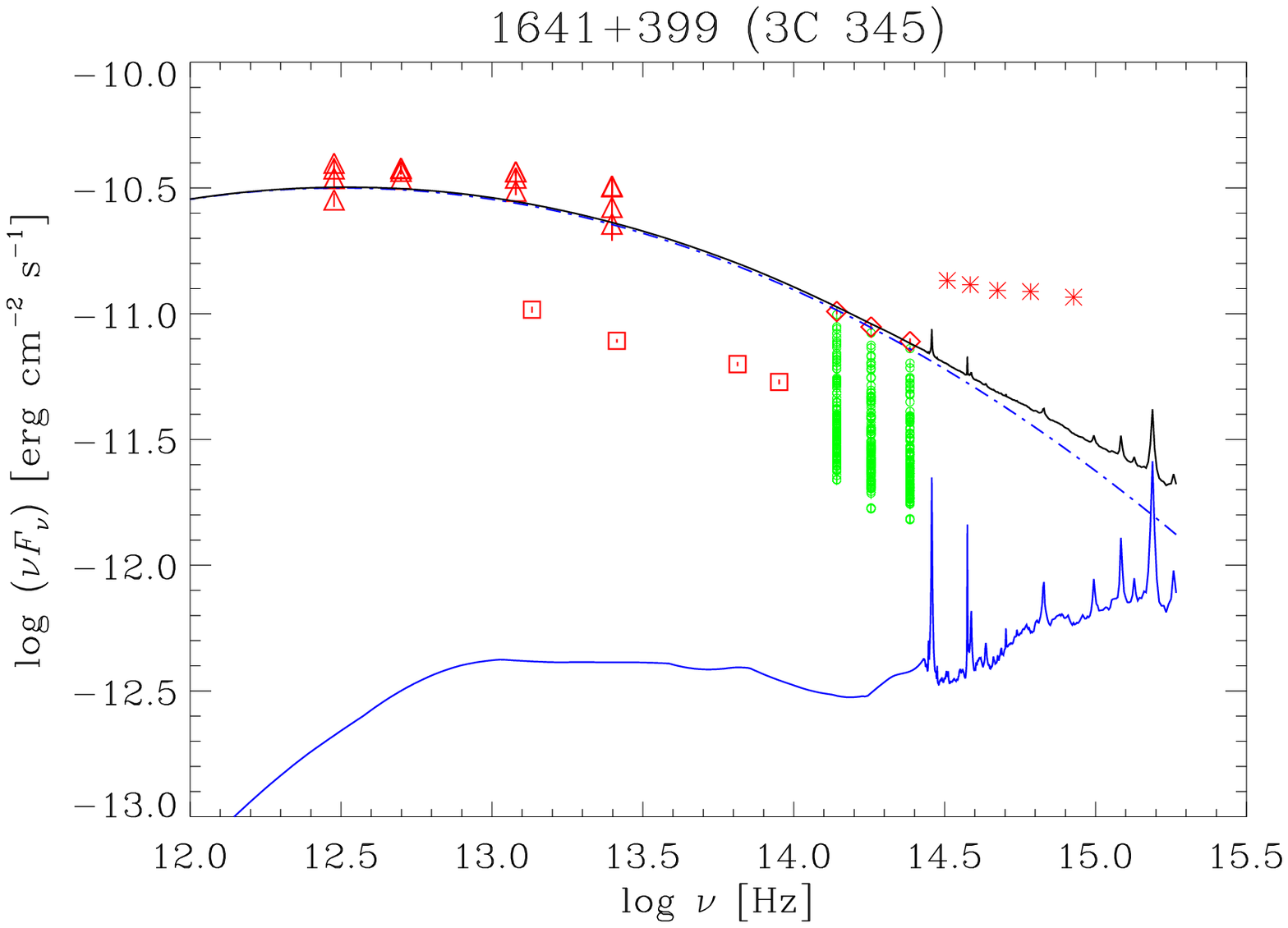,width=0.30\linewidth}
    \psfig{figure=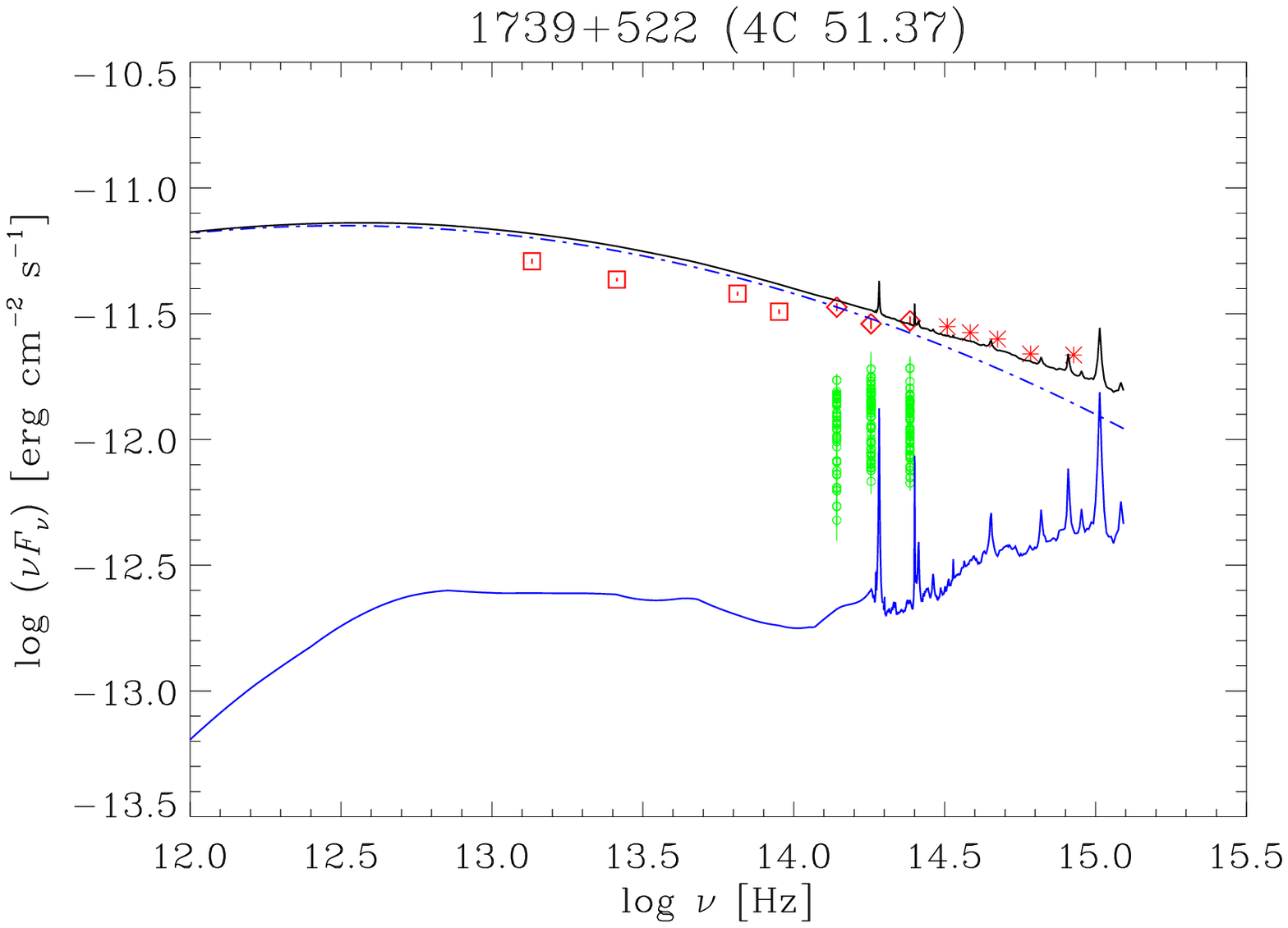,width=0.30\linewidth}}
    \centerline{
    \psfig{figure=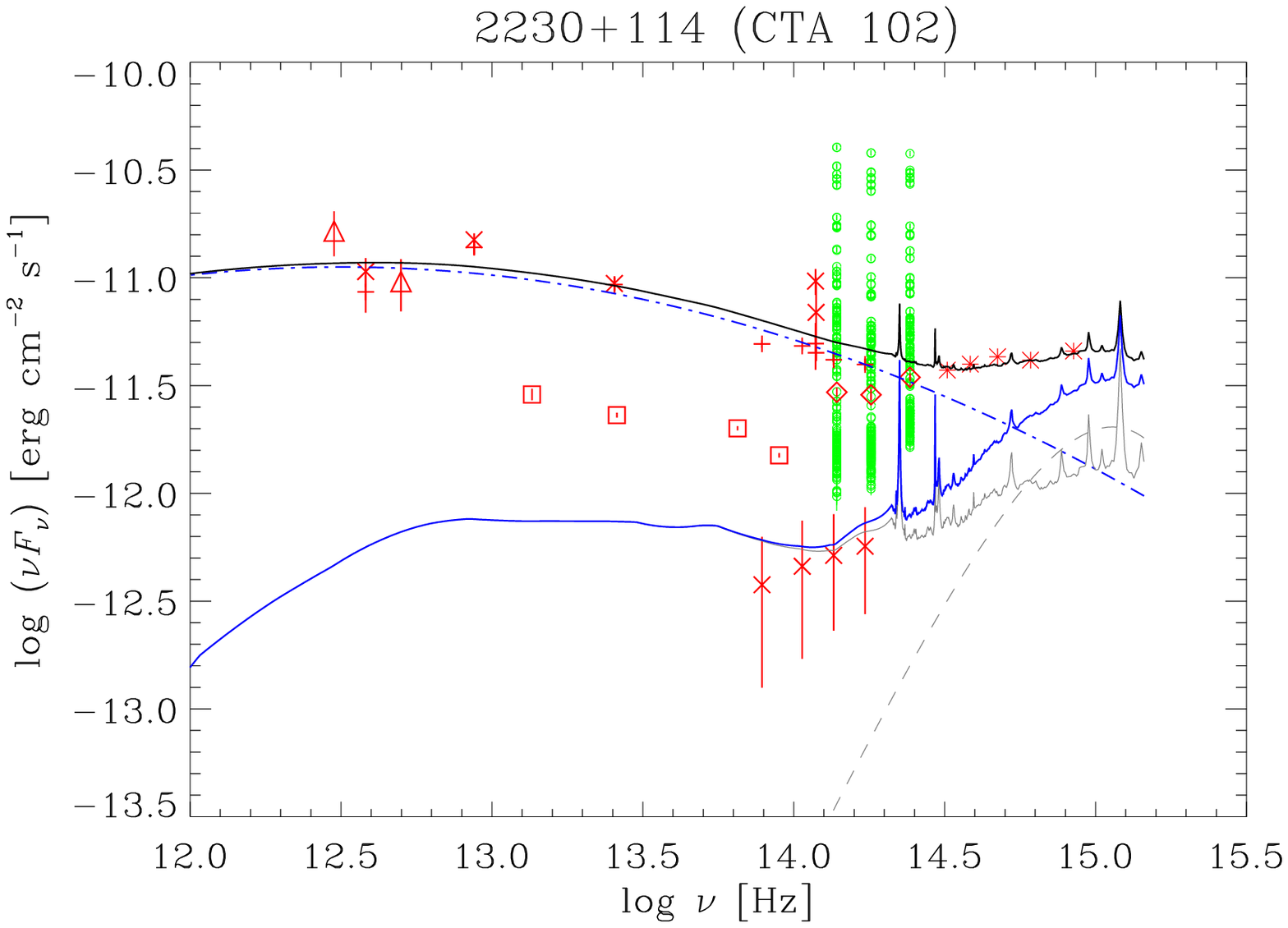,width=0.30\linewidth}
    \psfig{figure=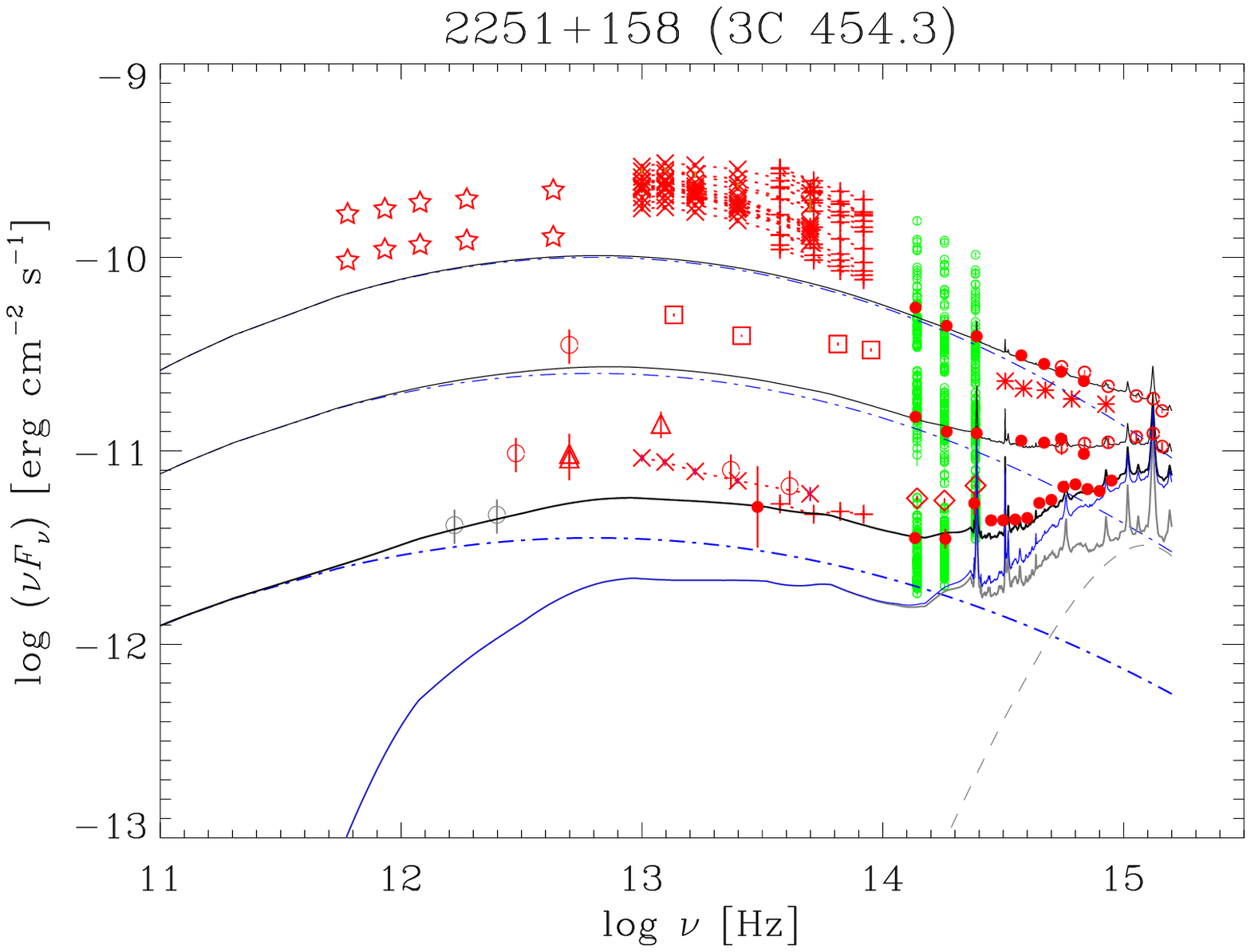,width=0.30\linewidth}}
    \caption{Spectral energy distribution of the GASP-WEBT FSRQs. Data from {\em ISO} (red circles), {\em IRAS} (triangles), {\em Herschel} (stars), {\em Spitzer} (plus signs and crosses), {\em WISE} (squares), 2MASS (diamonds), and SDSS (asterisks) are shown. The green circles represent the results of the GASP near-IR monitoring at the Campo Imperatore and Teide observatories. The optical spectrum of 1226+023 by \citet{but09} as well as near-IR-to-UV SEDs of 1510$-$089, 1633+382, and 2251+158 already published in papers by the GASP-WEBT collaboration are also plotted.
Contributions from the synchrotron (dot-dashed line) and QSO-like (solid line) emission are displayed in blue. The latter is derived from the SWIRE QSO1 templates, but when a brighter disc is required, a black-body component is added. In these cases we show in grey both the QSO1 template (solid line) and the black body (dashed line). The model fit to the 3C 273 SED requires further dust emission, obtained by adding another black-body component (grey dotted line). The black line represents the sum of all components.} 
    \label{sed_fsrqs}    
   \end{figure*}

For this very famous blazar, many infrared data are available, as shown in Fig.\ \ref{sed_fsrqs}. In particular, {\em ISO} observed the source in two epochs: 1996 December 6 (red circles) and 1997 December 18 (grey circles). According to \citet{haa04}, these data may reveal a thermal dust bump, which however is not confirmed by the {\em IRAS} data. {\em Spitzer} spectra are from \citet{ogl11}; the bright state was observed in 2005 July, while the low state was observed in 2006 December. Data from the IRAC detector are plotted as plus signs, those from IRS as crosses; the two instruments were not pointing at the source at the same time. Observations by {\em Herschel} on 2010 November 30 -- December 1 and 2011 January 8 \citep{weh12} are shown as red stars.
To better characterise the interplay between the jet and QSO-like emission, we also plotted three infrared-to-UV SEDs (small filled and empty red circles) taken from \citet{rai11}. 
The highest and intermediate SEDs were built with GASP-WEBT and {\em Swift} data acquired during a multifrequency campaign on this source; the lowest SED is originally from \citet{neu79}.
The QSO-like emission signature is best visible in this faint state, where the flux excess in the $J$/UV$m2$ bands is the signature of the H$\alpha$/Ly$\alpha$ broad emission line \citep[see also][]{rai08c}, and the spectral curvature in the optical band is due to the little blue bump.
The models displayed in Fig.\ \ref{sed_fsrqs} show possible fits to the SEDs, where a jet emission component is superimposed to a QSO-like contribution with enhanced disc emission.

\section{Conclusions}
In this paper we have investigated the near-IR properties of blazars.
The {\em WISE} and 2MASS catalogues allow us to study the statistical properties of the whole class and in particular to investigate the differences between the two blazar subtypes of FSRQs and BL Lac objects. The addition of the SDSS database makes it possible to complete the information on blazar SEDs around the peak of the synchrotron emission component.

However, blazars are characterised by strong variability at all wavelengths and a deeper understanding of the blazar phenomenon can be achieved only through long-term monitoring, which is possible only for a limited number of sources.

We have thus considered the 28 blazars (14 FSRQs and 14 BL Lac objects) that belong to the target list of the GASP project of the WEBT, and for which continuous monitoring was performed since the project birth in 2007. Data in the near-IR come from the Campo Imperatore and Teide observatories.

Among the blazars in the {\em WISE}, 2MASS, and SDSS catalogues, the GASP sources are well distributed in infrared colours and redshifts, but generally have high infrared--optical fluxes, so they represent a mini-sample of bright blazars.

We have shown $JHK$ light curves of these objects, revealing noticeable variability for most of them.
In average, the mean fractional flux variation is greater for FSRQs than for BL Lac objects, with the notable exception of AO 0235+164, which was found to be the most variable object in the considered period. Indeed, this blazar has often shown a behaviour more similar to FSRQs than to BL Lacs. We have compared variability in the three near-IR bands: while in general the BL Lacs show more variability at higher frequency, the reverse is true for FSRQs. This is the consequence of the different emission components overlapping in this band. Colour variability is very small in BL Lac objects, not exceeding a few percent. 
The nearly achromaticity of the near-IR emission of BL Lacs is a characteristic of the dominant synchrotron jet emission. In contrast, the larger spectral changes exhibited by FSRQs are due to the overlapping between synchrotron and quasar-like emission components.

We have built infrared SEDs of these sources, including the results of the GASP-WEBT near-IR monitoring plus archive and literature data from {\em WISE}, 2MASS, {\em IRAS}, {\em ISO}, {\em Spitzer}, and {\em Herschel}, as well as optical information from the SDSS and literature.
Although these SEDs do not contain simultaneous data, as would be desirable for these variable objects, they can nevertheless help us understand the interplay among the various emission contributions: synchrotron radiation from the jet, QSO-like emission including torus, disc, and BLR radiation, and radiation from the host galaxy.
We have modelled the synchrotron emission with a log-parabola, while for the host galaxy and QSO-like contributions we have adopted SWIRE templates, with the possible inclusion of black-body components to simulate additional dust emission or enhanced disc emission.
We have used single temperature black-bodies for these additional components instead of multi-temperature black-bodies or dusty galaxy templates to keep the interpretation as simple as possible.

A strong host galaxy signature has been found in the SED of Mkn 421, and especially in those of Mkn 501 and 1ES 2344+514, the three closest and high-energy peaked BL Lacs, which were the first ones to be detected at TeV energies. At nearly the same redshift of 1ES 2344+514 lies 3C 371, which shows an important host galaxy contribution too. A bit farther, BL Lacertae does not show evidence of host galaxy in its SED, even if we know that the host affects the source photometry in faint states \citep[e.g.][]{rai13}. Apart from the above cases, all other BL Lac SEDs are well fitted by a synchrotron component; the only indications for the presence of dust thermal emission are found for 3C 66A and Mkn 421, as previously reported \citep{imp88}. 

For many FSRQs, the near-IR band signs the transition from a non-thermal, synchrotron-dominated emission to a thermal, QSO-like dominated emission.
The cases of PKS 0528+134 and 4C 71.07, whose {\em WISE} data were acquired in very faint states, suggest that torus emission is relatively weak in FSRQs. A particularly bright disc is required to explain the SEDs of 9 out of the 14 GASP FSRQs. From the SED fits we can derive disc luminosities at the peak of $\log (\nu L_\nu) \sim 44.8$--46.6 [$\rm erg \, s^{-1}$], the highest values characterising 4C 71.07, PKS 0528+134, and 4C 38.41, which are the most distant objects. For comparison, various authors analysing different QSO samples obtained mean values ranging from $\sim 44.9$ to $\sim 45.8$, with large dispersion \citep[see][and references therein]{elv12}. Hence, our FSRQs have accretion discs with luminosities in the same range as QSO, up to the highest values.
In the case of 3C 273, an extra dust contribution peaking at $\log \nu \sim 14$ [Hz] is needed to explain the near-to-medium infrared SED.

\section*{Acknowledgments}
We are grateful to an anonymous referee for useful comments.
This article is partly based on observations made with the telescopes IAC80
and TCS operated by the Instituto de Astrofisica de Canarias in the Spanish
Observatorio del Teide on the island of Tenerife. Most of the observations
were taken under the rutinary observation programme. The IAC team
acknowledges the support from the group of support astronomers and
telescope operators of the Observatorio del Teide.
This work was supported by Russian RFBR foundation grant 12-02-00452. AZT-24 observations are made within an agreement between Pulkovo, Rome and Teramo observatories.
This research has made use of the NASA/ IPAC Infrared Science Archive, which is operated by the Jet Propulsion Laboratory, California Institute of Technology, under contract with the National Aeronautics and Space Administration.
This publication makes use of data products from the Wide-field Infrared Survey Explorer, which is a joint project of the University of California, Los Angeles, and the Jet Propulsion Laboratory/California Institute of Technology, funded by the National Aeronautics and Space Administration.This publication makes use of data products from the Two Micron All Sky Survey, which is a joint project of the University of Massachusetts and the Infrared Processing and Analysis Center/California Institute of Technology, funded by the National Aeronautics and Space Administration and the National Science Foundation.
Funding for SDSS-III has been provided by the Alfred P. Sloan Foundation, the Participating Institutions, the National Science Foundation, and the U.S. Department of Energy Office of Science. The SDSS-III web site is http://www.sdss3.org/.
SDSS-III is managed by the Astrophysical Research Consortium for the Participating Institutions of the SDSS-III Collaboration including the University of Arizona, the Brazilian Participation Group, Brookhaven National Laboratory, University of Cambridge, Carnegie Mellon University, University of Florida, the French Participation Group, the German Participation Group, Harvard University, the Instituto de Astrofisica de Canarias, the Michigan State/Notre Dame/JINA Participation Group, Johns Hopkins University, Lawrence Berkeley National Laboratory, Max Planck Institute for Astrophysics, Max Planck Institute for Extraterrestrial Physics, New Mexico State University, New York University, Ohio State University, Pennsylvania State University, University of Portsmouth, Princeton University, the Spanish Participation Group, University of Tokyo, University of Utah, Vanderbilt University, University of Virginia, University of Washington, and Yale University.

\bsp

\label{lastpage}

\end{document}